%% file: iclr2027_conference.tex
\documentclass{article} 
\usepackage{iclr2027_conference,times}

\input{math_commands.tex}

\usepackage{hyperref}
\usepackage{url}
\usepackage{booktabs}
\usepackage{graphicx}
\usepackage{subcaption}
\usepackage{tabularx}
\usepackage{xcolor}
\usepackage{placeins}
\usepackage{float}
\usepackage{adjustbox}
\usepackage{wrapfig}
\usepackage{wrapfig}
\usepackage{amssymb}
\usepackage{algorithm}
\usepackage{algorithmicx}
\usepackage{algcompatible}
\usepackage{algpseudocode}
\usepackage{capt-of}
\usepackage{longtable}
\usepackage{xspace}
\usepackage{needspace}
\usepackage{multirow}
\usepackage{titletoc}
\usepackage{xcolor}
\usepackage{enumitem}
\usepackage[table]{xcolor}
\usepackage{colortbl}
\definecolor{knitnetblue}{RGB}{230,242,255}
\definecolor{knitnetorange}{HTML}{60C1E5}

\let\cite\citep 

\definecolor{KnitCommentBlue}{RGB}{0,85,160}
\providecommand{\KnitComment}[1]{%
    \Statex {\color{KnitCommentBlue}\itshape\#\ #1}}

\newcommand{\methodrule}{%
    \arrayrulecolor{black!15}%
    \specialrule{0.25pt}{0.35ex}{0.35ex}%
    \arrayrulecolor{black}%
  }

\usepackage{tikz}
  \usetikzlibrary{arrows.meta}

  \makeatletter

  \newcommand{\knit@arrow}{%
      \begingroup
      \edef\knitrowheight{%
          \the\dimexpr\ht\@arstrutbox+\dp\@arstrutbox\relax
      }%
      \raisebox{0.45ex}[0pt][0pt]{%
          \tikz[baseline=0pt]{%
              \draw[
                  -{Stealth[length=2.5pt,width=2.5pt]},
                  line width=0.4pt
              ]
              (0pt,\knitrowheight)
              .. controls
                  (6pt,\knitrowheight)
                  and
                  (6pt,0pt)
              ..
              (0pt,0pt);
          }%
      }%
      \endgroup
  }

  \providecommand{\knitdelta}[1]{}
  \renewcommand{\knitdelta}[1]{%
      \makebox[8pt][l]{\knit@arrow}%
      \makebox[26pt][r]{\ensuremath{#1}}%
  }

  \providecommand{\knitmetric}[2][0]{}
  \renewcommand{\knitmetric}[2][0]{%
      #2%
      \makebox[8pt][r]{%
          \ifnum#1=1\relax
              \knit@arrow
          \fi
      }%
  }

  \makeatother

\title{\centering
KnitNet: Learning to Search at Inference Time for Exact RNA Inverse Design
}

\author{
\begin{minipage}{\textwidth}
    \centering
    \vspace{12pt}
    \normalsize
    \textbf{
        Antonia Panescu$^{1}$ \quad
        Xin Wang$^{1}$ \quad
        Haoran Xu$^{2,1}$
    } \\[4pt]
    \textbf{
        Qingyu Chen$^{1}$ \quad
        Carlos Oliver$^{3}$ \quad
        Rex Ying$^{1}$
    } \\[10pt]
    \normalfont
    $^{1}$Yale University \quad
    $^{2}$Sichuan University \quad
    $^{3}$Vanderbilt University
\end{minipage}
}

\iclrfinalcopy 
\begin{document}

\maketitle

\begin{abstract}
RNA inverse design seeks nucleotide sequences that fold into specified target structures. Existing methods rely on learned generators that can miss exact structural matches, especially beyond their training distributions, or on costly search over a vast, multimodal sequence space. Search is further complicated by globally coupled nucleotide interactions, as correcting one pairing error can create new errors beyond a fixed repair scope. To address this, we introduce KnitNet, an inference-time scaling framework that learns from verifier feedback to refine sequences for exact fold-back matching. Starting from candidates generated for a target secondary structure, KnitNet uses mismatched folds to infer nucleotide dependencies and adapt its joint repair strategy, while an online surrogate learns to predict which edits are promising. This makes search progressively more informed and compute-efficient. Crucially, KnitNet generalizes across RNA structural regimes by changing only the optimization signal to match the verifier required by each setting. Across five secondary-structure benchmarks, KnitNet increases exact solve rates by an average of $21$ percentage points over continued sampling with $70\%$ less time-to-solve. On OpenKnot57 specifically, KnitNet is the first method to achieve exact computational matches for all complex, pseudoknotted structures. Beyond 2D, KnitNet improves 3D structural self-consistency over per-target RL by $7.8\%$ while being $39.3\%$ faster, demonstrating generalization and effective inference-time scaling across RNA design regimes.
  \end{abstract}

\vspace{-4pt}

{\centering
    \small
    \bfseries
    Code:\enspace
    \href{https://github.com/antonia-panescu/KnitNet}
    {\textcolor{knitnetorange}{\texttt{github.com/antonia-panescu/KnitNet}}}
    \\[3pt]
    Correspondence:\enspace
    \href{mailto:antonia.panescu@yale.edu}
    {\textcolor{knitnetorange}{\texttt{antonia.panescu@yale.edu}}}
    \par
}

\vspace{8pt}

\begin{figure}[H]
    \centering
    \includegraphics[
        width=0.9\linewidth,
    ]{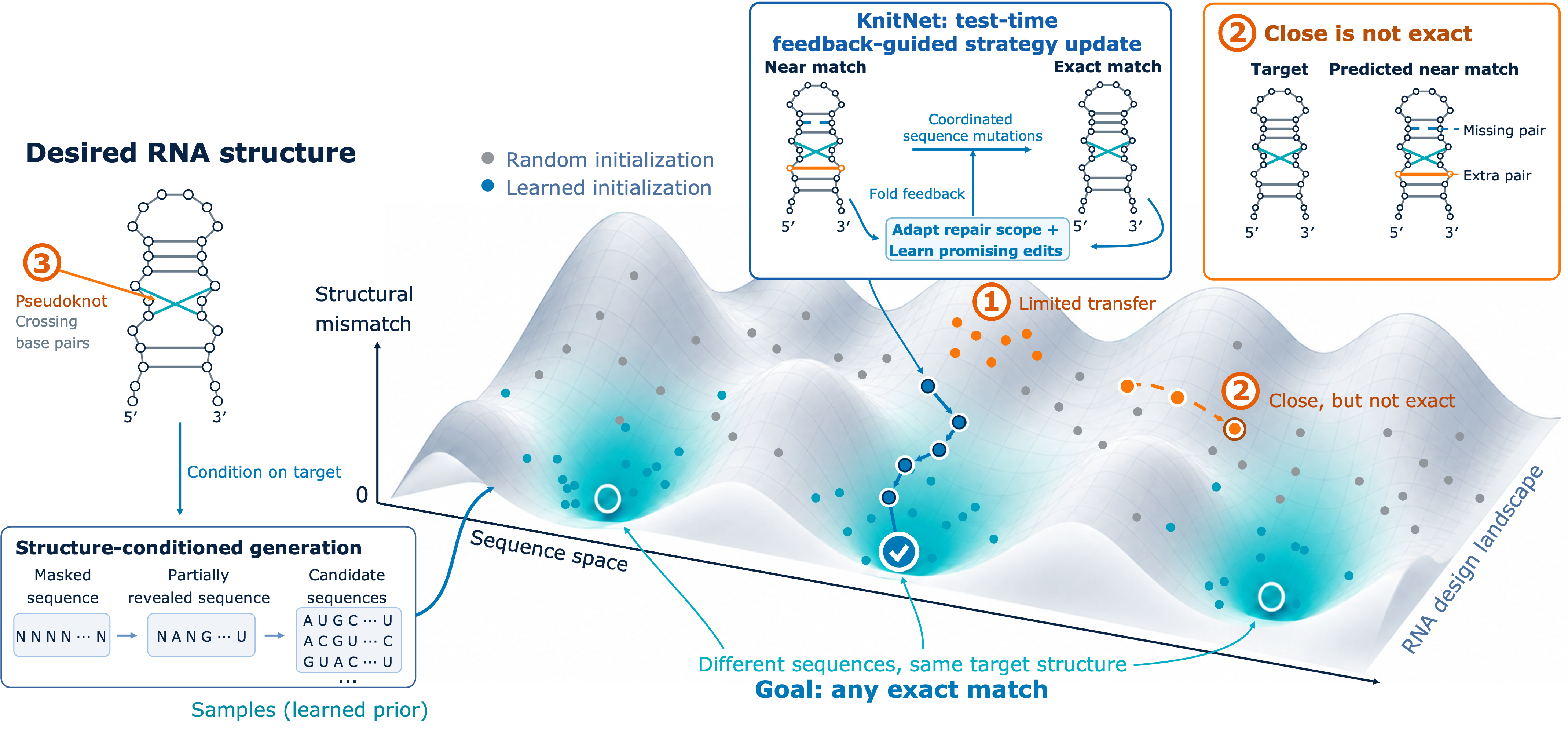}
    \caption{
      \textbf{RNA inverse design in a multimodal sequence landscape.}
      KnitNet turns test-time compute into exact RNA design solutions through feedback-guided learning and coordinated repair, starting from learned rather than random initializations.
      Numbered callouts highlight limitations of existing approaches:
      \textcircled{\scriptsize 1} limited transfer across design tasks
      and folding verifiers;
      \textcircled{\scriptsize 2} search that stalls when repairing
      one interaction disrupts another; and
      \textcircled{\scriptsize 3} restrictions to pseudoknot-free structures.
    }
    \label{fig:introduction}
    \vspace{-8pt}
\end{figure}

\section{Introduction}
\label{sec:introduction}

Designing RNA sequences that reliably match a desired structure remains challenging in RNA engineering. Controlling RNA structure could enable realizing desired biological functions, such as programmable regulation of translation or improving the stability and immunogenicity of mRNA vaccines~\cite{green2014toehold,zhang2023lineardesign}. While structure-conditioned sequence generation has become a powerful tool in protein engineering~\cite{dauparas2022proteinmpnn}, extending it to RNA three-dimensional (3D) design faces limitations in experimental data and computational validation~\cite{nithin2024rna3d}. Secondary structure (2D) offers a complementary, experimentally actionable design layer with substantially greater data coverage ~\cite{rnacentral2021,waymentsteele2022eternafold,he2024ribonanza} and models that support reliable computational fold-back evaluation~\cite{he2024ribonanza,lorenz2011viennarna}. Together, these properties make secondary structure a biologically meaningful and computationally testable setting for RNA inverse design: \emph{given a target 2D pairing pattern, design a sequence whose predicted fold matches it exactly}.

Given this setting, the central algorithmic challenge is to efficiently find exact solutions in a vast, multimodal sequence space (Figure~\ref{fig:introduction}). Supervised RNA generators learn sequence reconstruction given a target structure, but do not optimize exact fold-back correctness~\cite{patil2024rnainformer}, which could be satisfied by non-native sequences as well. While reinforcement learning can incorporate this folding feedback~\cite{he2026struct2seq}, it specializes policies to the training verifier and data distribution, which limits transfer to substantially different design tasks ~\cite{panescu2026orderagnostic}(Figure~\ref{fig:introduction}\,\textcircled{\scriptsize 1}). Inference-time search avoids retraining by adapting candidates directly to each target and verifier. However, repeated folding evaluations are costly, and greedy or narrowly scoped repairs can stall when correcting one interaction disrupts another~\cite{zhou2023samfeo}(Figure~\ref{fig:introduction}\,\textcircled{\scriptsize 2}). Adaptive methods coordinate repairs using structural feedback, but their thermodynamic formulations exclude pseudoknots~\cite{garciamartin2016rnaifold2t,zhou2026fastdesign, kikuchi2026fmqa}, leaving biologically important motifs, such as the CRISPR-Cas12a guide-RNA pseudoknot, outside their scope~\cite{ageely2021pseudoknot}(Figure~\ref{fig:introduction}\,\textcircled{\scriptsize 3}). Together, these limitations motivate a general RNA design framework that adapts to new targets and verifiers without retraining, uses folding feedback efficiently to resolve coupled structural errors, and supports both planar and pseudoknotted structures. 

To this end, we introduce \textbf{KnitNet}, a two-stage RNA inverse-design framework that learns from folding feedback at inference time to adapt its search and efficiently find sequences whose predicted structures match the target exactly. \textbf{Our main contribution} is \emph{counterexample-guided joint repair}~\textbf{(1)}, which derives an error graph from observed base-pairing errors to \textbf{infer nucleotide interdependencies and expand the editable scope for joint optimization}, overcoming the limitation of local repair. As new errors emerge, KnitNet also revises the repair priorities, while an online surrogate learns which joint mutations are promising and limits expensive folding evaluations to those, making search \textbf{compute-efficient}. Crucially, the surrogate is refitted as the repair problem evolves, while also retaining evidence from earlier evaluations. To initialize this repair process, we develop a diffusion generator~\textbf{(2)} conditioned on a pairing-graph representation of RNA 2D structure, which we construct to support both planar targets and \textbf{pseudoknots}. Pretraining on more than 20 million structure-sequence pairs equips the generator with a reusable biological prior, which the repair procedure complements through target-specific inference-time learning. At the framework level, we make this search \textbf{transferable} across generators, design regimes, and objectives~\textbf{(3)} by separating pairing-error-guided repair from verifier-specific candidate ranking. 

Across inverse design tasks, we benchmark KnitNet against Order-Agnostic Struct2SeQ~\cite{panescu2026orderagnostic}, the SOTA 2D-conditioned generator, alongside specialized methods for each setting (Appendix~\ref{app:related_work}). On OpenKnot57, a challenging pseudoknot benchmark where previous evaluations left targets unsolved despite large sampling budgets~\cite{he2026struct2seq,townley2026denovo}, KnitNet achieves $57/57$ exact computational matches in individual runs, averaging $56.3/57$ over three seeds versus $49.7/57$ for Order-Agnostic Struct2SeQ. On challenging planar targets, an extended KnitNet search campaign achieves a best-in-class coverage of $97/100$ on Eterna100-v2 (Appendix~\ref{app:eterna_extended}), highlighting its broad capabilities across RNA structural regimes. To further demonstrate the value of allocating inference-time computation to feedback-guided learning and repair, we show KnitNet Search achieves higher exact solve rates in $70\%$ less time-to-solve on average for all 20 generator-benchmark combinations in Figure~\ref{fig:inference_scaling} and Appendix~\ref{app:scaling_eterna}. By applying the same KnitNet procedure to design PDB-derived RNA structures, we improve mean 3D fold-back self-consistency over specialized 3D-conditioned generators, showing that our method can generalize to 3D use-cases, and could help alleviate the data limitations of 3D modeling. What's more, KnitNet search outperforms target-specific RL~\cite{hu_rider_2026}, the previously leading 3D method, while being 39.3\% faster. 

\FloatBarrier

\section{The KnitNet Pipeline}
\label{sec:knitnet_pipeline}
\begin{figure}[H]
    \centering
    \includegraphics[width=\linewidth]{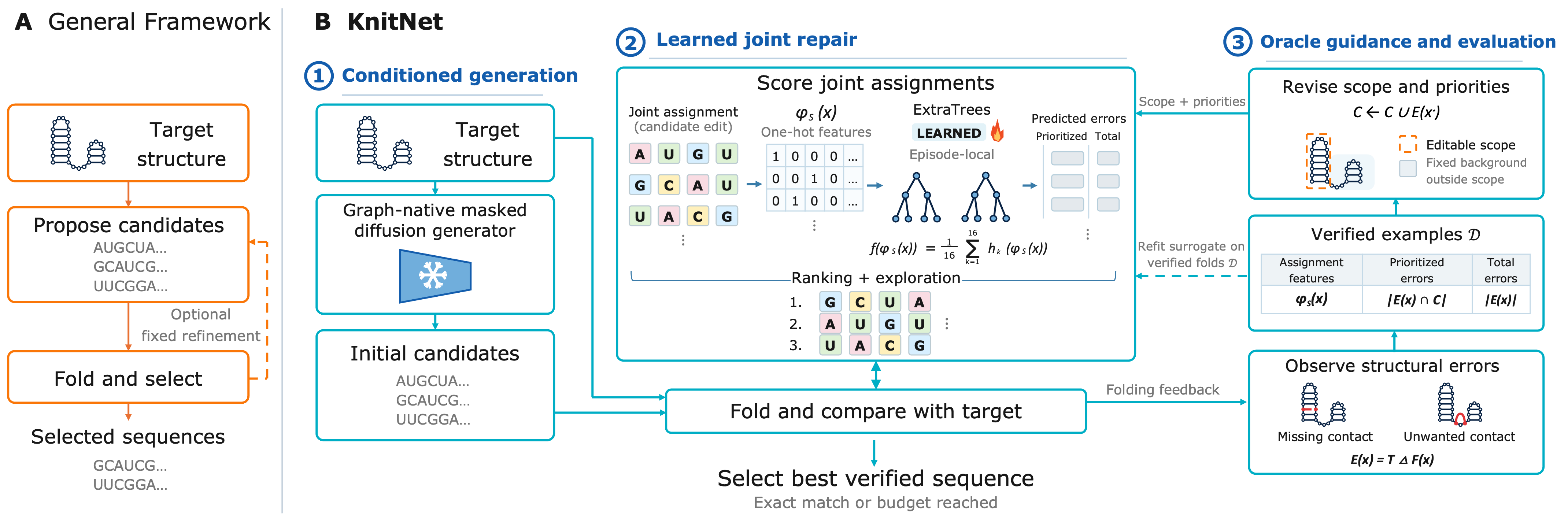}
    \caption{\textbf{(A) General RNA inverse design} proposes sequences, evaluates their predicted folds, and selects candidates against a target structure. \textbf{(B) KnitNet} combines \textbf{(1)} initialization with a frozen structure-conditioned generator, \textbf{(2)} coordinated nucleotide edits ranked by a local surrogate, and \textbf{(3)} folding feedback that updates the surrogate and revises editable positions and error priorities. This loop adapts subsequent repairs to observed folding errors. Search returns an exact predicted secondary-structure match when found, or the best evaluated sequence when the budget is exhausted.}
    \label{fig:method_details}
\end{figure}

In this section, we describe KnitNet, an RNA inverse-design framework that combines structure-conditioned sequence generation with adaptive inference-time repair, learning from folding feedback to guide coordinated nucleotide changes so that the predicted fold matches the input structure. For a target of length $L$, we represent the secondary structure as a pair set $T\subseteq\{(i,j):1\le i<j\le L\}$. Each nucleotide belongs to at most one target pair, and crossing pairs are permitted (i.e. pseudoknots). Over the RNA alphabet $\mathcal A=\{\mathrm{A},\mathrm{C},\mathrm{G},\mathrm{U}\}$, KnitNet seeks a candidate sequence $x\in\mathcal A^{L}$, with nucleotide $x_i$ at position $i$. Its predicted pair set $F(x)$ must exactly match the target, $F(x)=T$. Figure~\ref{fig:method_details} summarizes the KnitNet pipeline. First, a frozen structure-conditioned masked diffusion model generates a diverse set of candidate sequences ((Figure~\ref{fig:method_details}B\,\textcircled{\scriptsize 1}) (Section~\ref{sec:knitnet_pretraining}, Appendix \ref{app:output_characterization},~\ref{app:generator_details}). Then, the population search refines these candidates through contextual joint repair: an online local surrogate ranks coordinated nucleotide assignments (Figure~\ref{fig:method_details}B\,\textcircled{\scriptsize 2})(Section~\ref{sec:knitnet_search}), while verifier-specific scores provide graded search guidance (Figure~\ref{fig:method_details}B\,\textcircled{\scriptsize 3}). Exact agreement with the complete target pair set defines success (Section~\ref{sec:knitnet_metrics}). Appendix~\ref{app:notation} lists the notation.

\subsection{Generating initial candidates conditioned on the target structure}
\label{sec:knitnet_pretraining}

To generate the initial pool of candidate sequences, we use a GraphGPS ~\cite{rampasek2022graphgps} denoiser conditioned on a graph representation of the target RNA secondary structure. This enables combining local message passing with global attention to model RNA motif-level interactions alongside long-range dependencies between distant structural regions. We pretrain with a masked diffusion language modeling (MDLM) objective~\cite{sahoo2024mdlm} on a structure-sequence paired dataset we construct by applying the RibonanzaNet-SS 2D structure predictor \citep{he2024ribonanza} to the RNACentral sequence corpus \citep{rnacentral2021}. We choose this objective because masked diffusion allows each nucleotide prediction to use revealed sequence context on both sides, helping coordinate assignments across RNA’s coupled pairing sites. In contrast, left-to-right autoregressive designers cannot condition on subsequent sequence assignments~\cite{patil2024rnainformer,joshi2025grnade}. We detail dataset and split construction in Appendix~\ref{app:training-data}. Appendix~\ref{app:generator_details} details graph features and architecture, training objectives and hyperparameters, checkpoint selection, sampling, and training compute.

At inference time, we sample 96 candidates from the learned conditional distribution. For secondary-structure design, we fold each candidate with the selected oracle (RibonanzaNet-SS or ViennaRNA, detailed in Appendix~\ref{app:oracle_evaluation}) and compare its predicted base-pair set $F(x)$ with the target $T$. If any candidate satisfies $F(x)=T$, the target is recorded as solved; otherwise, the evaluated candidates initialize the search described in Section~\ref{sec:knitnet_search}. The generator remains frozen during subsequent stages. For 3D inverse design, KnitNet generates the 96-candidate pool conditioned on the target's DSSR-converted pair graph, and RhoFold+-based tertiary feedback guides subsequent search (Appendix~\ref{app:3d_protocol}).

\subsection{Learning to repair from folding counterexamples}
\label{sec:knitnet_search}

We introduce \emph{counterexample-guided joint repair}, an adaptive inference-time search that learns from folding feedback how to modify nucleotides together to produce sequences that satisfy the coupled pairing constraints of a target structure. This addresses a key challenge in RNA sequence design: interactions between distant nucleotides mean that a local, fixed edit can resolve one pairing error while introducing another elsewhere in the fold. Our contribution couples three adaptive decisions: (1) which interacting nucleotides to optimize jointly, (2) which coordinated edits merit expensive folding calls according to a learned evaluation model, and (3) which errors subsequent repairs should prioritize as new folding feedback becomes available. We call this \emph{counterexample-guided} because errors exposed by candidate folds revise the repair problem used in subsequent iterations. This feedback loop learns and adapts repairs to the target and sequence context at inference time. Algorithm~\ref{alg:knitnet_repair} summarizes this procedure.

\paragraph{(1) Constructing a contextual repair problem.} We identify a repair problem by folding a complete candidate sequence $x$ and comparing its predicted base-pair set $F(x)$ with the target $T$. A target pair absent from $F(x)$ is a \emph{missing pair}, and a predicted pair absent from $T$ is an \emph{unwanted pair}. Together, these discrepancies form the error set $E(x)=T\triangle F(x)$. If $E(x)=\varnothing$, the candidate solves the target (line~\ref{ln:knitnet:success}). Otherwise, the errors identify what is wrong, but do not themselves reveal which nucleotide changes will correct it. The population $\mathcal P$ retains candidate sequences across repair episodes. Each episode selects an anchor $a\in\mathcal P$ with unresolved errors $E(a)\neq\varnothing$. The target defines the \emph{assignment-unit set} $\mathcal V_T=\{\{i,j\}:(i,j)\in T\}\cup\{\{i\}:1\le i\le L,\ i\notin\bigcup_{(j,k)\in T}\{j,k\}\}$, consisting of target pairs and unpaired positions. Pair substitutions use the pair-assignment alphabet $\mathcal A_{\mathrm{pair}}=
  \{\mathrm{AU},\mathrm{UA},\mathrm{GC},\mathrm{CG},\mathrm{GU},\mathrm{UG}\}\subset\mathcal A^{2}$, read as $(x_i,x_j)$ for the pair $(i,j)$,
and unpaired-position substitutions use $\mathcal A$. Pair substitutions preserve nucleotide compatibility~\cite{runge2019learna,tang2026samplingdesign}. We check whether the pair forms by folding the modified sequence.

To identify which assignment units to repair together, we construct an \emph{error graph} $\mathcal G_a=(\mathcal V_T,\mathcal E_a)$, whose edges connect the units containing the endpoints of observed folding errors. For example, if a nucleotide pairs with an unintended partner from another target pair, the two assignment units become linked, suggesting that they may require coordinated repair. We include errors from the anchor and its parent lineage to retain evidence of interactions that may disappear and recur during search. The connected component containing the most recurrent current error initializes the editable \emph{scope} $S\subseteq\mathcal V_T$. For a singleton component, we add a nearest unit when one is available (Appendix~\ref{app:repair_details}).

A \emph{joint assignment} $z$ specifies nucleotide values for the units in $S$, producing a candidate sequence $x\in\mathcal A^{L}$ that matches the anchor $a$ outside those units. Pair values introduced by substitution or random sampling come from $\mathcal A_{\mathrm{pair}}$. Beam proposals may retain or inherit incompatible observed pair assignments. Keeping the surrounding sequence unchanged lets us compare the effects of different nucleotide assignments within $S$. We initialize the prioritized error set $C\subseteq\{(i,j):1\le i<j\le L\}$ with anchor errors having at least one endpoint in a unit of $S$ (line~\ref{ln:knitnet:scope}). Using the oracle merit $m_F(x)$ in Section~\ref{sec:knitnet_metrics}, we rank repairs by the lexicographic loss
\begin{equation}
    \ell_C(x)=
    \left(|E(x)\cap C|,\;|E(x)|,\;-m_F(x)\right)\in\mathbb{Z}_{\geq0}^{2}\times\mathbb{R}.
    \label{eq:knitnet_repair_loss}
\end{equation}
Candidates with fewer errors in $C$ rank first. Ties are broken by fewer total folding errors and then higher oracle merit. This prioritizes the diagnosed conflict while retaining a comparison of the complete folds.

\paragraph{(2) Learning to predict joint repair quality.} Evaluating every proposed combination of nucleotide changes would require expensive folding calls. We therefore learn a local surrogate that predicts how many prioritized and total pairing errors would remain after a joint edit, allowing us to rank proposals before selecting candidates for folding evaluation. We implement this with a small ExtraTrees ensemble~\cite{geurts2006extratrees}. Successive splits allow each tree to condition its prediction on choices at multiple assignment units, representing how the effect of one change depends on others. This matches the coupled nature of repair: a combination of changes can be useful even when its individual changes are not. A simpler alternative, such as an additive linear surrogate, would lack these interactions between units, which we show boost performance in Table~\ref{tab:pat4-ablations-rc500}.

We fit the surrogate within each repair episode to the current sequence context and repair priorities. The evaluated sequence set $\mathcal D\subseteq\mathcal A^{L}$ contains the anchor and all evaluated joint proposals from that episode, with their folding outputs cached. The feature vector $\phi_S(x)$ concatenates one-hot encodings of the assignments in $S$, with seven entries per pair unit and five per unpaired unit, counting one extra category per unit (Appendix~\ref{app:repair_details}). Each tree $h_k$ is fitted on $\mathcal D$ to predict the prioritized and total error counts, $(|E(x)\cap C|,\;|E(x)|)$. The local surrogate $f$ averages predictions from $K$ trees:
\begin{equation}
    f(\phi_S(x))
    =\frac{1}{K}\sum_{k=1}^{K}h_k(\phi_S(x))
    \in\mathbb{R}^{2}.
    \label{eq:knitnet_repair_surrogate}
\end{equation}
We use the two outputs to rank proposals lexicographically (lines~\ref{ln:knitnet:fit}-\ref{ln:knitnet:select}). We use $K=16$ trees in the original experiments to limit refitting costs as the repair scope and error priorities evolve.
  For the component ablations in
  Table~\ref{tab:pat4-ablations-external}, we use $K=64$ across all ExtraTrees variants, following improved Eterna100-v2 recovery with the larger ensemble.

To build on promising repairs while preserving alternative combinations, we maintain a proposal beam $\mathcal H\subseteq\mathcal D$, $|\mathcal H|\le 8$, selected for repair quality and assignment diversity. Its members can seed proposals even when rejected by the outer population $\mathcal P$, and all evaluated joint proposals remain available for surrogate fitting (lines~\ref{ln:knitnet:revision}--\ref{ln:knitnet:relabel}). When the product of the substitution alphabets is small, we enumerate its assignments. Otherwise, we generate candidates through mutation and crossover of beam members and random assignments (line~\ref{ln:knitnet:fit}). Because surrogate predictions are imperfect, we select proposals using both predicted lexicographic rank and uniform exploration, sampling uniformly until enough observations exist to fit the surrogate (line~\ref{ln:knitnet:select}). We also evaluate single-unit mutations from $\mathcal P$ to allow simple corrections across the population. These candidates are excluded from $\mathcal D$ because they may differ from the anchor outside the units in $S$. We fold selected candidates to check for exact solutions and record residual errors (lines~\ref{ln:knitnet:verify} and~\ref{ln:knitnet:success}).

\paragraph{(3) Refining the repair problem through counterexamples.} Resolving the prioritized errors may leave other errors unresolved or introduce new ones, whose repair may require changes outside the current scope (Figure~\ref{fig:pat4_dynamics}). These residual errors serve as \emph{counterexamples} to the current repair attempt, guiding updates to both repair priorities and the editable scope. For a nonempty evaluated joint batch $\mathcal J$, we choose the best candidate $x^\star\in\arg\min_{x\in\mathcal J}\ell_C(x)$ under the lexicographic ranking in Equation~\ref{eq:knitnet_repair_loss} (line~\ref{ln:knitnet:witness}) and update the prioritized set as $C\leftarrow C\cup E(x^\star)$ (line~\ref{ln:knitnet:revision}). This promotes errors already counted in the total fold error into the primary repair priority while preserving earlier priorities. Before refitting the surrogate, we relabel all observations in $\mathcal D$ under the updated $C$, using cached folds without additional oracle calls (line~\ref{ln:knitnet:relabel}).

We also expand $S$ when $x^\star$ resolves the previously prioritized errors but still misfolds, when search stalls, or when the proposal pool is exhausted, subject to the episode budget (line~\ref{ln:knitnet:expand}). We add assignment units whose nucleotides appear most often in residual errors across the beam $\mathcal H$, breaking ties by proximity along the sequence to the current scope. Previously evaluated sequences retain the anchor's nucleotides at newly editable positions, so we can re-encode them for the expanded scope and reuse their cached folds. At episode boundaries, we rotate anchors to explore different sequence contexts. Appendix~\ref{app:repair_details} specifies expansion rules, episode limits, and remaining search policies.

\definecolor{KnitCommentBlue}{RGB}{0,85,160}
\providecommand{\KnitComment}[1]{%
    \Statex {\color{KnitCommentBlue}\itshape\#\ #1}}

\subsection{Oracle guidance and evaluation}
\label{sec:knitnet_metrics}

Applying the search algorithm described above across different folding oracles requires a common way to translate their predictions into search guidance. We therefore use verifier outputs in two complementary ways: to identify which target constraints remain unsatisfied and to score candidate sequences for search. Pairing discrepancies between a candidate's predicted fold and the target structure guide updates to repair priorities and scope, while an oracle-specific score $m_F:\mathcal A^{L}\to\mathbb R$ ranks the population $\mathcal P$ and breaks ties between repairs with equal prioritized and total error counts. The score $m_F(x)$ depends on the design setting and its associated verifier. Section~\ref{sec:results} describes these setting-specific choices, with full definitions in Appendix~\ref{app:oracle_evaluation}.

\section{Experiments}
\label{sec:results}

We demonstrate KnitNet’s design performance and computational efficiency across pseudoknotted, planar, and 3D RNA inverse design tasks. KnitNet search improves designs from diverse generators, while ablations establish the complementary benefits of learned initialization and search. Finally, search converts additional inference time into exact solutions more effectively than continued sampling. We analyze how learned, counterexample-guided joint repair contributes to this gain.

\textbf{RNA inverse design tasks and datasets.} To assess KnitNet's general applicability, we evaluate three settings that challenge different aspects of RNA design. \emph{(1) Pseudoknotted design} requires resolving complex crossing interactions in functional RNA motifs. We test this capability on OpenKnot57, where previous methods left targets unsolved~\cite{townley2026denovo}, and the larger PseudoBase++ benchmark~\cite{taufer2009pseudobase}. \emph{(2) Planar design} tests
transfer to the established thermodynamic setting, where specialized search methods tackle challenging human-designed puzzles. We evaluate Eterna100-v2 from the Eterna community challenge~\cite{andersonlee2016eterna100}, alongside Rfam-Learn-Test and Rfam-Taneda to cover naturally derived structures~\cite{runge2019learna,taneda2011modena}. \emph{(3) 3D design} tests whether our framework, developed for the 2D regime, can transfer to 3D geometry restraints. We use 64 RNA monomers from the gRNAde Das test split~\cite{joshi_generative_2025}, matching the fold-back verifier's single-chain scope (Appendix~\ref{app:dataset_statistics}). We ensure that our diffusion model's training corpus has no sequence-homologous overlap with any evaluation benchmark (Appendix~\ref{app:training-data}). We further select 500 challenging held-out structures from this corpus, including pseudoknots and RNAs longer than 500 nucleotides (Appendix~\ref{app:dataset_statistics}), for component ablations. 

\textbf{Experimental setup.}
To separate generation quality from search effectiveness, we first evaluate KnitNet and off-the-shelf generators using best-of-96 candidates per target, alongside existing search methods with their native initialization and folding-oracle guidance. We then test both stages of KnitNet independently: applying our search to other generators' candidates, as well as initializing existing search methods with KnitNet candidates. Post-initialization budgets are 30 minutes per target for pseudoknotted design, 60 minutes for planar design and ablations, and up to 2,048 logical evaluations for 3D design, in addition to the initial 96 evaluations. To guide search under different folding objectives, we use RibonanzaNet-SS pairing margins, ViennaRNA target ensemble probabilities, and a RhoFold+ GDT/RMSD reward for pseudoknotted, planar, and 3D design, respectively (Appendix~\ref{app:oracle_evaluation}). In 3D search, secondary-structure errors continue to guide repair, while the tertiary reward guides candidate ranking. The main secondary-structure tables report full-benchmark exact coverage and normalized wall-time solve AUC, measuring final recovery and how early solutions are found, as means and sample standard deviations over three seeds. Additional metrics, including Jaccard, MCC, and capped time to solve, appear in Appendix Tables~\ref{tab:h130-full-metrics} and~\ref{tab:h131-full-metrics}. For 3D design, within each seed we report scTM, GDT$_{\mathrm{TS}}$, and RMSD for the highest-scTM candidate in each target's final 96-candidate pool. Table values are the mean and sample standard deviation of the 64-target averages over the same three seeds. Appendix~\ref{app:3d_protocol} details conditioning inputs, scoring, and search/RL budget accounting.

\textbf{Baselines.}
Because existing RNA design methods specialize in different structural inputs and folding objectives, we select generator and search baselines appropriate to each setting, for example aRNAque for pseudoknotted design~\cite{merleau2022arnaque}, RIFT-VAE for planar design~\cite{watanabe2026riftvae}, and tertiary-conditioned generators alongside RIDER's target-specific reinforcement learning for 3D design~\cite{hu_rider_2026}. Order-Agnostic Struct2SeQ~\cite{panescu2026orderagnostic}, a recent SOTA method, serves as a shared comparator across all three settings. Appendix~\ref{app:related_work} provides background on the complete baseline catalogue. Appendices~\ref{app:full-PK} and \ref{app:full-planar} show implementation details.

\subsection{Results}

\textbf{KnitNet increases exact solve rates on pseudoknotted RNA targets.}
KnitNet achieves $57/57$ exact computational fold-back matches on OpenKnot57 in individual runs and is the first method to find exact solutions for all targets in this benchmark. Over three seeds, KnitNet averages $56.3/57$ on OpenKnot57 and $335.7/339$ on PseudoBase++-339, outperforming generator and search baselines on both, with higher solve AUC than other search methods reflecting faster achievement of exact solutions (Table~\ref{tab:OK57_PseudoBase}). When modularly applying KnitNet Search to every off-the-shelf generator, we observe consistent improvements, averaging $30.2$ and $108.3$ additional exact solves, respectively, demonstrating usefulness beyond KnitNet's own initialization. Conversely, when initializing DesiRNA, aRNAque, and Enzymer in their original configurations with KnitNet Generator starting sequences, we again see improvements averaging $9.8$ and $67.7$ additional exact solves, respectively, supporting the value of sampling from a learned prior informed by biological sequence-structure relationships.

We further show that KnitNet's advantage persists when DesiRNA uses the same RibonanzaNet-SS structural-margin guidance and KnitNet Generator starting candidates, with statistically significant gains on OpenKnot57 and our held-out 500-target test set. The closer performance on PseudoBase++-339 primarily reflects a ceiling effect: $80.7\%$ of targets are already solved at initialization, and both methods approach complete recovery. Nevertheless, KnitNet ranks first in exact recovery, solve AUC, and capped time to solve. Appendix~\ref{app:full-PK} details the adaptation and comparisons.

\begin{table*}[t]
    \centering
    \scriptsize
    \setlength{\tabcolsep}{1.8pt}
    \renewcommand{\arraystretch}{0.76}
    \setlength{\aboverulesep}{0.25ex}
    \setlength{\belowrulesep}{0.25ex}
    \setlength{\cmidrulesep}{0.25ex}
    \caption{\textbf{Inverse-design performance on two pseudoknot benchmarks.} We evaluate each generator off-the-shelf and with KnitNet Search. For search methods, we compare their original implementation with initialization from KnitNet Generator starting candidates. DesiRNA-RNSS uses the same RibonanzaNet-SS structural-margin guidance as KnitNet. $\Delta$ denotes gain in solved-target count when applying a KnitNet component (search or initialization), relative to the original baseline. Higher Solve AUC indicates both more and earlier solutions. Values are mean $\pm$ sample SD over three seeds. \textbf{Bold} and \underline{underlined} indicate the best and second-best displayed means, respectively.}
    \label{tab:OK57_PseudoBase}
    \vspace{-2pt}
    \begin{adjustbox}{max width=\textwidth}
    \begin{tabular}{@{}llcccccc@{}}
        \toprule
        Method & Configuration & \multicolumn{3}{c}{OpenKnot57} & \multicolumn{3}{c}{PseudoBase++-339} \\
        \cmidrule(lr){3-5}
        \cmidrule(lr){6-8}
        & & Exact (/57) $\uparrow$ & $\Delta$ & Solve AUC $\uparrow$ & Exact (/339) $\uparrow$ & $\Delta$ & Solve AUC $\uparrow$ \\
        \midrule
        \multicolumn{8}{l}{\textit{A. Off-the-shelf generators}} \\
        Struct2Seq
          & Generation only
          & $30.0 \pm 0.00$ &
          & --
          & $265.0 \pm 3.61$ &
          & -- \\
          & + KnitNet search
          & $\underline{56.0 \pm 0.00}$
          & \knitdelta{+26.0}
          & $0.9583 \pm 0.0040$
          & $\mathbf{335.7 \pm 0.58}$
          & \knitdelta{+70.7}
          & $\underline{0.9882 \pm 0.0019}$ \\
        Struct2Seq OA
          & Generation only
          & $40.3 \pm 1.53$ &
          & --
          & $282.7 \pm 4.16$ &
          & -- \\
          & + KnitNet search
          & $\mathbf{56.3 \pm 0.58}$
          & \knitdelta{+16.0}
          & $\mathbf{0.9754 \pm 0.0032}$
          & $\underline{335.3 \pm 0.58}$
          & \knitdelta{+52.6}
          & $0.9879 \pm 0.0013$ \\
        gRNAde 2D
          & Generation only
          & $8.7 \pm 2.52$ &
          & --
          & $175.0 \pm 1.73$ &
          & -- \\
          & + KnitNet search
          & $52.7 \pm 1.15$
          & \knitdelta{+44.0}
          & $0.8740 \pm 0.0221$
          & $331.3 \pm 1.53$
          & \knitdelta{+156.3}
          & $0.9649 \pm 0.0038$ \\
        RNAinformer
          & Generation only
          & $9.7 \pm 3.51$ &
          & --
          & $178.7 \pm 5.03$ &
          & -- \\
          & + KnitNet search
          & $44.3 \pm 0.58$
          & \knitdelta{+34.6}
          & $0.7608 \pm 0.0064$
          & $332.3 \pm 2.89$
          & \knitdelta{+153.6}
          & $0.9676 \pm 0.0085$ \\
        \midrule
        \multicolumn{8}{l}{\textit{B. Existing search methods}} \\
        DesiRNA \textcolor{black}
          & Original init.
          & $40.7 \pm 1.15$ &
          & $0.6869 \pm 0.0223$
          & $288.3 \pm 4.04$ &
          & $0.8300 \pm 0.0119$ \\
          & KnitNet init.
          & $41.0 \pm 2.65$
          & \knitdelta{+0.3}
          & $0.6824 \pm 0.0436$
          & $300.3 \pm 2.89$
          & \knitdelta{+12.0}
          & $0.8719 \pm 0.0045$ \\
        {DesiRNA--RNSS}
          & {Original init.}
          & {$54.3 \pm 1.15$} &
          & {$0.9192 \pm 0.0170$}
          & {$334.7 \pm 0.58$} &
          & {$0.9805 \pm 0.0031$} \\
          & {KnitNet init.}
          & {$55.0 \pm 2.00$}
          & {\knitdelta{+0.7}}
          & {$0.9306 \pm 0.0148$}
          & {$335.0 \pm 0.00$}
          & {\knitdelta{+0.3}}
          & {$0.9853 \pm 0.0003$} \\
        aRNAque
          & Original init.
          & $23.0 \pm 2.65$ &
          & $0.2748 \pm 0.0448$
          & $185.3 \pm 4.04$ &
          & $0.4743 \pm 0.0059$ \\
          & KnitNet init.
          & $36.0 \pm 2.00$
          & \knitdelta{+13.0}
          & $0.6005 \pm 0.0278$
          & $301.7 \pm 6.03$
          & \knitdelta{+116.3}
          & $0.8685 \pm 0.0150$ \\
        Enzymer
          & Original init.
          & $5.0 \pm 1.73$ &
          & $0.0552 \pm 0.0135$
          & $204.7 \pm 3.51$ &
          & $0.5376 \pm 0.0122$ \\
          & KnitNet init.
          & $21.0 \pm 1.00$
          & \knitdelta{+16.0}
          & $0.3605 \pm 0.0211$
          & $279.3 \pm 0.58$
          & \knitdelta{+74.7}
          & $0.8162 \pm 0.0026$ \\
        \midrule
        \rowcolor{knitnetblue}
        \textbf{KnitNet (ours)}
          & Generation only
          & $31.3 \pm 1.53$ &
          & --
          & $273.7 \pm 5.03$ &
          & -- \\
        \rowcolor{knitnetblue}
          & + KnitNet search
          & $\mathbf{56.3 \pm 1.15}$
          & \knitdelta{+25.0}
          & $\underline{0.9677 \pm 0.0191}$
          & $\mathbf{335.7 \pm 0.58}$
          & \knitdelta{+62.0}
          & $\mathbf{0.9885 \pm 0.0021}$ \\
        \bottomrule
    \end{tabular}
    \end{adjustbox}
    \vspace{-2pt}
\end{table*}

\textbf{KnitNet transfers across target distributions and folding objectives.} We next apply KnitNet to planar targets, replacing RibonanzaNet-SS structural feedback with ViennaRNA thermodynamic feedback to guide search. Despite this change in objective, KnitNet Search improves exact solve rates for every tested generator on all three benchmarks (Table~\ref{tab:h131-native-search-auc}). On Eterna100-v2, it adds $45.9$ solves on average, reaching $82.0/100$ from KnitNet Generator candidates and $85.7/100$ from RIFT-VAE candidates. On the naturally derived RNA structures in the two Rfam benchmarks, it consistently reaches $100/100$ on Rfam-Learn-Test and achieves the second-highest mean exact solve rate on Rfam-Taneda at $24.3/29$ with RIFT-VAE candidates. While the leading method in this thermodynamic setting is NEMO \cite{portela2018nemo}, its formulation relies on a planar target representation, ViennaRNA-specific scoring, and hand-crafted heuristics. It therefore cannot support pseudoknotted targets and has limited generalization. Similarly, DesiRNA's replica-exchange Monte Carlo exploration appears particularly effective under the thermodynamic objective~\cite{wirecki2025desirna}. However, this advantage does not persist on the pseudoknot benchmarks, where KnitNet achieves higher exact solve rates (Table~\ref{tab:OK57_PseudoBase}), indicating that DesiRNA's relative strength may be due to setting-specific inductive biases. Taken together, the results across all five secondary-structure benchmarks show that KnitNet does not depend on a single target topology, folding model, reward formulation, data distribution, or initialization distribution. Rather than being specialized to one inverse-folding setting, the same learned repair framework transfers across pseudoknotted and planar targets, structural and thermodynamic verifiers, and synthetic and naturally derived RNAs. 

To assess achievable coverage beyond the initial 60-minute per-target budget, we extended the KnitNet search campaign on Eterna100-v2, reaching $97/100$ exact computational solves. This demonstrates that performance under the initial budget does not represent a ceiling on KnitNet's design capability. Appendix~\ref{app:eterna_extended} details the campaign and analyzes the three remaining unsolved targets.

\begingroup

  \Needspace{8\baselineskip}
  \setlength{\columnsep}{12pt}
  \setlength{\intextsep}{0pt}

  \begin{wraptable}{r}{0.48\textwidth}
      \centering
      \footnotesize
      \setlength{\tabcolsep}{2.5pt}
      \renewcommand{\arraystretch}{1.05}
      \captionsetup{
          font={footnotesize},
          labelfont={color=black},
          justification=justified,
          singlelinecheck=false,
          skip=4pt
      }

      \caption{%
      \textbf{Independent-predictor evaluation of Eterna100-v2 designs.}
      Sequences designed and selected using ViennaRNA are scored without reselection by five predictors unused in training-label
      generation, search, or selection. Metrics are averaged across predictors; arrows denote initialization followed by search.}
      \label{tab:independent_eterna}

      \begin{tabular}{@{}lccc@{}}
          \toprule
          Pipeline
              & Exact (\%) $\uparrow$
              & Jaccard $\uparrow$
              & MCC $\uparrow$ \\
          \midrule
          KnitNet (extended)
              & \textbf{32.27} & \textbf{0.7054} & \textbf{0.7798} \\
          DesiRNA (native)
              & 31.75 & 0.6846 & 0.7616 \\
          KnitNet $\to$ KnitNet
              & 25.60 & 0.6700 & 0.7517 \\
          KnitNet $\to$ NEMO
              & 25.13 & 0.6510 & 0.7334 \\
          \bottomrule
      \end{tabular}
  \end{wraptable}

  \noindent\textbf{KnitNet designs maintain advantage against baselines
  when scored using independent oracles.}
  To test whether KnitNet's performance advantage persists under
  independent folding predictors, we evaluated our frozen Eterna100-v2
  designs using NUPACK~\citep{zadeh2011nupack},
  MXFold2~\citep{sato2021mxfold2},
  RNAstructure~\citep{reuter2010rnastructure},
  EternaFold~\citep{waymentsteele2022eternafold}, and
  SPOT-RNA~\citep{singh2019spotrna}, none of which was used for
  training-label generation, search, or candidate selection.
  KnitNet's designs were generated and selected using ViennaRNA
  before panel scoring, with no reselection based on the panel
  results. KnitNet's extended campaign ranks first across all three
  panel-averaged metrics (Table~\ref{tab:independent_eterna}).
  Relative to \mbox{DesiRNA}, the strongest baseline in this
  comparison, it improves mean Jaccard by $3.0\%$ and mean MCC by
  $2.4\%$, with an exact-solve-rate gain of $0.52$ percentage points.
  These results show that the extended campaign's advantage persists
  under panel-averaged evaluation by predictors independent of the
  search process.

  \par
\endgroup

 \begin{table*}[t]
      \centering
      \fontsize{6.5}{7.5}\selectfont
      \setlength{\tabcolsep}{0.9pt}
      \renewcommand{\arraystretch}{0.88}

      \setlength{\aboverulesep}{0.25ex}
      \setlength{\belowrulesep}{0.25ex}
      \setlength{\cmidrulesep}{0.25ex}

      \captionsetup{
          font=footnotesize,
          skip=4pt
      }

      \caption{\textbf{Inverse-design performance on
      three planar benchmarks.}
      Configurations and reporting conventions follow
      Table~\ref{tab:OK57_PseudoBase}.}
      \label{tab:h131-native-search-auc}

      \begin{adjustbox}{max width=\textwidth}
        \begin{tabular}{@{}llccccccccc@{}}
            \toprule
            Method & Configuration &
            \multicolumn{3}{c}{Eterna100} &
            \multicolumn{3}{c}{Rfam-Learn-Test} &
            \multicolumn{3}{c}{Rfam-Taneda} \\
            \cmidrule(lr){3-5}
            \cmidrule(lr){6-8}
            \cmidrule(lr){9-11}
            & &
            Exact (/100) $\uparrow$ & $\Delta$ & Solve AUC $\uparrow$ &
            Exact (/100) $\uparrow$ & $\Delta$ & Solve AUC $\uparrow$ &
            Exact (/29) $\uparrow$ & $\Delta$ & Solve AUC $\uparrow$ \\
            \midrule

            \multicolumn{11}{l}{\textit{A. Off-the-shelf generators}} \\

            Struct2Seq
              & Generation only
              & $33.0 \pm 1.00$ &
              & --
              & $18.0 \pm 3.61$ &
              & --
              & $8.7 \pm 1.15$ &
              & -- \\
              & + KnitNet search
              & $81.0 \pm 0.00$
              & \knitdelta{+48.0}
              & $0.7620 \pm 0.0047$
              & $\mathbf{100.0 \pm 0.00}$
              & \knitdelta{+82.0}
              & $0.9803 \pm 0.0008$
              & $23.0 \pm 0.00$
              & \knitdelta{+14.3}
              & $0.7827 \pm 0.0011$ \\
            \addlinespace[1pt]

            Struct2Seq OA
              & Generation only
              & $40.0 \pm 1.73$ &
              & --
              & $65.0 \pm 2.65$ &
              & --
              & $17.0 \pm 1.00$ &
              & -- \\
              & + KnitNet search
              & $82.0 \pm 1.00$
              & \knitdelta{+42.0}
              & $0.7781 \pm 0.0042$
              & $\mathbf{100.0 \pm 0.00}$
              & \knitdelta{+35.0}
              & $0.9966 \pm 0.0006$
              & $23.7 \pm 0.58$
              & \knitdelta{+6.7}
              & $0.7982 \pm 0.0176$ \\
            \addlinespace[1pt]

            RIFT-VAE
              & Generation only
              & $38.0 \pm 0.00$ &
              & --
              & $94.0 \pm 0.00$ &
              & --
              & $17.0 \pm 0.00$ &
              & -- \\
              & + KnitNet search
              & $85.7 \pm 1.53$
              & \knitdelta{+47.7}
              & $0.8061 \pm 0.0076$
              & $\mathbf{100.0 \pm 0.00}$
              & \knitdelta{+6.0}
              & $\mathbf{0.9997 \pm 0.0000}$
              & $\underline{24.3 \pm 0.58}$
              & \knitdelta{+7.3}
              & $0.7982 \pm 0.0091$ \\

            \midrule
            \multicolumn{11}{l}{\textit{B. Existing search methods}} \\

            DesiRNA
              & Original init.
              & $87.7 \pm 2.52$ &
              & $0.8461 \pm 0.0184$
              & $\mathbf{100.0 \pm 0.00}$ &
              & $0.9827 \pm 0.0015$
              & $24.0 \pm 0.00$ &
              & $0.8243 \pm 0.0007$ \\
              & KnitNet init.
              & $80.7 \pm 2.08$
              & \knitdelta{-7.0}
              & $0.7773 \pm 0.0160$
              & $\mathbf{100.0 \pm 0.00}$
              & \knitdelta{0.0}
              & $0.9707 \pm 0.0011$
              & $23.3 \pm 1.15$
              & \knitdelta{-0.7}
              & $0.7768 \pm 0.0265$ \\
            \addlinespace[1pt]

            FastDesign
              & Original init.
              & $80.3 \pm 1.53$ &
              & $0.7793 \pm 0.0147$
              & $\mathbf{100.0 \pm 0.00}$ &
              & $0.9654 \pm 0.0010$
              & $24.0 \pm 0.00$ &
              & $0.8101 \pm 0.0016$ \\
              & KnitNet init.
              & $75.3 \pm 3.21$
              & \knitdelta{-5.0}
              & $0.7373 \pm 0.0333$
              & $\mathbf{100.0 \pm 0.00}$
              & \knitdelta{0.0}
              & $0.9722 \pm 0.0003$
              & $24.0 \pm 0.00$
              & \knitdelta{0.0}
              & $0.8132 \pm 0.0006$ \\
            \addlinespace[1pt]

            aRNAque
              & Original init.
              & $72.0 \pm 2.00$ &
              & $0.6970 \pm 0.0120$
              & $\underline{99.3 \pm 1.15}$ &
              & $0.8937 \pm 0.0078$
              & $23.3 \pm 0.58$ &
              & $0.7847 \pm 0.0034$ \\
              & KnitNet init.
              & $74.0 \pm 0.00$
              & \knitdelta{+2.0}
              & $0.7184 \pm 0.0038$
              & $\mathbf{100.0 \pm 0.00}$
              & \knitdelta{+0.7}
              & $0.9700 \pm 0.0008$
              & $23.3 \pm 0.58$
              & \knitdelta{0.0}
              & $0.7705 \pm 0.0188$ \\
            \addlinespace[1pt]

             NEMO
                & Native search
                & $\underline{89.0 \pm 1.00}$ &
                & $\underline{0.8579 \pm 0.0060}$
                & $\mathbf{100.0 \pm 0.00}$ &
                & $0.9908 \pm 0.0004$
                & $\mathbf{24.7 \pm 0.58}$ &
                & $\mathbf{0.8399 \pm 0.0130}$ \\
                & KnitNet init.
                & $\mathbf{89.7 \pm 0.58}$
                & \knitdelta{+0.7}
                & $\mathbf{0.8620 \pm 0.0055}$
                & $\mathbf{100.0 \pm 0.00}$
                & \knitdelta{0.0}
                & $\underline{0.9989 \pm 0.0006}$
                & $24.0 \pm 0.00$
                & \knitdelta{-0.7}
                & $\underline{0.8264 \pm 0.0006}$ \\

            \midrule

            \rowcolor{knitnetblue}
            \textbf{KnitNet (ours)}
              & Generation only
              & $36.0 \pm 1.73$ &
              & --
              & $29.7 \pm 1.53$ &
              & --
              & $11.0 \pm 1.00$ &
              & -- \\
            \rowcolor{knitnetblue}
              & + KnitNet search
              & $82.0 \pm 1.00$
              & \knitdelta{+46.0}
              & $0.7826 \pm 0.0024$
              & $\mathbf{100.0 \pm 0.00}$
              & \knitdelta{+70.3}
              & $0.9759 \pm 0.0032$
              & $23.7 \pm 0.58$
              & \knitdelta{+12.7}
              & $0.7780 \pm 0.0228$ \\

            \bottomrule
        \end{tabular}
        \end{adjustbox}
    \end{table*}

\begin{wrapfigure}{r}{0.48\textwidth}
      \centering
      \includegraphics[width=\linewidth]{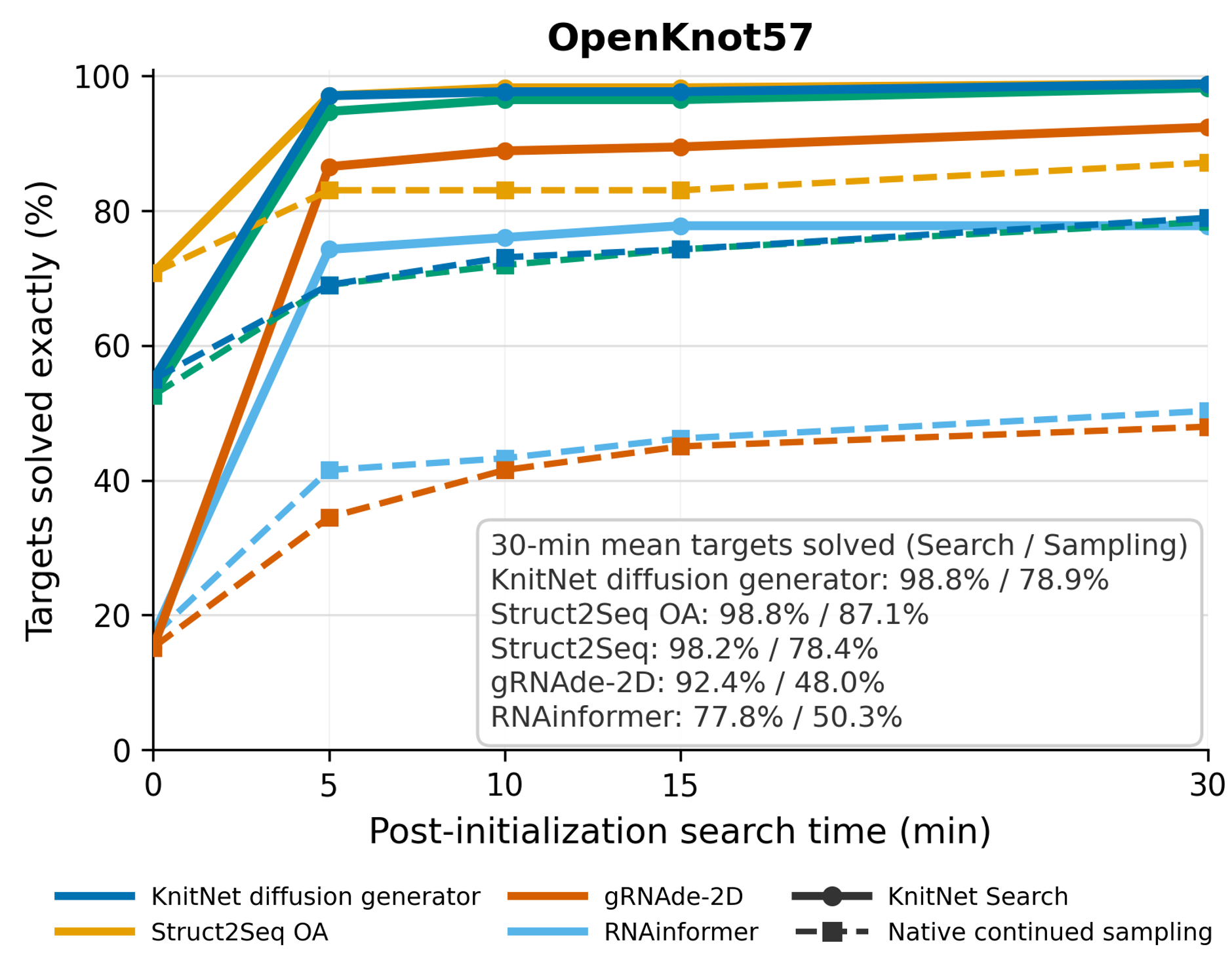}
      \caption{%
          \textbf{KnitNet search converts additional inference time into additional exact solutions.}
          Exact coverage is plotted against post-initialization wall-clock time.  Both methods start from identical pools of 96 candidates. Solid curves represent KnitNet search, while dashed curves represent native continued sampling. Curves report means over three seeds.
      }
      \label{fig:inference_scaling}
      \vspace{-8pt}
  \end{wrapfigure}
  \noindent\textbf{KnitNet Search uses inference-time compute more effectively than continued sampling.} Because continued sampling can itself improve solve
  coverage~\cite{brown2024large}, we evaluate inference-time scaling of our verifier-guided repair against drawing more
  samples from the generator. For every target, generator, and seed, both
  approaches begin from the same pool of 96 candidates. One applies KnitNet Search, whereas the matched control continues sampling from the corresponding source generator. Both receive the same post-initialization wall-clock budget and stop upon finding an exact solution or reaching the time limit. The KnitNet Search budget includes verifier calls, which dominate its runtime; the counterexample-guided repair operations themselves are comparatively inexpensive. KnitNet Search achieves higher exact coverage in every matched generator-benchmark configuration. On OpenKnot57, it improves exact coverage by $11.7$-$44.4$ percentage points over continued sampling (Figure~\ref{fig:inference_scaling}). Across the four remaining secondary-structure benchmarks, KnitNet Search improves final exact coverage by $16.6$ percentage points and reduces capped mean time to solve by $69\%$ on average (Appendix~\ref{app:scaling_eterna}). These results show that allocating additional inference-time compute to our counterexample-guided joint repair method is more effective than allocating it to continued generator sampling. 

\begingroup
  \noindent\textbf{KnitNet retains its advantage on longer RNA targets.}
  We also investigate how sequence length affects exact recovery and computational cost on the held-out 500-target test set. Within one hour, KnitNet solves $197/200$ targets of at most 100 nucleotides and $2/56$ targets of 401--511 nucleotides,
  while oracle-evaluation throughput falls from approximately 37 to 2 evaluations per second across these length bins. Extending search to two hours increases recovery from $8/127$ to $15/127$ among targets of 301-511 nucleotides. Although longer sequences remain challenging, KnitNet achieves
  higher exact recovery than the strongest tested baseline in every length bin, preserving its comparative advantage as target length increases (Appendix~\ref{app:length_scaling}).
  \par
  \endgroup

\textbf{Ablations.}
  On the held-out 500-target test set, the complete KnitNet framework achieves 306/500 exact solves within the 60-minute search budget. Learned initialization and search provide complementary gains: search from random roots and continued generator sampling reach only 210/500 and 209/500 solves, respectively. The full 500-target ablation results are reported in
  Appendix~\ref{app:knitnet-ablations}.

  \begingroup
  To quantify the contributions of adaptive repair and
  learned ranking, we additionally evaluate component ablations on OpenKnot57 and Eterna100-v2 at a matched budget of 15,000 logical oracle evaluations (Table~\ref{tab:pat4-ablations-external}). These ablations use a 64-tree ExtraTrees surrogate, which improved Eterna100-v2 recovery over the original 16-tree configuration. The complete framework achieves the highest exact coverage and solve AUC on both benchmarks, with every ablation reducing both metrics. Fixing the repair problem loses two exact solves on OpenKnot57 and 13 on Eterna100-v2, while removing surrogate ranking reduces solve AUC by $0.015$ and $0.025$, respectively. The full method achieves significantly higher solve AUC than every tested ablation on both benchmarks (paired 95\% confidence intervals for the differences exclude zero). Together, these results demonstrate that adaptive repair and learned ranking improve exact recovery and make more effective use of oracle evaluations on both benchmarks.
  \par
  \endgroup

  \begin{table}[H]
      \centering
      \color{black}
      \small
      \setlength{\tabcolsep}{2pt}
      \renewcommand{\arraystretch}{1.08}

\caption{\textbf{Component ablations on OpenKnot57
  and Eterna100-v2.} Each configuration starts with 96 candidates and runs for up to 15,000 folding-oracle evaluations. Higher solve AUC indicates more targets solved earlier in search. $\Delta$AUC is the
  full method's AUC minus the ablation's AUC; brackets show 95\% confidence intervals for this difference.}
  \label{tab:pat4-ablations-external}

      \begin{tabular*}{\textwidth}{
          @{\extracolsep{\fill}}lcccccc@{}
      }
          \toprule
          & \multicolumn{3}{c}{OpenKnot57}
          & \multicolumn{3}{c}{Eterna100-v2} \\
          \cmidrule(lr){2-4}\cmidrule(l){5-7}
          Configuration
          & \shortstack{Exact\\(/57) $\uparrow$}
          & \shortstack{Solve\\AUC $\uparrow$}
          & \shortstack{$\Delta$AUC\\95\% CI}
          & \shortstack{Exact\\(/100) $\uparrow$}
          & \shortstack{Solve\\AUC $\uparrow$}
          & \shortstack{$\Delta$AUC\\95\% CI} \\
          \midrule

          \textbf{Full KnitNet search}
          & \textbf{57} & \textbf{0.963} & --
          & \textbf{85} & \textbf{0.768} & -- \\

          W/o surrogate ranking
          & 56 & 0.948 & $+0.015\ [0.004, 0.029]$
          & 82 & 0.743 & $+0.025\ [0.006, 0.045]$ \\

          W/ fixed repair problem
          & 55 & 0.936 & $+0.027\ [0.009, 0.050]$
          & 72 & 0.701 & $+0.067\ [0.032, 0.105]$ \\

          W/ random repair scope
          & 54 & 0.928 & $+0.035\ [0.013, 0.061]$
          & 77 & 0.714 & $+0.054\ [0.020, 0.089]$ \\

          W/ linear surrogate ranking
          & 56 & 0.951 & $+0.012\ [0.002, 0.026]$
          & 81 & 0.736 & $+0.032\ [0.009, 0.057]$ \\

          Fixed problem + no surrogate
          & 55 & 0.930 & $+0.033\ [0.011, 0.057]$
          & 72 & 0.688 & $+0.080\ [0.044, 0.121]$ \\

          \bottomrule
      \end{tabular*}
  \end{table}

\begin{figure}[!t]
  \centering
  \includegraphics[height=0.22\textheight,keepaspectratio]{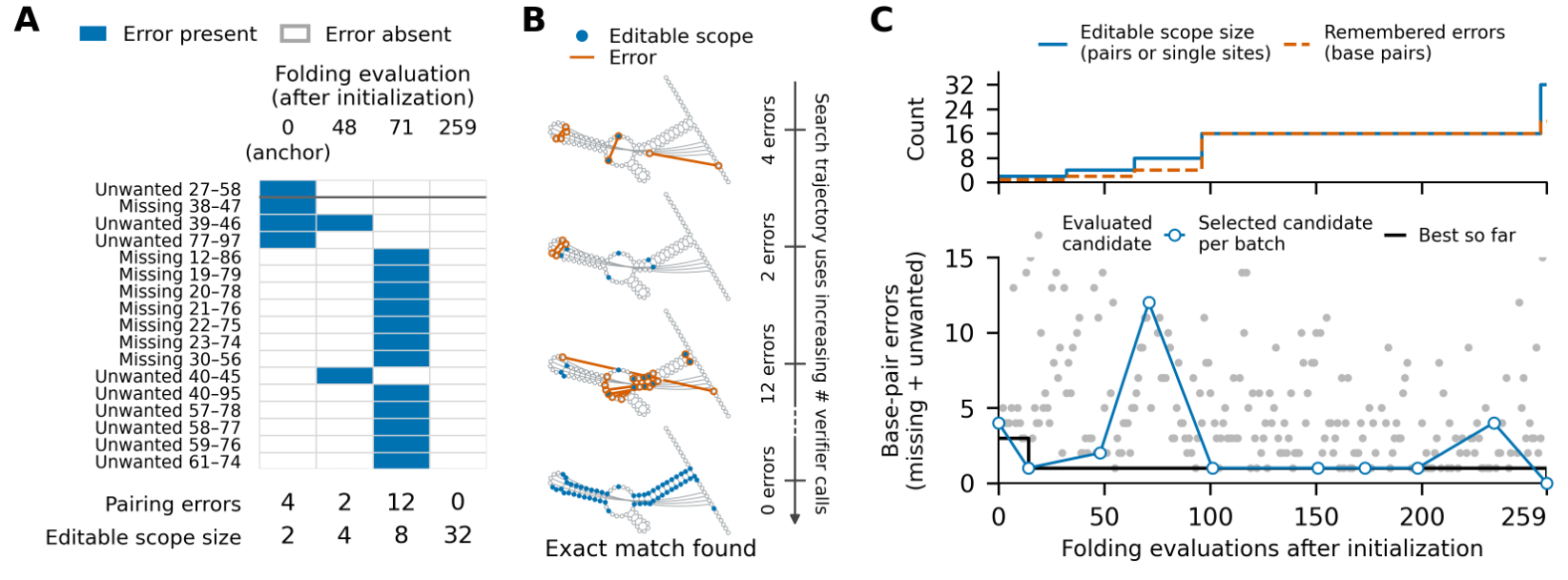}
  \caption{%
      \textbf{KnitNet updates joint repair using folding feedback.}
      \textbf{(A)} Evolution of pairing errors during a search trajectory, with snapshots labeled by the number of folding evaluations after initialization. 
      \textbf{(B)} Structural visualization of the repair evolution. 
      \textbf{(C)} Repair scope and error memory (top) and full-fold error count (bottom). The blue connecting line links successive selected assignments.}
  \label{fig:pat4_dynamics}
\end{figure}

\textbf{Mechanistic analysis shows how failed folds guide KnitNet's adaptive repair.}    
We investigate how learning and refining the repair strategy through verifier feedback enables KnitNet to solve more complex targets than other methods. The search trajectory from a solved PseudoBase++-339 target (Figure~\ref{fig:pat4_dynamics}A,B) illustrates the importance of this mechanism: correcting the initially targeted errors can leave other errors unresolved and introduce new ones, highlighting that a locally successful repair does not guarantee a globally correct fold. 
Figure~\ref{fig:pat4_dynamics}C (top) shows how KnitNet remembers  observed errors and progressively expands the editable scope to include additional nucleotides, allowing potentially correlated structural regions to be revised together. Importantly, the trajectory shows a temporary increase in error count before reaching an exact fold (Figure~\ref{fig:pat4_dynamics}C, bottom), suggesting that successful repair need not improve monotonically. More broadly, repairs that clear prioritized errors but leave errors outside the editable scope occur in 83.1\% of OpenKnot57 and 80.2\% of Eterna100 search trajectories on targets unsolved at initialization, underscoring the importance of KnitNet’s feedback-driven adaptation of the repair scope across distinct target regimes.

\begin{wraptable}{r}{0.50\textwidth}
      \centering
      \tiny
      \setlength{\tabcolsep}{2pt}
      \renewcommand{\arraystretch}{0.95}
      \input{tables/3d_inverse_design}
      \vspace{-6pt}
\end{wraptable}

  \textbf{KnitNet generalizes to 3D inverse design.} Finally, we test whether our learned repair strategy extends from secondary-structure matching to 3D inverse design. Guided by tertiary-structure similarity, KnitNet Search increases mean scTM and GDT$_{\mathrm{TS}}$ and reduces RMSD for all six generators across three seeds (Table~\ref{tab:3d_inverse_design}). These gains across 2D- and 3D-conditioned generators support transfer to tertiary-structure objectives, with our method improving even the strongest baseline, RiboDiffusion, from $0.679$ to $0.773$ mean scTM. We also compare KnitNet Search with RIDER's target-specific RL, the previous leading post-training approach for 3D inverse design~\cite{hu_rider_2026}. Starting from the same RIDE initial candidates, KnitNet Search achieves higher mean scTM (0.732 versus 0.679) and lower mean RMSD ($2.233$ versus $2.691$~\AA) under the budgets detailed in Appendix~\ref{app:3d_protocol}. Beyond improving structural self-consistency, KnitNet Search reduces runtime by $39.3\%$ relative to RIDER RL in a separate timing comparison with a matched oracle budget and identical hardware allocation. Together, these results support KnitNet search as a transferable inference-time refinement procedure that improves 3D structural self-consistency.
  \par

\section{Conclusion}
\label{sec:conclusion}
We introduced KnitNet, an RNA inverse-design framework that combines a learned, structure-conditioned sequence prior with counter-example-guided joint repair at inference time. KnitNet converts additional test-time computation into exact secondary-structure matches more efficiently than continued sampling  across generators, planar and pseudoknotted targets, and folding objectives. Its transfer to 3D design supports the broader applicability of our method. Mechanistic analysis illustrates how verifier feedback is used to expand the repair scope and address globally-coupled nucleotide errors, underpinning KnitNet’s distinctive advantage in adapting repair across RNA design regimes. These findings motivate greater emphasis on inference-time scaling through adaptive search, building on learned generators that provide effective starting candidates. Beyond structural objectives, it would be valuable to extend KnitNet to other RNA properties, such as RNA–protein interactions, mRNA stability, and translation. We discuss these future opportunities in Appendix~\ref{app:limitations_future_work}.

\clearpage

\subsection*{Acknowledgements}

A.P. and X.W. are supported by Yale University's Computational Biology and Biomedical Informatics (CBB) program. A.P. also acknowledges support from an International Fellowship from the American Association of
University Women (AAUW). This work was supported in part by the National Science Foundation (NSF) under award IIS-2403317. Computational resources were provided by the Yale Center for Research Computing through the Misha cluster, the Vanderbilt Center for Structural Biology (CSB) through the Bluefin Compute Cluster, and the National Center for Supercomputing Applications (NCSA) through DeltaAI.
A joint resource of NCSA and the University of Illinois Urbana-Champaign, DeltaAI receives funding from NSF award OAC-2320345 and the State of Illinois. Our DeltaAI allocation was provided through the Advanced
Cyberinfrastructure Coordination Ecosystem: Services \& Support (ACCESS) program, funded by NSF awards 2138259, 2138286, 2138307, 2137603, and 2138296.

\subsection*{AI use statement}

  We used generative AI tools to aid and polish manuscript writing, retrieve and discover related work, support research ideation and execution, and draft manuscript sections. We did not use these tools to generate synthetic datasets, prove mathematical claims, or for purposes beyond those described above. Separately, we used RibonanzaNet-SS to generate predicted secondary-structure labels for existing RNAcentral sequences, as described in Appendix~\ref{app:training-data}. We reviewed
  all AI-assisted content to ensure accuracy, preserve the intended meaning, and faithfully represent the methods and findings. We take responsibility for the final content of this work, including all text, claims, and artifacts.

\subsection*{Reproducibility statement}

We release the pretraining, generation, search, and evaluation code, pretrained KnitNet generator weights, and evaluation datasets at \url{https://github.com/antonia-panescu/KnitNet}. The release includes the five external secondary-structure benchmarks, the 500-target ablation panel with its frozen starting candidates, and the 64 reference structures used for 3D evaluation. The repository provides step-by-step installation and usage instructions, experiment configurations and random seeds, and guidance for obtaining third-party folding models and structure-analysis tools. Users can supply their own targets, select an optimization objective, customize scoring, and set time or oracle-evaluation budgets. Generation and ViennaRNA-guided design support CPU execution without a GPU, and unit and installation tests help users check their setup. The appendices detail the architecture, training procedure, evaluation protocols, and data construction and overlap checks.

\clearpage

\bibliography{iclr2027_conference}
\bibliographystyle{iclr2027_conference}

\clearpage
\appendix
\startcontents[appendices]
\section*{Appendix}
\printcontents[appendices]{}{1}[1]{}

 \clearpage

\section{Related Work}
\label{app:related_work}

\paragraph{Generative and diffusion models for RNA design.}
  Structure-conditioned generators learn reusable sequence-structure
  regularities. RNAinformer uses axial attention over nucleotide interaction
  matrices, supporting pseudoknots~\citep{patil2024rnainformer}. For tertiary
  design, RDesign learns hierarchical representations, gRNAde uses geometric
  graph networks for single- and multi-state design, and RiboDiffusion conditions
  diffusion on RNA backbones~\citep{tan2024rdesign,joshi2025grnade,
  huang2024ribodiffusion}; RDiffusion supports multiple conditions, including
  secondary and tertiary structure~\citep{wang2026rdiffusion}. These models
  provide useful priors, but sequence reconstruction does not directly optimize
  fold-back correctness, and higher native-sequence recovery need not imply
  better structural self-consistency~\citep{joshi2025grnade,hu_rider_2026}.
  KnitNet uses masked discrete diffusion for bidirectional completion of
  partially observed sequences~\citep{austin2021d3pm,sahoo2024mdlm}, complemented
  by joint paired-endpoint assignments to accommodate coupled RNA constraints.
  The generator supplies diverse initial candidates and remains frozen, while
  counterexample-guided repair uses failed folds to improve subsequent proposals.

  \paragraph{Combining learned generation with refinement.}
  SentRNA combines a prior learned from human-designed sequences with
  adaptive-walk refinement~\citep{shi2018sentrna}, while RIFT-VAE combines
  grammar-conditioned pretraining with latent-space optimization and tests
  generated sequences as starting points for existing
  solvers~\citep{watanabe2026riftvae}. These approaches place refinement in
  promising regions of sequence space. But their native refinement uses
  prescribed mutation moves or a model-specific latent representation, rather
  than rebuilding joint repair scopes from recurring folding errors. KnitNet
  builds on their generation-refinement strategy with adaptive repair scopes
  and error priorities, operating across generators without access to their
  latent spaces.

  \paragraph{Learning design policies from folding feedback.}
  LEARNA learns a construction policy per target; Meta-LEARNA transfers a
  policy trained across targets, and Meta-LEARNA-Adapt continues updating it
  on the new task~\citep{runge2019learna}. EternaBrain combines mutation choices
  learned from human play with refinement strategies, while DRAG learns mutation
  policies over hierarchical RNA graphs~\citep{koodli2019eternabrain,li2025drag}.
  Struct2SeQ uses deep Q-learning for secondary-structure-conditioned generation,
  including pseudoknots, and its order-agnostic extension removes the fixed
  left-to-right decoding order~\citep{he2026struct2seq,panescu2026orderagnostic}.
  For 3D design, RIDER fine-tunes RIDE with structural self-consistency
  rewards~\citep{hu_rider_2026}. These methods align design with structural
  outcomes, but adapting a policy to a different objective or verifier can
  require further optimization. KnitNet places target-specific learning in its
  local repair surrogate, retaining feedback-driven adaptation without updating
  the generator.

  \paragraph{Structure-guided exploration of sequence space.}
  RNA-SSD combines structural decomposition with stochastic local search, while
  INFO-RNA uses energy-guided initialization and neighbor
  selection~\citep{andronescu2004new,busch2006info}. Enzymer weights mutation
  locations by ensemble defect and adapts mutation
  extent~\citep{zandi2016adaptive}. Other exploration strategies include
  multi-objective genetic search (MODENA), greedy evolution (GREED-RNA), and
  heavy-tailed mutation sizes (aRNAque)~\citep{taneda2012modena,
  lozano2025greedrna,merleau2022arnaque}; ant-colony optimization
  (antaRNA)~\citep{kleinkauf2015antarna}; and tree or nested Monte Carlo search
  (MCTS-RNA, NEMO)~\citep{yang2017mcts,portela2018nemo}. DesiRNA combines
  targeted mutations with replica-exchange Monte
  Carlo~\citep{wirecki2025desirna}. These strategies help escape locally
  favorable regions. But broader exploration alone does not identify which
  interacting positions must change together: correcting one interaction can
  disrupt another. KnitNet's error-derived repair scopes and surrogate ranking
  address this coordination problem and prioritize joint edits for evaluation.

  \paragraph{Optimization objectives and coupled assignments.}
  SAMFEO combines structure-aware mutations with multifrontier search to optimize
  ensemble objectives~\citep{zhou2023samfeo}; SamplingDesign optimizes coupled
  sequence distributions using Monte Carlo gradient
  estimates~\citep{tang2026samplingdesign}; and PKProbDesign extends
  probability-guided design to pseudoknotted
  structures~\citep{otagaki2026pkprobdesign}. These objectives provide graded
  guidance, but improving them need not yield an exact decoded fold, and
  thermodynamic scores do not directly transfer to every learned or tertiary
  verifier. KnitNet separates oracle-specific ranking from pairing-error repair
  and exact secondary-structure verification. This supports different ranking
  objectives while preserving the repair procedure, including secondary-error
  guidance alongside tertiary ranking in 3D design.

  \paragraph{Coordinated repair and adaptation of the search problem.}
  RNAblueprint uses dependency graphs to sample assignments compatible with
  multiple structures~\citep{hammer2017rnablueprint}, but compatibility alone
  does not ensure that the prescribed pairs form. RNAiFold provides
  large-neighborhood constraint solving, while RNAiFold2T adapts which positions
  remain fixed according to structural
  agreement~\citep{garciamartin2013rnaifold,garciamartin2016rnaifold2t}.
  FastDesign uses misfolded rivals to identify differential positions and
  accumulate joint assignment constraints from local energy
  differences~\citep{zhou2026fastdesign}. These establish direct precedents for
  adaptive, failure-guided repair. But their thermodynamic formulations do not
  directly supply equivalent repair machinery under a learned pseudoknot-capable
  verifier; FastDesign's pruning specifically relies on a pseudoknot-free energy
  decomposition. KnitNet's counterexample-guided scope and priority revision
  instead use empirical complete-fold errors, extending coordinated repair
  beyond this analytic setting. Its surrogate provides empirical rankings,
  rather than certified thermodynamic exclusions.

  \paragraph{Learning to allocate expensive folding evaluations.}
  Montparnasse adapts rollout policies online through generalized nested rollout
  policy adaptation~\citep{cazenave2026montparnasse}, while RNA FMQA learns a
  factorization-machine surrogate and optimizes it with quadratic-optimization
  annealing~\citep{kikuchi2026fmqa}. These methods learn from search evaluations,
  but place adaptation in a rollout policy or a surrogate for a fixed
  whole-sequence objective. KnitNet couples its local surrogate with revision
  of the repair scope and prioritized errors. Our method differs primarily in
  where learning enters the search loop: rather than directly parameterizing
  mutation actions or replacing search with a learned policy, we retain an
  explicit structure-aware proposal mechanism and learn which proposals merit
  expensive folding evaluations. This separates proposal generation from
  proposal valuation, while uniform exploration and the protected proposal beam
  preserve alternatives when surrogate predictions are imperfect.
  \clearpage
\section{Pretraining Corpus Statistics and Quality Control}
  \label{app:training-data}

  The structure-conditioned generator uses an RNAcentral release 26 corpus
  with 21,619,209 training examples, 10,000 validation examples, and 2,000
  held-out test examples. Each sequence is paired with a RibonanzaNet-SS
  predicted structure, represented as a canonical base-pair set. Design
  correctness is assessed separately by fold-back verification.

  \paragraph{Structural supervision.}
  RNAcentral combines annotations from multiple databases, including
  R2DT template-based and family-consensus structures
  \cite{rnacentral2026,sweeney_r2dt_2021}. Native sequences did not always
  reproduce their annotated targets in our preliminary fold-back checks.
  Because sequence--structure relationships are degenerate and depend on
  molecular context, we refold all retained sequences with RibonanzaNet-SS
  to align supervision with the verifier. We chose RibonanzaNet as the fold-back oracle because it incorporates experimental
  SHAPE and DMS data training. Its SS variant is fine-tuned on PDB structures and
  achieves a cluster-averaged base-pair $F_1$ of 0.890 on its held-out PDB
  benchmark \cite{he2024ribonanza}. We use evaluation mode to disable
  dropout, fold equal-length groups without padding, and keep the resulting
  labels fixed throughout training.

  We retain sequences of 20-512 nt to prioritize reliable supervision.
  Reduced prediction accuracy on longer RNAs has been reported in
  secondary-structure benchmarks \cite{zhang2026secondary}; 512 nt is our
  pretraining length limit. We plan to extend this range as improved predictors are
  validated on longer RNAs.

  \paragraph{Split construction.}
  After exact sequence deduplication, we form MMseqs2 clusters at 80\%
  identity and 80\% coverage of both sequences. Held-out examples come from
  cluster representatives absent from training and undergo direct
  cross-split screening at the same thresholds. The test split combines
  1,000 targets sampled uniformly over eligible unique pair sets with 1,000
  sampled across 30 strata: five length bins, three paired-fraction bins,
  and pseudoknot presence or absence. Validation uses remaining eligible
  representatives and covers the same strata. The test set contains 2,000 distinct target pair sets. The validation
  set contains 10,000 distinct sequences spanning 6,033 target pair sets,
  with multiple sequence examples for some structures. Both sets are
  selected using deterministic hash rankings with seed 20260901.

  Due to current computational-resource constraints, for the computationally intensive ablations we use a fixed, outcome-blind subset of 500 of the 2,000 test targets. We construct this subset through two deterministic nested halvings, from 2,000 to 1,000 and then from 1,000 to 500 targets, preserving equal contributions from the uniformly sampled and stratified panels while balancing length, paired fraction, and pseudoknot presence.

  \subsection{Split statistics and quality control}

  We scan every example in all three splits. All sequences contain only
  A/C/G/U, sequence and structure lengths agree, and all structures pass
  the strict extended-dot-bracket parser. Internal audits find no
  cross-split sequence matches at the specified thresholds, exact sequence
  containments, or shared identifiers, sequences, or target pair sets.
  Table~\ref{tab:training-splits} summarizes the complete corpus.

  \begin{table}[!htb]
      \centering
      \caption{%
          \textbf{RNAcentral split statistics.}
          Length is min / median / max; GC is nucleotide-weighted.
          Paired fraction is median [IQR]. PK denotes targets with
          crossing base pairs.
      }
      \label{tab:training-splits}
      \small
      \setlength{\tabcolsep}{4pt}
      \renewcommand{\arraystretch}{1.08}

      \begin{adjustbox}{max width=\linewidth}
      \begin{tabular}{@{}l r c c c r@{}}
          \toprule
          \textbf{Split}
          & \textbf{$N$}
          & \textbf{Length (nt)}
          & \textbf{GC}
          & \textbf{Paired fraction}
          & \textbf{PK (\%)} \\
          \midrule
          Train
          & 21,619,209
          & 20 / 255 / 512
          & 0.523
          & 0.506 [0.441, 0.560]
          & 82.94 \\
          Val
          & 10,000
          & 20 / 71 / 512
          & 0.444
          & 0.345 [0.144, 0.491]
          & 44.09 \\
          Test
          & 2,000
          & 20 / 187.5 / 512
          & 0.444
          & 0.415 [0.286, 0.520]
          & 67.25 \\
          \bottomrule
      \end{tabular}
      \end{adjustbox}
  \end{table}

  Held-out targets have shorter median lengths, lower GC content and
  pairing density, and fewer pseudoknots than training. All splits span
  20-512 nt and retain both planar and pseudoknotted targets.
  Figure~\ref{fig:training-split-distributions} shows these shifts.

  \par\smallskip
  \noindent
  \begin{minipage}{\linewidth}
      \captionsetup{type=figure}
      \centering
      \includegraphics[width=0.85\linewidth]{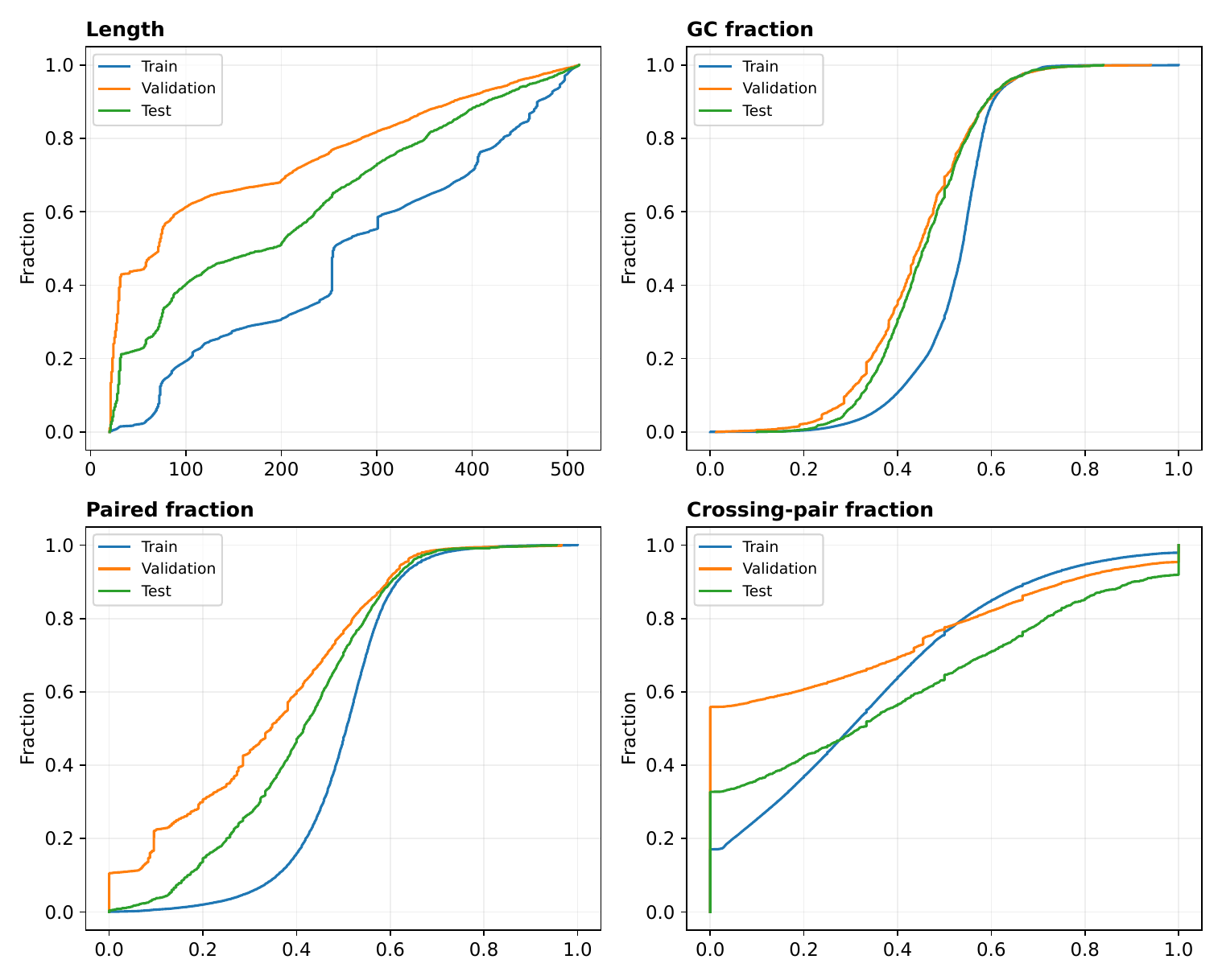}
      \caption{%
          \textbf{Training and held-out sequence distributions.}
      }
      \label{fig:training-split-distributions}
  \end{minipage}
  \par\smallskip

  Structural statistics use actual crossings in the canonical pair set,
  independent of bracket labels. These checks establish valid, internally
  separated training pairs; the predicted structures remain pseudolabels.

\clearpage
 
  \section{Evaluation Dataset Statistics and Exposure Audit}
  \label{app:dataset_statistics}
  \suppressfloats[t]

  To assess generalization across RNA design tasks, we evaluate on benchmarks
  spanning different lengths and structural characteristics. Here we establish
  their complementary coverage and audit their overlap with generator
  pretraining, providing context for interpreting performance across datasets.

  \subsection{Benchmark composition and structural diversity}

  Table~\ref{tab:dataset-audit} summarizes five external secondary-structure
  benchmarks and the frozen 500-target internal ablation panel.
  OpenKnot57 and PseudoBase++-339 test pseudoknotted design, whereas
  Eterna100-v2, Rfam-Learn-Test, and Rfam-Taneda test planar structures.
  Their median lengths range from 57 to 250 nucleotides, covering both
  short and long targets. PseudoBase++-339 contains 339 distinct targets
  after deduplicating 416 valid records by length and nucleotide-pair set.
  We identify pseudoknots by actual pair crossings, $i<k<j<l$,
  independently of dot-bracket notation.

  \begin{table}[!htb]
      \centering
      \caption{Benchmark composition and overlap with generator pretraining.
      Length: min/median/max in nucleotides. PK: percentage with pseudoknots.
      Exact: targets matching our train set in length and pair set.
      NN: nearest-training equal-length pair-set Jaccard, p95/max;
      undefined scores are excluded.}
      \label{tab:dataset-audit}
      \scriptsize
      \setlength{\tabcolsep}{3pt}
      \renewcommand{\arraystretch}{1.0}
      \begin{tabular}{@{}l r l c c c c@{}}
          \toprule
          \textbf{Dataset} & \textbf{$N$} & \textbf{Verifier} &
          \textbf{Length} & \textbf{PK (\%)} & \textbf{Exact} &
          \textbf{NN p95 / max} \\
          \midrule
          OpenKnot57
          & 57 & RibonanzaNet-SS & 48 / 100 / 240
          & 100.0 & 0/57 & 0.484 / 0.606 \\
          PseudoBase++-339
          & 339 & RibonanzaNet-SS & 21 / 57 / 418
          & 100.0 & 0/339 & 0.750 / 0.960 \\
          Eterna100-v2
          & 100 & ViennaRNA & 12 / 104 / 400
          & 0.0 & 0/100 & 0.595 / 0.889 \\
          Rfam-Learn-Test
          & 100 & ViennaRNA & 51 / 250 / 447
          & 0.0 & 0/100 & 0.550 / 0.892 \\
          Rfam-Taneda
          & 29 & ViennaRNA & 55 / 152 / 452
          & 0.0 & 0/29 & 0.740 / 0.955 \\
          \midrule
          Ablation-500 (internal)
          & 500 & RibonanzaNet-SS & 20 / 182 / 511
          & 67.2 & 0/500 & 0.893 / 0.986 \\
          \bottomrule
      \end{tabular}
  \end{table}

\begin{figure}[!htb]
      \centering
      \includegraphics[width=\linewidth]{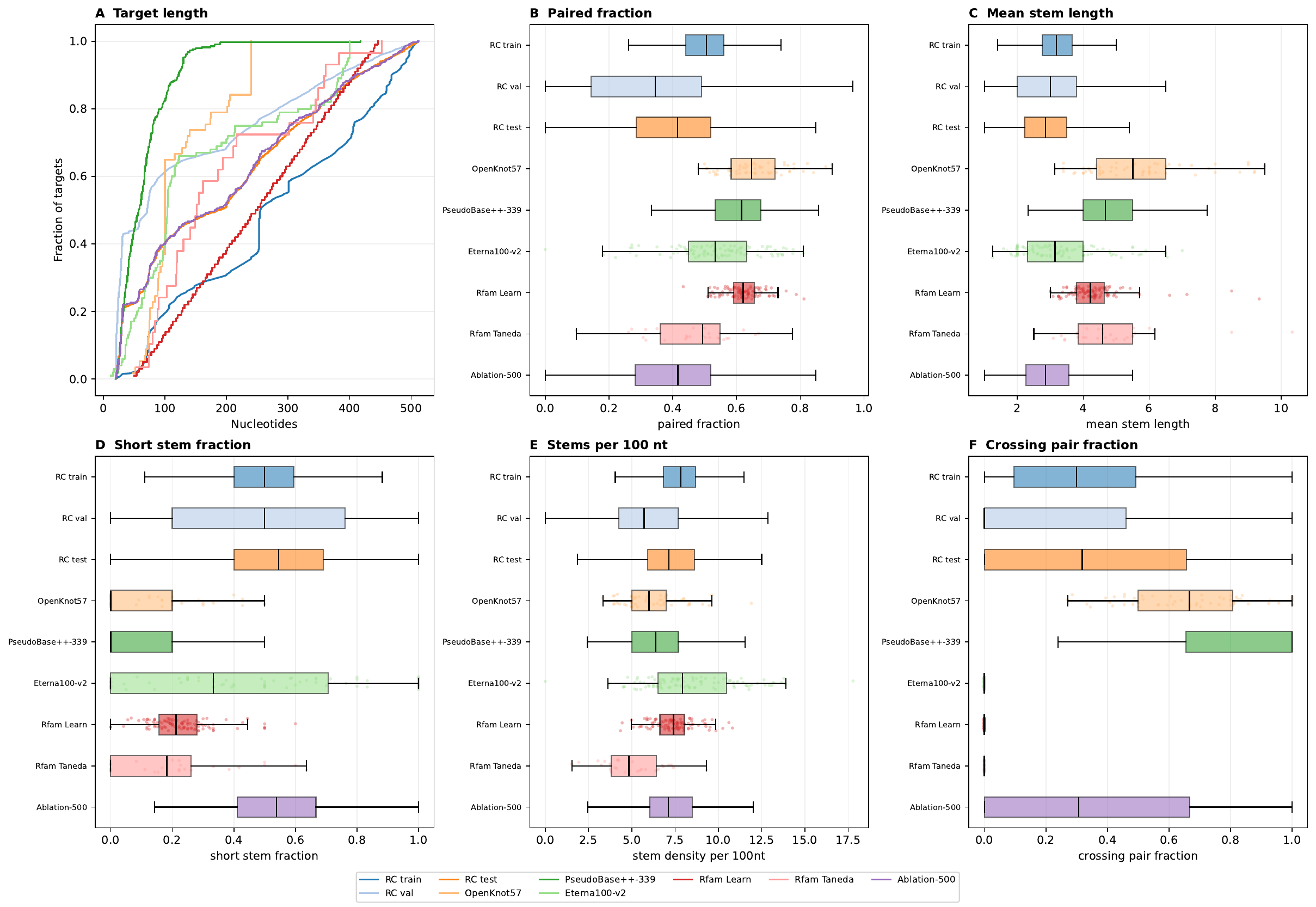}
      \caption{\textbf{Structural diversity across evaluation benchmarks.}
      Target length and five pairing and stem descriptors, with RNAcentral
      splits as references.}
      \label{fig:dataset-complexity}
  \end{figure}

  Figure~\ref{fig:dataset-complexity} extends this comparison.
  The differing distributions show that the panels cover complementary
  combinations of target size, pairing density, and stem organization,
  motivating evaluation across all five external benchmarks.

  \subsection{Overlap with generator pretraining}

  To evaluate generalization beyond memorized targets, we exclude benchmark
  matches from generator pretraining and audit the retained corpus.
  We summarize the checks here; corpus construction is detailed in Appendix~\ref{app:training-data}. Structural comparisons use canonical nucleotide-pair
  sets: an exact match requires equal length and identical pairs.
  To quantify similarity beyond exact matches, we compute
  \begin{equation}
      J_{\mathrm{pair}}(T,Q)
      = \frac{|T\cap Q|}{|T\cup Q|},
  \end{equation}
  for target pair set $T$, equal-length training pair set $Q$, and nonempty
  union $T\cup Q$. Each target's nearest-training score is the maximum
  over equal-length training candidates. Scores are undefined when no such
  candidate exists, including three Eterna100-v2 targets below the minimum
  training length.

\begin{figure}[!htb]
      \centering
      \includegraphics[
          width=\linewidth,
          height=0.20\textheight,
          keepaspectratio
      ]{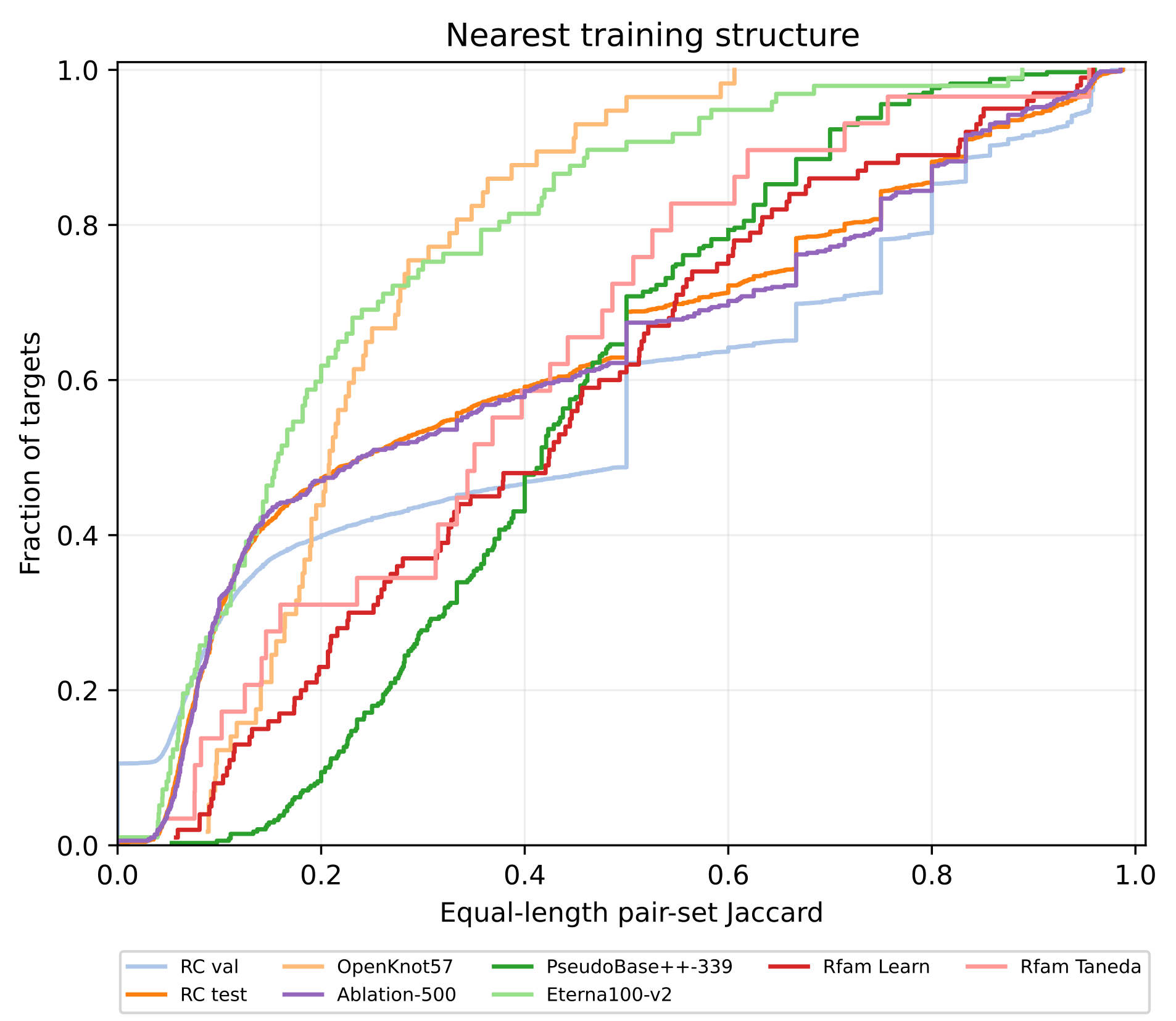}
      \caption{\textbf{Structural similarity to generator pretraining.}
      Cumulative distributions of each target's maximum pair-set Jaccard
      to an equal-length training structure. Lower Jaccard scores indicate less
      overlap; targets without an equal-length neighbor are omitted.}
      \label{fig:dataset-independence}
  \end{figure}

  None of the five external benchmarks or Ablation-500 contains an exact
  training target (Table~\ref{tab:dataset-audit}).
  Figure~\ref{fig:dataset-independence} shows the distribution of
  nearest-training similarities, revealing different degrees of structural
  proximity despite this absence of duplicates. OpenKnot57 is particularly
  distinct under this measure: its nearest-training Jaccard never exceeds
  0.606, whereas some targets in other panels have close neighbors, but still not exact matches.

  We also screen available reference or native sequences, following the use
  of sequence-similarity filtering in RNA inverse-design
  benchmarks~\citep{huang2024ribodiffusion}.
  Our audit detects no exact full-sequence matches or MMseqs2
  hits~\citep{steinegger2017mmseqs2} meeting both $\geq80\%$ identity and
  $\geq80\%$ coverage of each sequence; exclusions extend to complete
  sequence clusters containing matched members.
  These checks establish separation under the stated criteria, not
  the absence of shared RNA families or motifs. Because multiple sequences
  can realize the same target structure, sequence checks apply to supplied
  references rather than all possible design solutions.

  \subsection{3D Evaluation Set Curation}

  We use the Das test split released in the gRNAde repository~\citep{joshi2025grnade}, following its use by RIDER as a common benchmark for 3D RNA inverse design~\citep{hu_rider_2026}. We select the first
  stored conformer for each of its 98 records. We exclude 20 multi-chain records
  and 14 single-chain records with DSSR-detected backbone breaks, leaving 64
  continuous single-chain targets.
  These criteria are applied before inspecting any model output. DSSR
  v2.9.3~\citep{lu_dssr_2015}, with default annotations, provides the
  secondary-structure targets, including pseudoknots. We standardize inputs to
  meet each generator's atom requirements while retaining the original
  experimental coordinates for reward computation and final evaluation.  
 \clearpage

\clearpage
  \section{Composition and diversity of generated sequences}
  \label{app:output_characterization}

 \begin{wrapfigure}{r}{0.65\textwidth}
      \centering
      \includegraphics[width=\linewidth]
          {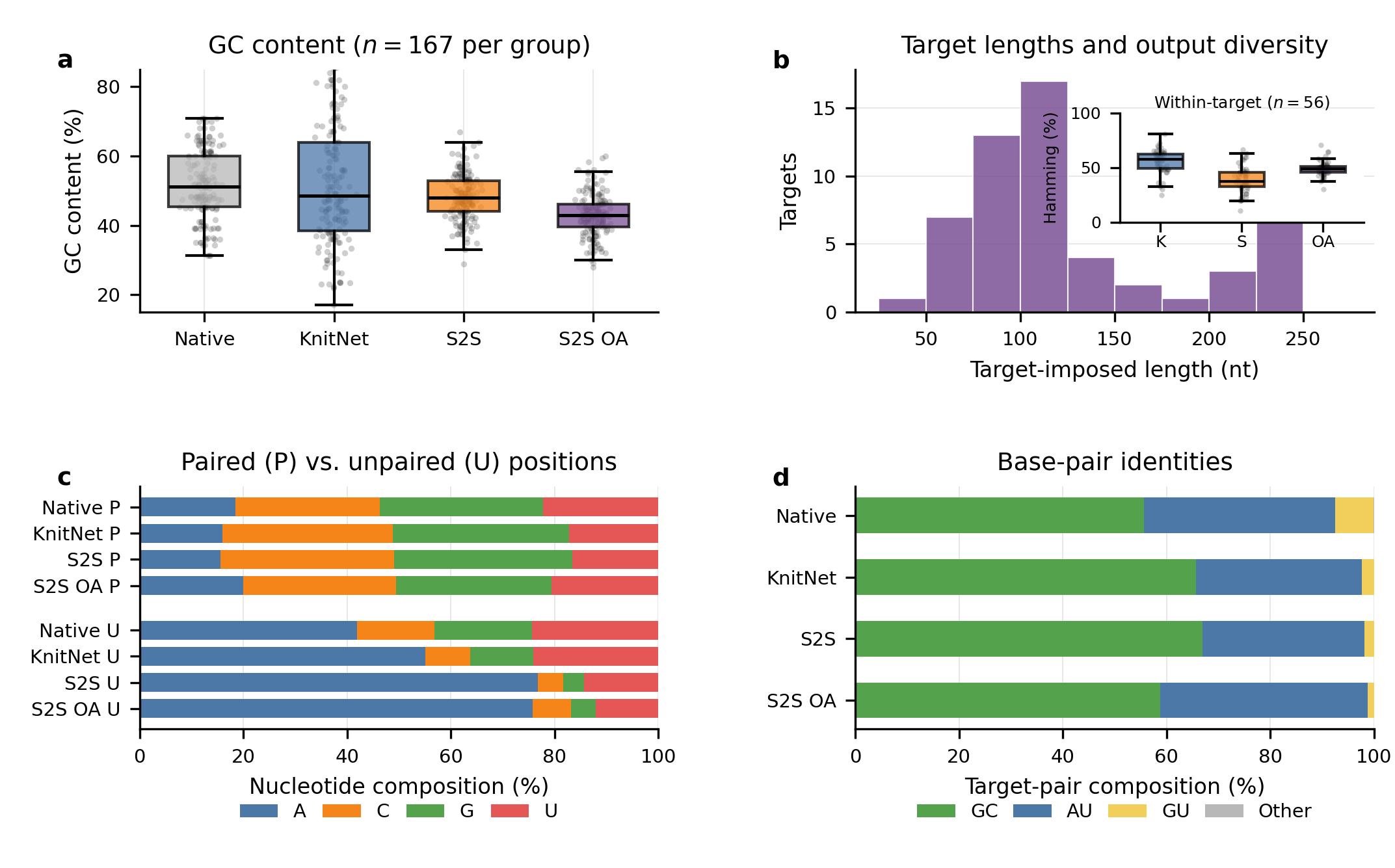}
      \caption{\textbf{Composition and diversity of matched OpenKnot57 outputs.}}
      \label{fig:knitnet-output-characterization}
  \end{wrapfigure}

  We characterize the sequences designed by KnitNet, Struct2Seq, and
  Struct2Seq OA using the first successful design for each target,
  restricting the comparison to 167 target-seed cases solved by all
  three methods across three seeds on OpenKnot57
  (Figure~\ref{fig:knitnet-output-characterization}).
  KnitNet produces more diverse sequences for the same target across
  seeds, with a median target-level Hamming distance of 57.7\%, compared with 37.7\% for Struct2Seq and 48.6\% for Struct2Seq OA. Its mean GC content (51.8\%) is also closest to the native reference
  (51.6\%), compared with 48.1\% and 43.0\%, respectively, although
  its GC distribution is wider and other composition differences
  are mixed. These statistics characterize successful computational designs
  and suggest the methods output biologically-feasible sequences.

  \subsection{Composition and diversity of 3D generator outputs}

\begin{wrapfigure}{r}{0.65\textwidth}
      \centering
      \includegraphics[width=\linewidth]
          {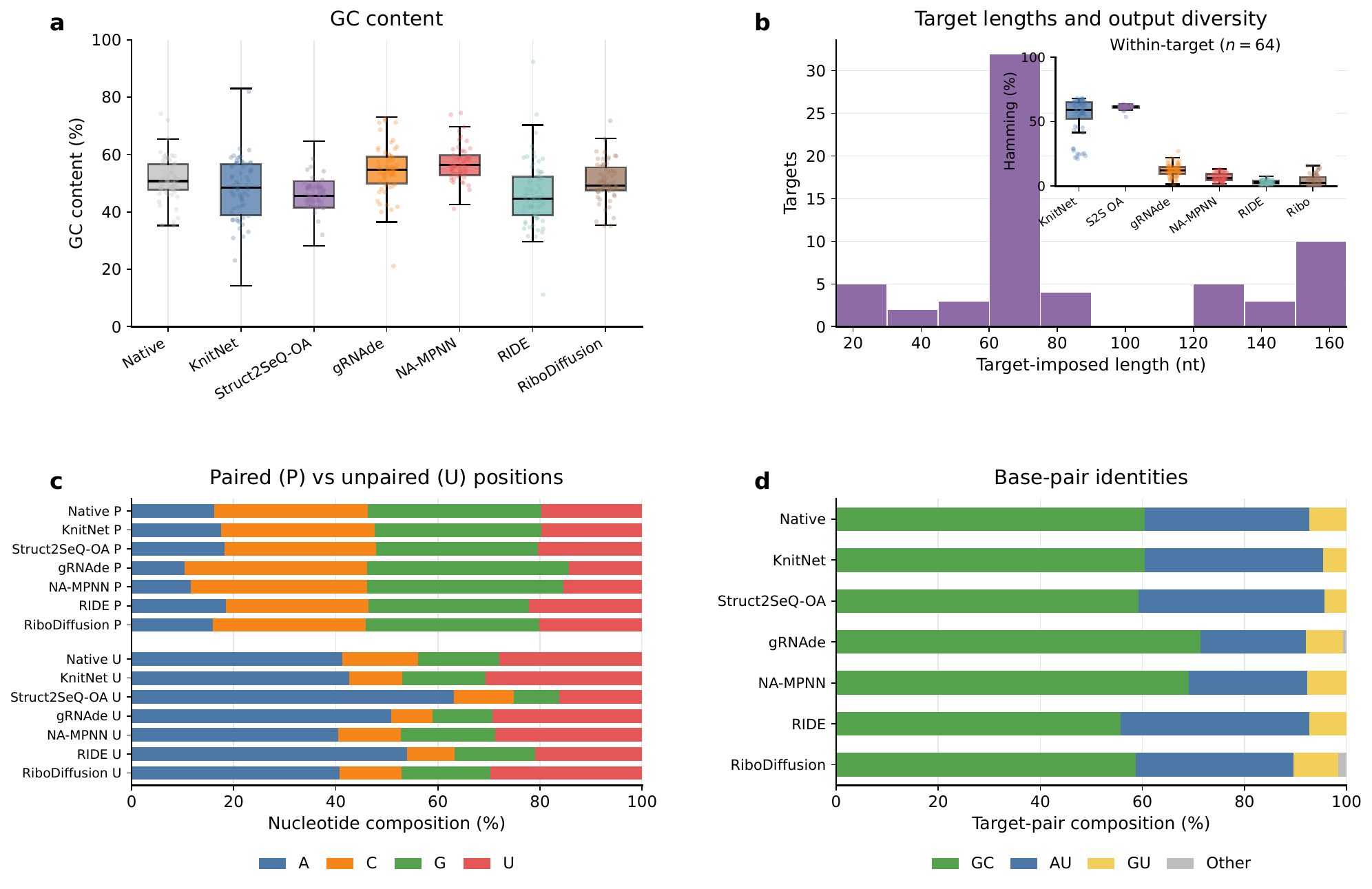}
      \caption{\textbf{Composition and diversity of 3D generator outputs.}}
      \label{fig:knitnet-3d-output-characterization}
  \end{wrapfigure}

  We characterize all 96 initial sequences generated for each of the 64
  three-dimensional targets at seed 42, giving 6,144 sequences per generator
  and 36,864 generated sequences in total. Every generated sequence is included
  regardless of its fold-back score. Search and target-specific RL outputs are
  excluded, so this analysis describes the original generator distributions
  rather than selected final designs.

  KnitNet produces a broad initial sequence distribution, with 54.1\% mean
  within-target Hamming diversity, second to Struct2SeQ-OA at 61.2\%.
  The other four generators yield substantially more concentrated pools, with
  mean diversity between 3.0\% and 12.2\%. KnitNet's mean GC content is 48.8\%,
  compared with 51.3\% for the native references. The paired-position,
  unpaired-position, and target-pair compositions in
  Figure~\ref{fig:knitnet-3d-output-characterization} further show that the
  generators occupy distinct sequence distributions even when conditioned on
  the same targets.
\clearpage
\section{Generator Architecture, Pretraining, and Sampling }
\label{app:generator_details}

  \subsection{Graph representation and features}
  \label{app:graph_features}
  Each nucleotide is a node, with bidirectional backbone and target
  base-pair edges, including crossing pairs for pseudoknots. Each node
  has 31 structural features: pairing status, normalized sequence
  position, eight Laplacian positional encodings, 16 random-walk
  return probabilities, and five descriptors of loop size, stem size,
  nesting depth, junction degree, and distance to a sequence end.
  Each directed edge has 18 features: edge type (2), binned sequence
  separation (10), bracket layer (1), and indicators of stem, loop,
  and crossing relationships (5). Structural features depend only
  on the target structure; nucleotide-or-mask and diffusion-noise
  embeddings are added separately.
  
  \subsection{Denoiser architecture}
  The denoiser has 12 GraphGPS layers, hidden dimension $d=512$, eight
  attention heads, and approximately 36.2 million parameters. Each layer
  combines edge-aware GATv2 message passing with global self-attention,
  normalization, and residual connections. For each nucleotide, we sum a
  projected structural feature vector, a nucleotide-or-mask embedding,
  and an embedding of diffusion noise level $\sigma(t)$.

  Dropout rates are $0.15$ within
  GraphGPS blocks, $0.2$ for global attention, and $0.1$ after each block
  and in the pair head.

  With generator parameters $\theta$, a linear nucleotide head maps node
  embeddings $\mathbf h_i\in\mathbb R^{d}$ to four logits. Their softmax
  gives the nucleotide distribution $p^{(4)}_{\theta,i}:\mathcal A\to[0,1]$
  at position $i$. For an ordered target pair $(i,j)$ with $i<j$, the
  pair-head multilayer perceptron $g_\theta:\mathbb R^{4d}\to\mathbb R^{6}$
  predicts the joint assignment distribution $p^{(6)}_{\theta,ij}:\mathcal A_{\mathrm{pair}}\to[0,1]$:
  \begin{equation}
      p^{(6)}_{\theta,ij}
      =\operatorname{softmax}\!\left(
        g_\theta([\mathbf h_i,\mathbf h_j,\mathbf h_i\odot \mathbf h_j,|\mathbf h_i-\mathbf h_j|])
      \right).
      \label{eq:knitnet_pair_head}
  \end{equation}
  The pair head has widths
  $2048\!\rightarrow\!256\!\rightarrow\!512\!\rightarrow\!6$, using SiLU
  and dropout after the first linear layer and layer normalization
  before the final projection. Its classes are the six assignments
  in $\mathcal{A}_{\mathrm{pair}}$.

  \subsection{Masked diffusion objective}
  For each training sequence, we sample diffusion time $t$ uniformly from
  $[\epsilon,1]$ and clip it to $[\epsilon,1-\epsilon]$, with
  clipping constant $\epsilon=10^{-3}$. The log-linear noise schedule is
  $\sigma(t)=-\log[1-(1-\epsilon)t]$. Nucleotides are independently masked
  with probability
  $\mu(t)=1-\exp[-\sigma(t)]$~\cite{sahoo2024mdlm}.
  The time weight is
  $w(t)=\min\{100,\sigma'(t)/(\exp[\sigma(t)]-1)\}
  =\min\{100,1/t\}$.

  For a minibatch $\mathcal B$, each example $b\in\mathcal B$ has a
  sequence $x_b$, a sampled diffusion time $t_b$, and a corrupted sequence
  $\widetilde x_{b,t_b}$ whose positions hold nucleotides or mask tokens.
  The masked-position set $\mathcal M_{\mathcal B}$ contains indices $(b,i)$.
  The supervised-pair set $\mathcal P_{\mathcal B}$ contains indices $(b,i,j)$
  for target pairs with both endpoints masked and observed chemistry in
  $\mathcal A_{\mathrm{pair}}$. Both heads condition on the corrupted
  sequence, target structure, and diffusion time. We add the example index
  $b$ to each head's distribution and minimize
  \begin{equation}
  \begin{aligned}
      \mathcal{L}_{\mathrm{nt}}
      &=-\frac{1}{|\mathcal{M}_{\mathcal{B}}|}
        \sum_{(b,i)\in\mathcal{M}_{\mathcal{B}}}
        w(t_b)\log p^{(4)}_{\theta,b,i}(x_{b,i}),\\
      \mathcal{L}_{\mathrm{pair}}
      &=-\frac{1}{|\mathcal{P}_{\mathcal{B}}|}
        \sum_{(b,i,j)\in\mathcal{P}_{\mathcal{B}}}
        w(t_b)\log p^{(6)}_{\theta,b,ij}(x_{b,i}x_{b,j}),\\
      \mathcal{L}_{\mathrm{pre}}
      &=\mathbb{E}
        [\mathcal{L}_{\mathrm{nt}}+0.25\mathcal{L}_{\mathrm{pair}}].
  \end{aligned}
  \label{eq:knitnet_pretraining}
  \end{equation}
  The expectation averages over minibatches, diffusion times, and masks.
  An empty supervision set contributes zero. Training uses independent
  nucleotide masking; joint endpoint revelation is used during sampling.

  \subsection{Training recipe and checkpoint selection}
  The released model was trained on the synthetic RNAcentral corpus containing
21,619,209 training sequences with a 512-nucleotide length cap, with
  10,000 validation and 2,000 test examples. Corpus construction and split
  definitions are described in Appendix~\ref{app:training-data}.
  Table~\ref{tab:generator_recipe} summarizes optimization. We use the
  checkpoint at update 802,294 (epoch 38) from a run configured for
  50 epochs; the learning-rate schedule spans the full configured duration.

  Every two epochs, we generate 16 candidates for each target in a fixed
  1,000-target validation subset and fold them with RibonanzaNet-SS.
  We also evaluate the full 10,000-target validation split every ten epochs.
  Checkpoints are ranked by the trailing two-evaluation mean of best-of-16
  pair Jaccard, averaged over the fixed subset. Validation sampling uses
  a fixed base seed of 1234 across checkpoints. Folding is used for
  validation, rather than in the training loss.

  \begin{table}[t]
      \centering
      \caption{Optimization settings for the released generator.}
      \label{tab:generator_recipe}
      \begin{tabular}{ll}
          \toprule
          Setting & Value \\
          \midrule
          Optimizer & AdamW, $\beta_1=0.9$, $\beta_2=0.98$ \\
          Weight decay & $0.01$ \\
          Peak learning rate & $3\times10^{-4}$ \\
          Warmup & Linear from $10^{-6}$ for 105,565 updates (10\%) \\
          Decay & Cosine, minimum learning rate $10^{-5}$ \\
          Effective batch size & 1,024 sequences \\
          Per-GPU batch / accumulation & 256 / 1 \\
          Training hardware & $4\times$ NVIDIA H200, distributed data parallel \\
          Numerical precision & Mixed bfloat16 \\
          Gradient-norm clipping & $1.0$ \\
          Configured duration & 50 epochs / 1,055,650 updates \\
          Training seed & 42 \\
          \bottomrule
      \end{tabular}
  \end{table}

  \subsection{Candidate sampling}
  We generate 96 initial candidates per target. Each trajectory starts
  fully masked and uses 32 reverse transitions on a uniform time grid
  from $1$ to $\epsilon$. At each transition from $t$ to $s<t$, an
  unrevealed unit remains masked with probability $\mu(s)/\mu(t)$;
  otherwise its assignment is sampled. A unit is either an unpaired
  nucleotide or a target pair, whose endpoints are revealed together
  using the joint head. Paired assignment follows its established use
  in RNA design~\cite{runge2019learna,tang2026samplingdesign}.
  We apply temperature $0.8$ and top-$k=2$ truncation separately to the
  four-way nucleotide and six-way pair distributions. Revealed assignments
  remain fixed. An additional denoiser pass fills any remaining masks
  by argmax, retaining joint pair assignments. Validation uses the same
  sampler with 16 candidates per target. Evaluataion seeds are specified
  with each benchmark protocol.

  \subsection{Training compute}
  Training to the selected checkpoint used four H200 GPUs, totaling
  552 H200 GPU-hours.

  \clearpage
  \section{Learned Search Algorithm Pseudocode}
  \label{app:search_pseudocode}

  Algorithm~\ref{alg:knitnet_repair} states the search for secondary-structure design, where a candidate solves the target when $F(x)=T$. The budget $B$ and spent count $b$ measure logical folding evaluations after initialization. The batches $\mathcal J$ and $\mathcal U$ contain evaluation requests, and their sizes count repeated requests separately. The operation $\operatorname{uniq}(\cdot)$ extracts a batch's set of distinct sequences. The unset-surrogate marker $f=\bot$ triggers uniform joint-proposal sampling. For 3D design, Appendix~\ref{app:3d_protocol} keeps these secondary-structure repairs but ranks and exports candidates by a tertiary reward and uses a different stopping rule.

  \begin{algorithm}[H]
      \caption{Counterexample-guided joint repair for secondary-structure design}
      \label{alg:knitnet_repair}
      \footnotesize
      \begin{algorithmic}[1]

          \Require Target $T$, oracle $(F,m_F)$, evaluated candidates $\mathcal R$,
                   post-initialization budget $B$

          \State Cache $\mathcal R$;
                 $\mathcal P\gets\Call{SelectPopulation}{\mathcal R}$;
                 $b\gets0$; close episode

          \While{$b<B$ and no cached candidate solves $T$}

              \KnitComment{Start a scoped repair episode}
              \State \textbf{if} episode closed,
                     $(a,S,C,\mathcal D,\mathcal H)\gets\Call{StartEpisode}{\mathcal P}$
                     \label{ln:knitnet:scope}

              \KnitComment{Propose and rank joint assignments}
              \State $\mathcal Q\gets\Call{JointProposals}{a,S,\mathcal H}\setminus \mathcal D$;
                     $f\gets\Call{FitExtraTrees}{\mathcal D,S,C}$
                     if $\mathcal Q\ne\varnothing$ and $|\mathcal D|\ge 8$; else $f\gets\bot$
                     \label{ln:knitnet:fit}
              \State $(\mathcal J,\mathcal U)\gets\Call{SelectBatch}{\mathcal Q,f,\mathcal P,B-b}$
                     \label{ln:knitnet:select}

              \KnitComment{Verify complete folds}
              \State Fold and cache $\mathcal J$ and $\mathcal U$;
                     $b\gets b+|\mathcal J|+|\mathcal U|$
                     \label{ln:knitnet:verify}

              \KnitComment{Refine counterexamples and relabel observations}
              \If{$\mathcal J\ne\varnothing$}
                  \State Choose $x^\star\in
                         \arg\min_{x\in \mathcal J}^{\mathrm{lex}}\ell_C(x)$;
                         $r\gets\Call{RecordProgress}{x^\star,\mathcal J,\mathcal D,C}$
                         \label{ln:knitnet:witness}
                  \State $\mathcal D\gets \mathcal D\cup\operatorname{uniq}(\mathcal J)$;
                         $C\gets C\cup E(x^\star)$
                         \label{ln:knitnet:revision}
                  \State Relabel $\mathcal D$ from cached folds;
                         $\mathcal H\gets\Call{QualityDiversityBeam}{\mathcal D,S,C}$
                         \label{ln:knitnet:relabel}
              \EndIf

              \KnitComment{Update the population and expand the repair scope}
              \State $\mathcal P\gets\Call{SelectPopulation}{\mathcal P\cup\operatorname{uniq}(\mathcal J)\cup\operatorname{uniq}(\mathcal U)}$;
                     close episode if episode budget exhausted
                     or ($\mathcal Q=\varnothing$ and $S$ cannot grow)
              \State \textbf{if} episode open and
                     $\bigl(\mathcal Q=\varnothing
                     \lor (\mathcal J\ne\varnothing\land r)\bigr)$,
                     $S\gets\Call{ExpandScope}{S,\mathcal H}$
                     \label{ln:knitnet:expandtest}
                     \label{ln:knitnet:expand}

          \EndWhile

          \State \Return an exact cached candidate, or the best cached
                 candidate marked unsolved
                 \label{ln:knitnet:success}

      \end{algorithmic}
  \end{algorithm}

\clearpage

  \section{Joint repair implementation details}
  \label{app:repair_details}

  \paragraph{Population and episode budget.}
  We retain up to 32 distinct sequences, ranked by oracle merit, then pair Jaccard and fewer pairing errors. Anchors rotate through this population. We cap each episode at 512 joint evaluations to limit search in one sequence context. Each batch contains at most 32 evaluations. For a batch of $n=|\mathcal J|+|\mathcal U|$ evaluation requests with $\operatorname{uniq}(\mathcal J)\subseteq\mathcal Q$, we reserve at least $\max(1,\lfloor n/4\rfloor)$ slots for ordinary single-unit mutations to preserve simple corrections outside the current repair scope. Unused joint slots also go to these mutations. Both types consume the overall search budget.

  \paragraph{Scope construction and expansion.}
  In $\mathcal G_a$, an error whose endpoints belong to the same assignment unit produces a self-loop. Repeated observations contribute to the error's recurrence count rather than parallel graph edges. We count each anchor error twice and each occurrence in its preceding 15 ancestors once, emphasizing current errors while retaining recurring conflicts. We initialize the scope with the error-graph component containing the highest-count current error. For a singleton component, we add the nearest other assignment unit along the sequence, when available, to permit a coupled edit. We expand after four consecutive joint batches without lexicographic improvement over the best preceding episode observation, comparing both under the current prioritized error set $C$. We also expand when the previously prioritized errors are resolved but others remain, or when proposals are exhausted. In Algorithm~\ref{alg:knitnet_repair}, \textsc{RecordProgress} returns $r=\mathrm{true}$ when one of the first two conditions holds, and exhausted proposals appear as $\mathcal Q=\varnothing$. Each expansion adds up to $\max(2,|S|)$ units, prioritizing their endpoint incidence in residual errors across the proposal beam and breaking ties by sequence distance to $S$. Exhaustion at full scope ends the episode.

  \paragraph{Proposal generation.}
  We enumerate the product of the pair-substitution and nucleotide alphabets when it contains at most 4096 assignments.
  For larger spaces, we combine all single-unit neighbors of beam members, 1024 unit-wise crossover proposals with mutations at up to two units, and 256 random assignments. We remove duplicates and previously evaluated episode sequences before selection; constructing the pool requires no folding calls.

  \paragraph{Surrogate and exploration.}
  We fit a multi-output ExtraTrees regressor after eight distinct observations, including the anchor, using 16 trees
  in the original experiments,
  with maximum depth 10, minimum leaf size one, all features eligible at each split, and random seed 42.
  The component ablations in
  Table~\ref{tab:pat4-ablations-external} use 64 trees for the ExtraTrees surrogate.
  One-hot features follow a fixed unit order and include an additional category for incompatible observed pair assignments. For a joint quota $q>0$, we reserve $\max(1,\lfloor q/4\rfloor)$ evaluations for uniform
  exploration and select the remainder by predicted lexicographic rank. Before fitting, selection is entirely uniform. This preserves exploration when predictions are inaccurate.

  \paragraph{Protected proposal beam.}
  We retain up to six assignments by repair loss, then add up to two that greedily maximize minimum assignment Hamming distance to those retained. This preserves alternative combinations independently of outer-population selection. Only the best newly evaluated joint proposal contributes errors to the prioritized error set $C$, preventing arbitrary exploratory failures from expanding it. All joint observations remain available for refitting, with labels recomputed from cached folds without additional evaluations.

\clearpage

 \clearpage
  \section{Ablations on the Held-Out Test Set}
  \label{app:knitnet-ablations}

  \begingroup

  We report the complete ablation results on the held-out 500-target
  panel described in Appendix~\ref{app:training-data}.
  Each configuration starts with 96 candidates and receives up to
  60 minutes of additional sampling or search, using RibonanzaNet-SS
  evaluation and the original 16-tree ExtraTrees surrogate.
  Exact counts include targets solved at initialization; solve AUC
  is normalized over the 60-minute horizon.

  \begin{table}[htbp]
      \centering
      \small
      \setlength{\tabcolsep}{5pt}
      \renewcommand{\arraystretch}{1.08}

      \caption{
      \textbf{KnitNet ablations on 500 held-out targets.}
      Search TTS is the median post-initialization time among targets
      newly solved during search. Jaccard measures endpoint
      structural agreement.}
      \label{tab:pat4-ablations-rc500}

      \begin{adjustbox}{max width=\linewidth}
      \begin{tabular}{@{}lcccc@{}}
          \toprule
          Configuration
          & \shortstack{Exact\\(/500) $\uparrow$}
          & \shortstack{Solve\\AUC $\uparrow$}
          & \shortstack{Search TTS\\(min) $\downarrow$}
          & Jaccard $\uparrow$ \\
          \midrule

          \textbf{Full KnitNet search}
          & \textbf{306} & \textbf{0.5698} & 1.37 & 0.9230 \\

          \midrule
          \multicolumn{5}{l}{\textit{Sampling without search}} \\
          Best-of-96
          & 139 & 0.2780 & -- & 0.8102 \\
          Additional sampling
          & 209 & 0.3954 & -- & 0.8933 \\

          \midrule
          \multicolumn{5}{l}{\textit{Initialization control}} \\
          Random init.\ then search
          & 210 & 0.4067 & 0.17 & 0.7365 \\

          \midrule
          \multicolumn{5}{l}{\textit{Search component ablations}} \\
          W/ random repair scope
          & 293 & 0.5435 & 0.93 & 0.9133 \\
          W/o surrogate ranking
          & 297 & 0.5515 & 0.99 & 0.9178 \\
          W/ fixed repair problem
          & 301 & 0.5598 & 1.01 & 0.9164 \\
          W/o protected proposal beam
          & 301 & 0.5600 & 1.03 & 0.9184 \\
          W/ linear surrogate ranking
          & 301 & 0.5637 & 0.87 & \textbf{0.9265} \\

          \bottomrule
      \end{tabular}
      \end{adjustbox}
  \end{table}

  \paragraph{Component controls.}
  Random scope selection preserves the numbers of paired and
  unpaired assignment units while replacing error-guided localization.
  Removing surrogate ranking selects joint proposals uniformly.
  The fixed-problem variant freezes both the editable scope and
  error priorities within each episode. Removing the protected beam
  restricts proposal parents to outer-population survivors while
  retaining observations for surrogate fitting. The linear-surrogate
  variant replaces ExtraTrees with additive ridge regression.

  \paragraph{Results.}
  Full KnitNet solves 97 more targets than continued sampling and 96 more than search from random initialization. Every component ablation reduces exact coverage and solve AUC. Linear ranking achieves slightly higher Jaccard despite fewer exact solves, illustrating the distinction between partial agreement and exact recovery. Search TTS excludes unsolved targets, so a method can appear faster by solving only easier targets. Exact coverage and solve AUC account for the full panel.

  \par
  \endgroup

\clearpage

  \section{Folding oracles and evaluation protocol}
  \label{app:oracle_evaluation}

  \subsection{Two-dimensional fold-back evaluation}
  \label{app:2d_protocol}

  We use complete predicted pair sets to construct repair problems and
  determine exactness. Pair discrepancies provide a shared interface across
  the secondary-structure oracles, while continuous, oracle-specific merits
  distinguish candidates with similar residual errors. These merits rank
  candidates within each oracle; their numerical values are not compared
  across oracles.

  \smallskip
  \noindent\textbf{OpenKnot57 (RibonanzaNet-SS).}
  RibonanzaNet-SS supplies directed pair scores and a pseudoknot-capable
  fold~\cite{he2024ribonanza,he2026struct2seq}. We use Arnie's Hungarian
  decoder with threshold $\tau=0.5$ and minimum helix length one.
  For candidate sequence $x$, the directed pair scores $q_{ij}\in[0,1]$
  form an $L\times L$ matrix after masking $|i-j|\leq3$. We suppress
  their dependence on $x$ and define
  $b_{ij}=\operatorname{logit}(q_{ij})$,
  $\bar b_{ij}=\operatorname{logit}((q_{ij}+q_{ji})/2)$, and
  $b_\tau=\operatorname{logit}(\tau)$, clipping logit arguments to
  $[10^{-7},1-10^{-7}]$. We define the margins $c_{ij},c_i:\mathcal A^{L}\to\mathbb R$ by
  \begin{equation}
  \begin{aligned}
   c_{ij}(x)
   &=\bar b_{ij}-\max\!\left\{
   b_\tau,\max_{k\notin\{i,j\}}b_{ik},
   \max_{k\notin\{i,j\}}b_{jk}\right\},
   &&(i,j)\in T,\\
   c_i(x)&=b_\tau-\max_{k\ne i}b_{ik},
   &&i\text{ unpaired in }T,
  \end{aligned}
  \label{eq:knitnet_margins}
  \end{equation}
  and use the weakest constraint as the merit:
  \begin{equation}
   m_F(x)=\min_{c\in\mathcal C(T)}c(x).
   \label{eq:knitnet_margin_merit}
  \end{equation}
  The constraint-margin set $\mathcal C(T)=\{c_{ij}:(i,j)\in T\}\cup\{c_i:i\text{ unpaired in }T\}$ contains the margin functions for all
  target-pair and target-unpaired constraints. Positive margins favor target pairs over the threshold
  and competing partners, while suppressing pairing at unpaired sites.
  Maximizing the minimum focuses search on the least satisfied
  requirement; exactness is checked separately on the decoded fold.

  \smallskip
  \noindent\textbf{Eterna100 (ViennaRNA).}
  We use ViennaRNA 2.6.4 with Turner 2004 parameters at
  $37\,^{\circ}\mathrm{C}$~\cite{andersonlee2016eterna100,lorenz2011viennarna,mathews2004turner,turner2010nndb}.
  Following thermodynamic inverse-design
  objectives~\cite{zhou2023samfeo,tang2026samplingdesign}, we rank candidates
  by target ensemble probability:
  \begin{equation}
   \log P_{\mathrm V}(T\mid x)
   =-\frac{G(x,T)-G_{\mathrm{ens}}(x)}{R\mathcal T},
   \qquad G_{\mathrm{ens}}(x)=-R\mathcal T\log Z(x),
   \label{eq:knitnet_vienna}
  \end{equation}
  Here the target free energy $G(x,T)$ and ensemble free energy
  $G_{\mathrm{ens}}(x)$ are measured in kcal/mol. The partition function
  $Z(x)$, gas constant $R$ in kcal\,mol$^{-1}$\,K$^{-1}$, and absolute
  temperature $\mathcal T=310.15\,\mathrm K$ determine the ensemble free energy. This
  rewards the target relative to competing folds, rather than lowering
  its energy alone. The incompatibility count $n_{\mathrm{inc}}$ is the
  number of target pairs assigned nucleotides outside
  $\mathcal A_{\mathrm{pair}}$. We use
  \begin{equation}
   m_F(x)=
   \begin{cases}
    1+[1-\log P_{\mathrm V}(T\mid x)]^{-1}, & n_{\mathrm{inc}}=0,\\
    -n_{\mathrm{inc}}/|T|, & n_{\mathrm{inc}}>0.
   \end{cases}
   \label{eq:knitnet_vienna_merit}
  \end{equation}
  Since $P_{\mathrm V}(T\mid x)\in(0,1]$, the denominator $1-\log P_{\mathrm V}(T\mid x)$ is at least one.
  This preserves probability ordering among compatible sequences and
  ranks them above incompatible ones. Exactness requires the returned
  minimum-free-energy pair set to equal $T$.

  \smallskip
  \noindent\textbf{Structural agreement and exact coverage.}
  We evaluate structural fold-back rather than recovery of a particular
  reference sequence~\cite{joshi2025grnade}. Pair Jaccard similarity
  diagnoses partial progress by penalizing missing and unwanted pairs:
  \begin{equation}
   J(x,T)=\frac{|F(x)\cap T|}{|F(x)\cup T|}.
  \label{eq:knitnet_jaccard}
  \end{equation}
  We set $J(x,T)=1$ when both pair sets are empty.
  Our primary endpoint is exact recovery, $F(x)=T$, equivalent to $J=1$
  under this convention. For a panel of $N$ targets, define the
  first exact-solve time $t_j^\star$ as elapsed wall time after initialization
  for target $j$, with $t_j^\star=0$ for initial solves and
  $t_j^\star=\infty$ for completed nonsolves. At elapsed search time $t$
  and time horizon $t_{\max}>0$, we report
  \begin{align}
   \mathrm{Coverage}(t)
   &=\frac1N\sum_{j=1}^N\mathbf1[t_j^\star\leq t],
   \label{eq:knitnet_coverage}\\
   \mathrm{AUC}(t_{\max})
   &=\frac1{t_{\max}}\int_0^{t_{\max}}\mathrm{Coverage}(t)\,dt.
   \label{eq:knitnet_auc}
  \end{align}
  Coverage includes every target in the evaluated panel and counts
  initial solves. The horizon is 30 minutes for the pseudoknot benchmarks
  and 60 minutes for the planar benchmarks and ablations. AUC rewards
  earlier recovery within the same time horizon.
  Incomplete panels yield coverage lower bounds, with unfinished or
  failed targets reported separately rather than treated as nonsolves.
  Final AUC and comparisons require complete, valid panels.
  Sequence composition and diversity are characterized in
  Appendix~\ref{app:output_characterization}.

  \subsection{Three-dimensional fold-back evaluation}
  \label{app:3d_protocol}

  \smallskip
  \noindent\textbf{Conditioning and sampling.}
  KnitNet and Struct2SeQ-OA condition on the target's DSSR pair graph \cite{lu_dssr_2015};
  the tertiary generators condition on 3D coordinates. We use their
  pretrained checkpoints and the settings in Table~\ref{tab:3d_sampling}.
  For each of seeds 42, 731, and 2718, we generate a fresh 96-sequence pool
  for every generator, evaluate it as the frozen arm, and use that same pool
  to initialize the corresponding search arm. Generation and search randomness
  vary with the run seed. Generator weights remain fixed during search.

  \input{tables/3d_sampling}

  \smallskip
  \noindent\textbf{Search objective and budgets.}
  We retain RibonanzaNet-SS pairing residuals for repair construction and
  local surrogate learning, but use RhoFold+ to guide population ranking
  and terminal-pool selection toward tertiary
  agreement~\citep{shen2024rhofoldplus}. Search and RL share the released
  RIDER reward $R:\mathcal A^{L}\to\mathbb R$~\citep{hu_rider_2026}:
  \begin{equation}
   R(x)=25g^2-0.25r^2+
   \begin{cases}
    100(g-0.45), & g>0.45,\\
    20(3-r), & g\leq0.45\ \text{and}\ r<3,\\
    0, & \text{otherwise}.
   \end{cases}
   \label{eq:3d_reward}
  \end{equation}
  With fixed residue correspondences and a proper Kabsch fit on observed
  reference C4$'$ atoms, the distance-threshold agreement $g$ averages the
  fractions at distances below 1, 2, 4, and 8\,\AA. The RMSD $r$ is measured
  in \AA. These released bonus
  thresholds do not define exact tertiary recovery. Optimization accesses
  the target coordinates; changing to C3$'$ for final scoring does not
  make evaluation independent.

  Search allows 2,048 post-initialization logical evaluations plus up to
  96 root evaluations. Its retained stopping rule can terminate when
  all active candidates match the secondary pair graph, without implying
  tertiary recovery. We export up to 96 distinct evaluated sequences
  ranked by $R(x)$, without forced retention of roots.
  RIDER instead restarts from RIDE for each target, uses 2,048 adaptation
  reward calls ($34\times60+8$ rollouts), then generates 96 candidates.
  Terminal slot counts therefore match, but total design-time allowances
  differ; final evaluation folds are additional for every arm.

  RIDER adaptation uses 30 diffusion steps, temperature $0.5$, learning
  rate $5\times10^{-5}$, policy clipping range $0.5$, gradient-norm
  limit $1.0$, two policy passes per batch, and mixed precision.
  Final sampling uses the RIDE settings in Table~\ref{tab:3d_sampling}.

  \smallskip
  \noindent\textbf{Final scoring and aggregation.}
  We fold terminal candidates with a common RhoFold+ checkpoint in
  single-sequence mode, without MSA search, templates, or relaxation.
  US-align fits the RNA TM objective on the observed reference C3$'$
  positions $\mathcal O_{C3'}\subseteq\{1,\dots,L\}$ with fixed residue correspondences~\citep{zhang_us-align_2022}:
  \begin{equation}
   \mathrm{scTM}(x)=\frac1L\sum_{i\in\mathcal O_{C3'}}
   \frac{1}{1+(d_i/d_{0,\mathrm{RNA}}(L))^2}.
   \label{eq:3d_sctm}
  \end{equation}
  The aligned distance $d_i$ between predicted and reference atoms at
  position $i$ is measured in \AA\ after US-align's heuristic rigid-body fit.
  The RNA length scale $d_{0,\mathrm{RNA}}(L)$ is also measured in \AA.
  Both this scale and the normalization denominator use the original
  target length $L$, so missing atoms cannot shorten the normalization.

  We select the highest-scTM candidate among each target's 96 slots,
  breaking ties by sample order. RMSD and GDT$_{\mathrm{TS}}$ describe
  that same candidate under a separate proper C3$'$ Kabsch fit;
  GDT$_{\mathrm{TS}}$ averages fractions within 1, 2, 4, and 8\,\AA.
  We retain duplicate native draws and failed or unfilled slots without
  replacement. Targets with no scored candidate receive zero scTM and
  undefined RMSD/GDT$_{\mathrm{TS}}$, omitted from those means.
  All 39 method-seed runs have scored candidates for all 64 targets, so every
  seed-level mean weights the full panel equally. We average each metric over
  the 64 targets within each seed, then report the mean and sample standard
  deviation of those averages over seeds 42, 731, and 2718.

  \smallskip
  \noindent\textbf{Small-budget runtime comparison.}
  We time sequential seed-42 runs on the same H200 GPU with eight CPU
  cores, using 3GES chain A (67 nt), the lower median target sorted by
  length and identifier. Each method receives 240 logical reward calls:
  96 root plus 144 search calls, or four 60-rollout RIDER epochs with
  eight policy updates. Process-launch-to-exit timing includes loading,
  reward evaluation, updates, and output writing; it excludes root
  generation, RL terminal export, and final fold-back evaluation.
  Native caching remains enabled, so this measures total runtime rather
  than isolated optimizer overhead. It does not establish a full-budget
  or benchmark-wide speedup.

  \smallskip
  \noindent\textbf{Shared computation accounting.}
  Across both regimes, logical calls include inner-repair and cache-served
  requests. We count all evaluations executed in a batch and record the
  first exact candidate's ordered call index. Reported solve AUC uses elapsed
  wall time. We distinguish
  initialization from search/adaptation and final evaluation, and report
  physical oracle evaluations, neural forwards, and wall time because
  logical-call matching alone does not establish equal computational cost.

 \clearpage

\section{Limitations and Future Work }
\label{app:limitations_future_work}

  \noindent\textbf{Wet-lab validation.}
  Our study establishes computational generalization across several RNA
  inverse-design settings via exact fold-back matches and tertiary self-consistency measure agreement with computational oracles. While this is appropriate for the computational contribution we are submitting to this venue, we acknowledge that experimental structural validation and functional assays are essential to establish KnitNet’s practical utility
  in the laboratory and assess its potential for integration into drug-development pipelines. We intend to participate in future OpenKnot challenges to evaluate our designs experimentally, following the chemical-mapping and structural-validation framework demonstrated by \citet{townley2026denovo}. These experiments would test how improvements under our folding objectives translate to physical RNA structures.

  \noindent\textbf{Optimizing biological properties.}
  Having demonstrated KnitNet’s ability to optimize sequences under different structural objectives and target distributions, a natural next step is to replace or augment the folding-based objective with additional biological properties, such as binding, stability, or translation. Predictors trained on SELEX measurements or RNA-ligand binding labels could guide aptamer optimization, building on activity-guided generation such as
  RaptGen~\citep{iwano2022raptgen}. Other directions include RNA-protein
  binding, mRNA stability, and translation through coding-sequence and UTR
  design~\citep{zhang2023lineardesign,sample2019utr}. These applications would
  combine structural requirements with functional objectives and sequence
  constraints, such as preserving the encoded protein. Adapting repair to
  scalar biological feedback would also require learning which positions to
  modify when explicit pairing discrepancies are unavailable.

  \noindent\textbf{Preserving functional motifs.}
  A central challenge in de novo functional RNA design is to combine sequence
  exploration with control over motifs essential to the intended biological
  activity~\citep{joshi_generative_2025,townley2026denovo}.
  High native-sequence recovery is often used to assess RNA inverse folding:
  for example, gRNAde reports $56\%$ mean recovery on its 14-target 3D
  benchmark~\citep{joshi2025grnade}. When examining our method's native-sequence recovery on OpenKnot57, we observed that the generator produces sequences with an average native-sequence recovery of $32.67\%$, which KnitNet search slightly reduces to $31.98\%$.These figures illustrate why native recovery should be distinguished from structural fidelity. Successful designs can differ substantially from a reference sequence, consistent with the sequence exploration reported in Struct2SeQ~\citep{he2026struct2seq} and demonstrated experimentally in OpenKnot~\citep{townley2026denovo} This freedom supports de novo exploration, but functional motifs may need to remain intact. Both stages of KnitNet offer a route to preserving them: masked generation can condition on prescribed nucleotides, and joint repair already modifies selected positions while preserving the surrounding sequence. These capabilities could support design tasks where known functional motifs must be retained while the rest of the RNA is redesigned.

  \noindent\textbf{Designing conformational behavior.}
  Our current endpoint specifies one target structure. Although ViennaRNA
  guidance accounts for competing folds through target ensemble probability,
  we do not explicitly evaluate folding kinetics or robustness across
  conditions. Extending repair to desired and undesired conformational states
  could support RNA switches and other dynamic constructs, building on
  multi-state design in gRNAde and
  NUPACK~\citep{joshi2025grnade,wolfe2017multistate}. This would let search
  optimize a specified pattern of structural behavior rather than agreement
  with a single fold.

  \noindent\textbf{Richer structural feedback.}
  Our 3D extension already benefits from tertiary-similarity ranking, while
  2D pairing discrepancies guide repair. Geometric errors in residue distances,
  orientations, or contacts could provide more direct guidance for correcting
  tertiary packing. Richer interaction graphs could additionally represent
  noncanonical contacts and molecular interfaces beyond the current
  canonical/wobble pair assignments. Together, these extensions could connect
  joint repair to backbone and complex-design pipelines such as
  RFDpoly~\citep{favor2025rfdpoly}.

  \noindent\textbf{Scaling to longer RNAs.}
  Repeated full-fold evaluations remain a computational cost, and our
  512-nucleotide pretraining limit leaves kilobase-scale design unexplored.
  Hierarchical repair, inexpensive screening before full verification, and
  adaptive allocation between generation and search could extend practical
  reach. Differentiable folding methods such as
  JAX-RNAfold~\citep{krueger_jax-rnafold_2025} could also supply proposals where
  their folding assumptions apply, with complete-sequence verification
  retained to assess interactions between repaired regions.

\clearpage
  \onecolumn

  \section{Full Pseudoknot Benchmark Results}
  \label{app:full-PK}

\input{tables/appendix/pseudoknot_benchmark}

\clearpage
\section{Full Planar RNA Structure Benchmark Results}
\label{app:full-planar}

\input{tables/appendix/planar_benchmark}

\clearpage

\section{Inference-Time Scaling}
  \label{app:scaling_eterna}

  \begin{figure}[H]
      \centering
      \includegraphics[
          width=\linewidth,
          height=0.65\textheight,
          keepaspectratio
      ]{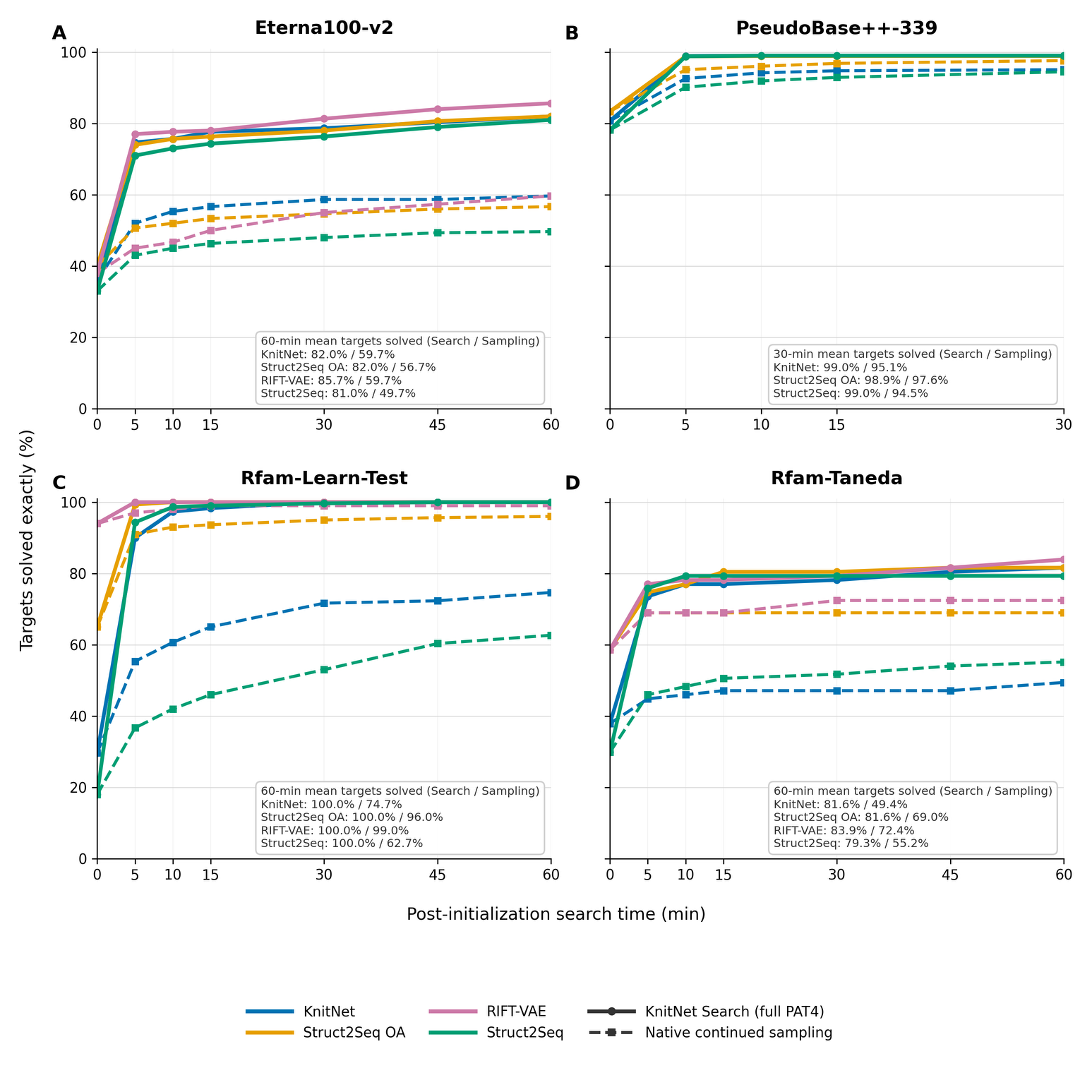}
      \caption{%
          \textbf{KnitNet search converts additional inference time into additional exact solutions.}
          Exact coverage is measured against post-initialization wall-clock time from identical frozen B96 generator pools. Solid curves use KnitNet search, while dashed curves allocate the same time to native continued
          sampling. Curves report means over three seeds.
      }
  \end{figure}
\clearpage

\clearpage
  \section{{Scaling with RNA Sequence Length}}
  \label{app:length_scaling}

  \begingroup

  \paragraph{Evaluation setup.}
  We investigate how target length affects exact recovery and
  computational cost using the held-out 500-target test set spanning
  20-511 nucleotides
  (Appendices~\ref{app:training-data}
  and~\ref{app:dataset_statistics}).
  For each target, we generate 96 initial candidates and run KnitNet
  for up to one hour after initialization. We compare against
  Struct2SeQ-OA, the strongest baseline on this dataset, under the
  same one-hour budget. We group targets into five length bins and
  report exact recovery and realized folding-oracle throughput.
  Exact-solve counts include targets solved at initialization.

  \begin{table}[htbp]
      \centering
      \small
      \setlength{\tabcolsep}{8pt}
      \renewcommand{\arraystretch}{1.12}

      \caption{{%
      \textbf{Exact recovery across RNA sequence lengths.}
      Results on the held-out 500-target test set after one hour.
      Entries give solved targets over all targets in each length
      bin, followed by the corresponding percentage.}}
      \label{tab:length-scaling}

      \begin{tabular}{@{}lcc@{}}
          \toprule
          Target length (nt)
          & KnitNet
          & Struct2SeQ-OA \\
          \midrule
          $\leq 100$
          & $\mathbf{197/200}$ ($98.5\%$)
          & $176/200$ ($88.0\%$) \\
          $101$--$200$
          & $\mathbf{53/61}$ ($86.9\%$)
          & $24/61$ ($39.3\%$) \\
          $201$--$300$
          & $\mathbf{48/112}$ ($42.9\%$)
          & $6/112$ ($5.4\%$) \\
          $301$--$400$
          & $\mathbf{6/71}$ ($8.5\%$)
          & $2/71$ ($2.8\%$) \\
          $401$--$511$
          & $\mathbf{2/56}$ ($3.6\%$)
          & $1/56$ ($1.8\%$) \\
          \midrule
          Overall
          & $\mathbf{306/500}$ ($61.2\%$)
          & $209/500$ ($41.8\%$) \\
          \bottomrule
      \end{tabular}
  \end{table}

  \paragraph{Recovery and computational cost.}
  KnitNet outperforms Struct2SeQ-OA in every evaluated length bin
  (Table~\ref{tab:length-scaling}), achieving 97 additional exact
  solves overall. The largest percentage-point gain occurs for targets
  of 101-200 nucleotides, where exact recovery increases from
  $39.3\%$ to $86.9\%$.
  Nevertheless, exact recovery declines substantially for both
  methods as target length increases.
  For KnitNet, realized folding-oracle throughput falls from
  approximately 37 evaluations per second for targets of at most
  100 nucleotides to 2 evaluations per second for targets of
  401-511 nucleotides. Thus, the same wall-time budget permits
  far fewer candidate evaluations for longer targets.

  \paragraph{Additional compute for longer targets.}
  To assess whether longer targets benefit from additional search,
  we extend the budget from one to two hours for the 127 targets
  of 301--511 nucleotides. Exact recovery increases from $8/127$
  ($6.3\%$) to $15/127$ ($11.8\%$), yielding seven additional
  solutions. This improvement shows that the one-hour results do
  not exhaust KnitNet's ability to solve these targets, although
  absolute exact recovery remains limited.

  \paragraph{Evaluation beyond the training length range.}
  We additionally evaluate ten held-out targets of 759--956
  nucleotides, exceeding the generator's 512-nucleotide pretraining
  limit. One-hour KnitNet search achieves no exact solves
  ($0/10$), but increases mean best pair Jaccard from $0.668$
  at initialization to $0.764$, a $14.4\%$ relative improvement.
  Matched continued sampling reaches $0.718$.
  These results demonstrate improved approximate structural
  recovery beyond the training length range, while exact recovery
  remains challenging.

  \paragraph{Interpretation.}
  Longer sequences increase the search space, while more expensive folding evaluations reduce the number of candidates explored within a fixed time budget. Less reliable structural guidance may also contribute. Chen et al.~\citep{chen2025comprehensive}
  report length-dependent prediction accuracy across secondary-structure methods, with substantial degradation for several methods on very long RNAs. Their evaluation does not include RibonanzaNet-SS, so it provides context rather than a direct
  explanation of our observed scaling.
  Taken together, our results show that KnitNet retains an
  exact-recovery advantage over the strongest tested baseline throughout the evaluated 20--511-nucleotide range and improves approximate recovery on substantially longer RNAs, while identifying long-sequence exact design as a remaining limitation.

  \par
  \endgroup

\clearpage

\section{Math Notation}
\label{app:notation}

Table~\ref{tab:notation} summarizes the main notation for the KnitNet pipeline and evaluation.

{\small
\renewcommand{\arraystretch}{1.05}
\begin{longtable}{@{}>{\raggedright\arraybackslash}p{0.27\linewidth}>{\raggedright\arraybackslash}p{0.68\linewidth}@{}}
\caption{Principal notation for the KnitNet pipeline and evaluation.}
\label{tab:notation}\\
\toprule
Symbol & Meaning \\
\midrule
\endfirsthead
\toprule
Symbol & Meaning \\
\midrule
\endhead
\bottomrule
\endfoot

\multicolumn{2}{@{}l}{\textit{Problem and initialization}}\\
$L$, $\mathcal A$, $x\in\mathcal A^L$ & Target length, RNA alphabet $\{\mathrm A,\mathrm C,\mathrm G,\mathrm U\}$, and a candidate sequence \\
$T$ & Target secondary structure as a set of nucleotide pairs, including crossing pairs \\
$F(x)$ & Pair set predicted for $x$ by the folding oracle \\
$E(x)=T\triangle F(x)$ & Missing target pairs and unwanted predicted pairs \\
$\mathcal R$ & Evaluated initial sequences sampled from the frozen generator \\
\midrule

\multicolumn{2}{@{}l}{\textit{Counterexample-guided joint repair}}\\
$\mathcal A_{\mathrm{pair}}$ & Six pair-substitution values: AU, UA, GC, CG, GU, and UG \\
$\mathcal P$, $a\in\mathcal P$ & Outer candidate population and the anchor of a repair episode \\
$\mathcal V_T$, $\mathcal G_a$, $\mathcal E_a$ & Target-defined assignment units, their error graph, and its edges from anchor and ancestor errors \\
$S\subseteq\mathcal V_T$ & Editable scope of assignment units. $|S|$ counts units \\
$z$, $x$ & Joint nucleotide assignment on $S$ and the resulting sequence, which matches $a$ outside those units \\
$C$ & Remembered, prioritized error-pair set in the current episode \\
$m_F(x)$ & Oracle-specific merit used to rank evaluated sequences, with larger values preferred \\
$\ell_C(x)$ & Lexicographic repair loss $\bigl(|E(x)\cap C|,|E(x)|,-m_F(x)\bigr)$ \\
$\mathcal D$, $\mathcal H\subseteq\mathcal D$ & Evaluated joint proposals in an episode, including its anchor, and the protected proposal beam \\
$\phi_S(x)$, $f(\phi_S(x))$ & Scope-assignment features and the surrogate's predicted prioritized and total error counts \\
$\mathcal Q$, $\mathcal J$, $\mathcal U$ & Joint proposals not yet evaluated in the episode, selected joint batch, and single-unit mutation batch. Batch sizes count evaluation requests \\
$x^\star$ & Best evaluated proposal in a nonempty joint batch, whose errors update $C$ \\
$b$, $B$ & Logical folding evaluations spent after initialization and the allowed budget \\
\midrule

\multicolumn{2}{@{}l}{\textit{Evaluation and 3D design}}\\
$J(x,T)$ & Pair-set Jaccard similarity of the predicted fold and target \\
$N$, $t_j^\star$, $\mathrm{Coverage}(t)$, $\mathrm{AUC}(t_{\max})$ & Panel size, first exact-solve time, solve coverage, and its normalized area over time horizon $t_{\max}$ \\
$R(x)$ & Tertiary reward used to rank candidates during 3D search \\
$\mathrm{scTM}(x)$ & RNA TM-score of the predicted 3D fold against the reference \\
\end{longtable}
}

\end{document}

%% file: math_commands.tex
\usepackage{amsmath,amsfonts,bm}

\def\eqref#1{equation~\ref{#1}}

\def\1{\bm{1}}

\DeclareMathAlphabet{\mathsfit}{\encodingdefault}{\sfdefault}{m}{sl}
\SetMathAlphabet{\mathsfit}{bold}{\encodingdefault}{\sfdefault}{bx}{n}



%% file: tables/3d_inverse_design.tex
\begin{minipage}[t]{\linewidth}
      \vspace{0pt}
      \centering
      \tiny
      \captionsetup{
        font=footnotesize,
        skip=4pt,
        justification=justified,
        singlelinecheck=false
      }
      \captionof{table}{\textbf{3D inverse-design performance.}
        Values are mean $\pm$ sample standard deviation of 64 3D targets averages over three seeds.}
      \label{tab:3d_inverse_design}

      \setlength{\tabcolsep}{2pt}
      \renewcommand{\arraystretch}{0.98}

      \begin{tabular*}{\linewidth}{@{\extracolsep{\fill}}lrrr@{}}
        \toprule
        Method &
        scTM $\uparrow$ &
        GDT$_{\mathrm{TS}}$ $\uparrow$ &
        RMSD (\AA) $\downarrow$ \\
        \midrule

        \multicolumn{4}{@{}l}{\textit{A. 2D generators}} \\

        KnitNet
          & $0.463 \pm 0.005$
          & $0.388 \pm 0.005$
          & $8.864 \pm 0.123$ \\
        \hspace{0.65em}+ KnitNet search
          & $0.660 \pm 0.005$
          & $0.660 \pm 0.008$
          & $2.964 \pm 0.049$ \\

        \addlinespace[1pt]
        \noalign{\vskip1.5pt
          \hbox to\linewidth{%
            \leaders\hbox{\rule{2pt}{0.3pt}\hskip2pt}\hfill
          }\vskip1.5pt}

        Struct2SeQ-OA
          & $0.465 \pm 0.006$
          & $0.404 \pm 0.011$
          & $7.555 \pm 0.372$ \\
        \hspace{0.65em}+ KnitNet search
          & $0.644 \pm 0.008$
          & $0.645 \pm 0.007$
          & $3.028 \pm 0.049$ \\

        \addlinespace[1pt]
        \midrule
        \multicolumn{4}{@{}l}{\textit{B. 3D generators}} \\

        gRNAde
          & $0.609 \pm 0.003$
          & $0.577 \pm 0.002$
          & $4.651 \pm 0.210$ \\
        \hspace{0.65em}+ KnitNet search
          & $0.746 \pm 0.005$
          & $0.752 \pm 0.008$
          & $2.218 \pm 0.101$ \\

        \addlinespace[1pt]
        \noalign{\vskip1.5pt
          \hbox to\linewidth{%
            \leaders\hbox{\rule{2pt}{0.3pt}\hskip2pt}\hfill
          }\vskip1.5pt}

        NA-MPNN
          & $0.663 \pm 0.001$
          & $0.646 \pm 0.003$
          & $3.778 \pm 0.103$ \\
        \hspace{0.65em}+ KnitNet search
          & $\underline{0.771 \pm 0.006}$
          & $\underline{0.780 \pm 0.007}$
          & $\underline{1.977 \pm 0.063}$ \\

        \addlinespace[1pt]
        \noalign{\vskip1.5pt
          \hbox to\linewidth{%
            \leaders\hbox{\rule{2pt}{0.3pt}\hskip2pt}\hfill
          }\vskip1.5pt}

        RIDE
          & $0.553 \pm 0.001$
          & $0.531 \pm 0.001$
          & $5.344 \pm 0.113$ \\
        \hspace{0.65em}+ KnitNet search
          & $0.732 \pm 0.005$
          & $0.739 \pm 0.005$
          & $2.233 \pm 0.051$ \\
        \hspace{0.65em}+ RIDER RL
          & $0.679 \pm 0.002$
          & $0.687 \pm 0.002$
          & $2.691 \pm 0.046$ \\

        \addlinespace[1pt]
        \noalign{\vskip1.5pt
          \hbox to\linewidth{%
            \leaders\hbox{\rule{2pt}{0.3pt}\hskip2pt}\hfill
          }\vskip1.5pt}

        RiboDiffusion
          & $0.679 \pm 0.002$
          & $0.662 \pm 0.003$
          & $3.404 \pm 0.085$ \\
        \hspace{0.65em}+ \textbf{KnitNet search}
          & $\mathbf{0.773} \pm \mathbf{0.004}$
          & $\mathbf{0.793} \pm \mathbf{0.001}$
          & $\mathbf{1.909} \pm \mathbf{0.133}$ \\

        \addlinespace[1pt]
        \bottomrule
      \end{tabular*}
    \end{minipage}

%% file: tables/3d_sampling.tex
\begin{table}[ht]
\centering\small
\caption{Sampling settings for the 96 initial sequences per target.}
\label{tab:3d_sampling}
\begin{tabularx}{\linewidth}{@{}lX@{}}
\toprule
Generator & Sampling \\
\midrule
KnitNet & 32 diffusion steps, temperature 0.8, top-$k=2$, confidence-ordered pairwise commits with Watson--Crick and wobble pairs, final argmax. \\
Struct2SeQ-OA & Random decoding orders. 32 samples each from $\epsilon$-greedy decoding at $\epsilon=0.05$ and $0.10$, and Q-softmax decoding. \\
gRNAde & Autoregressive sampling at temperature 0.1. \\
NA-MPNN & Random decoding orders at temperature 0.1. \\
RIDE & 50 stochastic DDIM steps at temperature 0.1. \\
RiboDiffusion & 50 ancestral diffusion steps with a cosine variance-preserving schedule and self-conditioning, using EMA weights. \\
\bottomrule
\end{tabularx}
\end{table}

%% file: tables/appendix/pseudoknot_benchmark.tex
 \begingroup
      \scriptsize
      \setlength{\tabcolsep}{1.4pt}
      \renewcommand{\arraystretch}{0.97}
      \setlength{\LTpre}{0.4em}
      \setlength{\LTpost}{0.4em}
      \setlength{\LTcapwidth}{\textwidth}

      \begin{longtable}{@{}llccccc@{}}
        \caption{\textbf{Pseudoknot benchmark results.}
        Entries are mean $\pm$ sample SD over seeds 42, 731, and 2718.
        Exact is the endpoint number of solved targets; solve AUC is normalized over
        the 30-minute post-initialization horizon. Jaccard and MCC are endpoint
        structural diagnostics. Capped TTS is
        $\operatorname{mean}[\min(T_{\mathrm{first\ exact}},30\,\mathrm{min})]$
        over the full panel, in minutes. Native initialization is untimed.
        The continued-sampling rows extend the same frozen native-initialization
        candidate pool using the source generator; this matched control was
        registered for Struct2Seq, Struct2Seq OA, and KnitNet.
        ``Original implementation'' means that
        the method's native structural oracle supplies online search guidance and
        RibonanzaNet-SS (RNSS) is used only for the common benchmark endpoint.
        ``Adapted to RNSS'' denotes our in-house variant in which RNSS also supplies
        online search guidance; these variants are not the authors' released
        implementations.}
        \label{tab:h130-full-metrics}\\

        \toprule
        Method & Variant &
        Exact $\uparrow$ &
        Solve AUC $\uparrow$ &
        Jaccard $\uparrow$ &
        MCC $\uparrow$ &
        \shortstack{Capped TTS\\(min) $\downarrow$} \\
        \midrule
        \endfirsthead

        \multicolumn{7}{c}{%
          \tablename\ \thetable\ continued from previous page
        } \\
        \addlinespace[2pt]
        \toprule
        Method & Variant &
        Exact $\uparrow$ &
        Solve AUC $\uparrow$ &
        Jaccard $\uparrow$ &
        MCC $\uparrow$ &
        \shortstack{Capped TTS\\(min) $\downarrow$} \\
        \midrule
        \endhead

        \midrule
        \multicolumn{7}{r}{Continued on next page} \\
        \endfoot

        \bottomrule
        \endlastfoot

        \multicolumn{7}{l}{\textit{A. OpenKnot57 (57 targets)}} \\*
        \addlinespace[1pt]
        \multicolumn{7}{l}{\textit{A.1 Generator baselines}} \\*

        \multirow{3}{*}{Struct2Seq}
          & Generation only
          & $30.0 \pm 0.00$ & --
          & $0.9683 \pm 0.0016$ & $0.9579 \pm 0.0049$ & -- \\*
          & + Continued sampling
          & $44.7 \pm 0.58$ & $0.7221 \pm 0.0199$
          & $0.9913 \pm 0.0004$ & $0.9884 \pm 0.0003$
          & $8.34 \pm 0.60$ \\*
          & + KnitNet search
          & $56.0 \pm 0.00$ & $0.9583 \pm 0.0040$
          & $0.9989 \pm 0.0006$ & $0.9986 \pm 0.0003$
          & $1.25 \pm 0.12$ \\
        \methodrule

        \multirow{3}{*}{Struct2Seq OA}
          & Generation only
          & $40.3 \pm 1.53$ & --
          & $0.9796 \pm 0.0017$ & $0.9724 \pm 0.0006$ & -- \\*
          & + Continued sampling
          & $49.7 \pm 1.15$ & $0.8346 \pm 0.0105$
          & $0.9948 \pm 0.0004$ & $0.9923 \pm 0.0003$
          & $4.96 \pm 0.31$ \\*
          & + KnitNet search
          & $56.3 \pm 0.58$ & $0.9754 \pm 0.0032$
          & $0.9996 \pm 0.0004$ & $0.9996 \pm 0.0004$
          & $0.74 \pm 0.10$ \\
        \methodrule

        \multirow{2}{*}{gRNAde 2D}
          & Generation only
          & $8.7 \pm 2.52$ & --
          & $0.8744 \pm 0.0101$ & $0.8347 \pm 0.0019$ & -- \\*
          & + KnitNet search
          & $52.7 \pm 1.15$ & $0.8740 \pm 0.0221$
          & $0.9899 \pm 0.0025$ & $0.9874 \pm 0.0034$
          & $3.78 \pm 0.66$ \\
        \methodrule

        \multirow{2}{*}{RNAinformer}
          & Generation only
          & $9.7 \pm 3.51$ & --
          & $0.7313 \pm 0.0087$ & $0.7113 \pm 0.0120$ & -- \\*
          & + KnitNet search
          & $44.3 \pm 0.58$ & $0.7608 \pm 0.0064$
          & $0.7886 \pm 0.0009$ & $0.7886 \pm 0.0010$
          & $7.18 \pm 0.19$ \\
        \methodrule
        \addlinespace[2pt]

        \multicolumn{7}{l}{\textit{A.2 Iterative solvers}} \\*

        \multirow{2}{*}{%
          \shortstack[l]{DesiRNA\\(original implementation)}
        }
          & Original init.
          & $40.7 \pm 1.15$ & $0.6869 \pm 0.0223$
          & $0.9807 \pm 0.0013$ & $0.9725 \pm 0.0018$
          & $9.39 \pm 0.67$ \\*
          & KnitNet init.
          & $41.0 \pm 2.65$ & $0.6824 \pm 0.0436$
          & $0.9826 \pm 0.0027$ & $0.9750 \pm 0.0035$
          & $9.53 \pm 1.31$ \\
        \methodrule

        \multirow{2}{*}{%
          \shortstack[l]{DesiRNA\\(adapted to RNSS)}
        }
          & Original init.
          & $54.3 \pm 1.15$ & $0.9192 \pm 0.0170$
          & $0.9976 \pm 0.0002$ & $0.9971 \pm 0.0004$
          & $2.42 \pm 0.51$ \\*
          & KnitNet init.
          & $55.0 \pm 2.00$ & $0.9306 \pm 0.0148$
          & $0.9987 \pm 0.0011$ & $0.9984 \pm 0.0014$
          & $2.08 \pm 0.44$ \\
        \methodrule

        \multirow{2}{*}{%
          \shortstack[l]{aRNAque\\(original implementation)}
        }
          & Original init.
          & $23.0 \pm 2.65$ & $0.2748 \pm 0.0448$
          & $0.7099 \pm 0.0048$ & $0.6957 \pm 0.0020$
          & $21.75 \pm 1.34$ \\*
          & KnitNet init.
          & $36.0 \pm 2.00$ & $0.6005 \pm 0.0278$
          & $0.8433 \pm 0.0010$ & $0.8376 \pm 0.0016$
          & $11.98 \pm 0.84$ \\
        \methodrule

        \multirow{2}{*}{%
          \shortstack[l]{aRNAque\\(adapted to RNSS)}
        }
          & Original init.
          & $49.7 \pm 0.58$ & $0.7392 \pm 0.0144$
          & $0.9256 \pm 0.0030$ & $0.9234 \pm 0.0029$
          & $7.83 \pm 0.43$ \\*
          & KnitNet init.
          & $53.0 \pm 0.00$ & $0.8916 \pm 0.0081$
          & $0.9298 \pm 0.0000$ & $0.9298 \pm 0.0000$
          & $3.25 \pm 0.24$ \\
        \methodrule

        \shortstack[l]{antaRNA\\(adapted to RNSS)}$^\dagger$
          & Original init.
          & $16.3 \pm 3.21$ & $0.2354 \pm 0.0355$
          & $0.8924 \pm 0.0028$ & $0.8587 \pm 0.0038$
          & $22.94 \pm 1.06$ \\
        \methodrule

        \shortstack[l]{MODENA\\(adapted to RNSS)}$^\dagger$
          & Original init.
          & $48.0 \pm 2.00$ & $0.8076 \pm 0.0193$
          & $0.9933 \pm 0.0028$ & $0.9903 \pm 0.0034$
          & $5.77 \pm 0.58$ \\
        \methodrule

        \multirow{2}{*}{%
          \shortstack[l]{Enzymer\\(original implementation)}
        }
          & Original init.
          & $5.0 \pm 1.73$ & $0.0552 \pm 0.0135$
          & $0.3473 \pm 0.0276$ & $0.3392 \pm 0.0035$
          & $28.34 \pm 0.40$ \\*
          & KnitNet init.
          & $21.0 \pm 1.00$ & $0.3605 \pm 0.0211$
          & $0.5176 \pm 0.0010$ & $0.5153 \pm 0.0019$
          & $19.18 \pm 0.63$ \\
        \methodrule

        \shortstack[l]{PKProbDesign\\(original implementation)}$^\dagger$
          & Original init.
          & $35.7 \pm 1.53$ & $0.5896 \pm 0.0169$
          & $0.7843 \pm 0.0006$ & $0.7822 \pm 0.0004$
          & $12.31 \pm 0.51$ \\
        \methodrule
        \addlinespace[2pt]

        \multicolumn{7}{l}{\textit{A.3 Our method}} \\*

        \multirow{3}{*}{\textbf{KnitNet (ours)}}
          & Generation only
          & $31.3 \pm 1.53$ & --
          & $0.9587 \pm 0.0029$ & $0.9455 \pm 0.0026$ & -- \\*
          & + Continued sampling
          & $45.0 \pm 1.73$ & $0.7326 \pm 0.0149$
          & $0.9892 \pm 0.0006$ & $0.9851 \pm 0.0006$
          & $8.02 \pm 0.45$ \\*
          & + KnitNet search
          & $56.3 \pm 1.15$ & $0.9677 \pm 0.0191$
          & $0.9996 \pm 0.0008$ & $0.9996 \pm 0.0007$
          & $0.97 \pm 0.57$ \\

        \midrule
        \multicolumn{7}{l}{\textit{B. PseudoBase++-339 (339 targets)}} \\*
        \addlinespace[1pt]
        \multicolumn{7}{l}{\textit{B.1 Generator baselines}} \\*

        \multirow{3}{*}{Struct2Seq}
          & Generation only
          & $265.0 \pm 3.61$ & --
          & $0.9833 \pm 0.0015$ & $0.9783 \pm 0.0022$ & -- \\*
          & + Continued sampling
          & $320.3 \pm 1.53$ & $0.9211 \pm 0.0059$
          & $0.9968 \pm 0.0003$ & $0.9966 \pm 0.0003$
          & $2.37 \pm 0.18$ \\*
          & + KnitNet search
          & $335.7 \pm 0.58$ & $0.9882 \pm 0.0019$
          & $0.9993 \pm 0.0001$ & $0.9993 \pm 0.0000$
          & $0.35 \pm 0.06$ \\
        \methodrule

        \multirow{3}{*}{Struct2Seq OA}
          & Generation only
          & $282.7 \pm 4.16$ & --
          & $0.9893 \pm 0.0011$ & $0.9871 \pm 0.0015$ & -- \\*
          & + Continued sampling
          & $331.0 \pm 1.00$ & $0.9622 \pm 0.0014$
          & $0.9984 \pm 0.0001$ & $0.9983 \pm 0.0001$
          & $1.13 \pm 0.04$ \\*
          & + KnitNet search
          & $335.3 \pm 0.58$ & $0.9879 \pm 0.0013$
          & $0.9993 \pm 0.0001$ & $0.9992 \pm 0.0001$
          & $0.36 \pm 0.04$ \\
        \methodrule

        \multirow{2}{*}{gRNAde 2D}
          & Generation only
          & $175.0 \pm 1.73$ & --
          & $0.9273 \pm 0.0013$ & $0.9179 \pm 0.0006$ & -- \\*
          & + KnitNet search
          & $331.3 \pm 1.53$ & $0.9649 \pm 0.0038$
          & $0.9970 \pm 0.0000$ & $0.9970 \pm 0.0003$
          & $1.05 \pm 0.11$ \\
        \methodrule

        \multirow{2}{*}{RNAinformer}
          & Generation only
          & $178.7 \pm 5.03$ & --
          & $0.9391 \pm 0.0029$ & $0.9315 \pm 0.0004$ & -- \\*
          & + KnitNet search
          & $332.3 \pm 2.89$ & $0.9676 \pm 0.0085$
          & $0.9951 \pm 0.0003$ & $0.9952 \pm 0.0003$
          & $0.97 \pm 0.25$ \\
        \methodrule
        \addlinespace[2pt]

        \multicolumn{7}{l}{\textit{B.2 Iterative solvers}} \\*

        \multirow{2}{*}{%
          \shortstack[l]{DesiRNA\\(original implementation)}
        }
          & Original init.
          & $288.3 \pm 4.04$ & $0.8300 \pm 0.0119$
          & $0.9878 \pm 0.0015$ & $0.9861 \pm 0.0008$
          & $5.10 \pm 0.36$ \\*
          & KnitNet init.
          & $300.3 \pm 2.89$ & $0.8719 \pm 0.0045$
          & $0.9917 \pm 0.0003$ & $0.9901 \pm 0.0003$
          & $3.84 \pm 0.13$ \\
        \methodrule

        \multirow{2}{*}{%
          \shortstack[l]{DesiRNA\\(adapted to RNSS)}
        }
          & Original init.
          & $334.7 \pm 0.58$ & $0.9805 \pm 0.0031$
          & $0.9992 \pm 0.0001$ & $0.9991 \pm 0.0002$
          & $0.59 \pm 0.09$ \\*
          & KnitNet init.
          & $335.0 \pm 0.00$ & $0.9853 \pm 0.0003$
          & $0.9993 \pm 0.0001$ & $0.9991 \pm 0.0002$
          & $0.44 \pm 0.01$ \\
        \methodrule

        \multirow{2}{*}{%
          \shortstack[l]{aRNAque\\(original implementation)}
        }
          & Original init.
          & $185.3 \pm 4.04$ & $0.4743 \pm 0.0059$
          & $0.9094 \pm 0.0041$ & $0.9031 \pm 0.0051$
          & $15.77 \pm 0.18$ \\*
          & KnitNet init.
          & $301.7 \pm 6.03$ & $0.8685 \pm 0.0150$
          & $0.9845 \pm 0.0016$ & $0.9839 \pm 0.0016$
          & $3.94 \pm 0.45$ \\
        \methodrule

        \multirow{2}{*}{%
          \shortstack[l]{aRNAque\\(adapted to RNSS)}
        }
          & Original init.
          & $334.0 \pm 0.00$ & $0.9546 \pm 0.0020$
          & $0.9931 \pm 0.0002$ & $0.9930 \pm 0.0003$
          & $1.36 \pm 0.06$ \\*
          & KnitNet init.
          & $333.3 \pm 0.58$ & $0.9801 \pm 0.0013$
          & $0.9934 \pm 0.0001$ & $0.9933 \pm 0.0001$
          & $0.60 \pm 0.04$ \\
        \methodrule

        \shortstack[l]{antaRNA\\(adapted to RNSS)}$^\dagger$
          & Original init.
          & $228.0 \pm 3.61$ & $0.6401 \pm 0.0134$
          & $0.9494 \pm 0.0012$ & $0.9419 \pm 0.0011$
          & $10.80 \pm 0.40$ \\
        \methodrule

        \shortstack[l]{MODENA\\(adapted to RNSS)}$^\dagger$
          & Original init.
          & $326.3 \pm 0.58$ & $0.9402 \pm 0.0027$
          & $0.9980 \pm 0.0003$ & $0.9974 \pm 0.0003$
          & $1.79 \pm 0.08$ \\
        \methodrule

        \multirow{2}{*}{%
          \shortstack[l]{Enzymer\\(original implementation)}
        }
          & Original init.
          & $204.7 \pm 3.51$ & $0.5376 \pm 0.0122$
          & $0.8054 \pm 0.0088$ & $0.7983 \pm 0.0081$
          & $13.87 \pm 0.37$ \\*
          & KnitNet init.
          & $279.3 \pm 0.58$ & $0.8162 \pm 0.0026$
          & $0.8853 \pm 0.0003$ & $0.8845 \pm 0.0003$
          & $5.51 \pm 0.08$ \\
        \methodrule

        \shortstack[l]{PKProbDesign\\(original implementation)}$^\dagger$
          & Original init.
          & $294.0 \pm 1.00$ & $0.8462 \pm 0.0019$
          & $0.8870 \pm 0.0001$ & $0.8869 \pm 0.0001$
          & $4.61 \pm 0.06$ \\
        \methodrule
        \addlinespace[2pt]

        \multicolumn{7}{l}{\textit{B.3 Our method}} \\*

        \multirow{3}{*}{\textbf{KnitNet (ours)}}
          & Generation only
          & $273.7 \pm 5.03$ & --
          & $0.9823 \pm 0.0013$ & $0.9796 \pm 0.0016$ & -- \\*
          & + Continued sampling
          & $322.3 \pm 1.53$ & $0.9384 \pm 0.0032$
          & $0.9972 \pm 0.0001$ & $0.9965 \pm 0.0003$
          & $1.85 \pm 0.10$ \\*
          & + KnitNet search
          & $335.7 \pm 0.58$ & $0.9885 \pm 0.0021$
          & $0.9993 \pm 0.0002$ & $0.9992 \pm 0.0003$
          & $0.34 \pm 0.06$ \\

      \end{longtable}

      \vspace{0.35em}

      \noindent
      \begin{minipage}{0.98\textwidth}
        \scriptsize
        $^\dagger$ The released antaRNA, MODENA, and PKProbDesign
        implementations expose no legitimate external-sequence initialization
        interface. Their native results are retained as source-faithful baselines.
        All TTS values measure post-initialization search, not strict
        process-launch-to-answer wall time.
      \end{minipage}

      \par
  \endgroup

  \subsection{Comparison with RNSS-adapted DesiRNA}
  \label{app:rnss-adapted-baselines}

  \begingroup

  \paragraph{Aligning search guidance and initialization.}
  To assess whether differences in search guidance explain KnitNet's performance advantage, we adapt DesiRNA to use the same RibonanzaNet-SS structural-margin objective
  (Equation~\ref{eq:knitnet_margin_merit}). We evaluate this adaptation with both DesiRNA's original initialization and KnitNet Generator starting candidates. The latter comparison controls for both initialization and search guidance. Both methods receive a 30-minute post-initialization search budget per target on the pseudoknot benchmarks, with exact recovery assessed using the same RibonanzaNet-SS fold-back criterion.
  Table~\ref{tab:h130-full-metrics} reports results over all targets and three seeds, including targets solved at initialization.

  \paragraph{OpenKnot57.}
  Aligning the search objective substantially improves DesiRNA: with KnitNet initialization, its mean exact recovery increases from $41.0/57$ to $55.0/57$. KnitNet nevertheless achieves higher mean exact recovery ($56.3/57$), higher solve AUC ($0.9677$ versus $0.9306$), and lower capped mean TTS ($0.97$ versus $2.08$ minutes) in Table~\ref{tab:h130-full-metrics}. Thus, our observed advantage persists when the competing search method receives the same
  starting candidates and structural-margin guidance.

  \paragraph{Ceiling effects on PseudoBase++-339.}
  The close performance on PseudoBase++-339 primarily reflects a
  saturation effect. KnitNet Generator already solves $80.7\%$ of
  targets at initialization, compared with approximately $55.0\%$
  on OpenKnot57. After search, KnitNet and RNSS-adapted DesiRNA
  with the same initialization both exceed $98.8\%$ exact recovery,
  leaving little remaining room to distinguish their performance.
  Nevertheless, KnitNet ranks first in mean exact recovery
  ($335.7/339$ versus $335.0/339$), solve AUC ($0.9885$ versus
  $0.9853$), and capped mean TTS ($0.34$ versus $0.44$ minutes)
  in this comparison.

  \paragraph{Paired statistical comparisons.}
  Tables~\ref{tab:rnss-paired-reference}
  and~\ref{tab:rnss-paired-differences} report the reference outcomes
  and paired differences from the statistical comparison.
  The reference summaries for this analysis are reported explicitly:
  they differ from the original benchmark summaries in
  Table~\ref{tab:h130-full-metrics}, and the paired differences
  correspond to the reference values in
  Table~\ref{tab:rnss-paired-reference}.

  Against RNSS-adapted DesiRNA with KnitNet initialization,
  KnitNet achieves $1.67$ additional exact solves on OpenKnot57
  ($95\%$ CI: $[0.33,\,3.67]$) and a solve-AUC improvement of
  $0.0436$ ($[0.0072,\,0.0914]$). Its capped mean TTS decreases
  by $1.309$ minutes, corresponding to a $2.69\times$ speedup.
  On the held-out 500-target test set, KnitNet achieves $15$
  additional exact solves ($[4,\,27]$; $p=0.0167$), improves
  solve AUC by $0.03256$ ($[0.01347,\,0.05225]$), and reduces
  capped mean TTS by $0.977$ minutes.
  The gains on both benchmarks remain significant against
  RNSS-adapted DesiRNA with native initialization.

  On PseudoBase++-339, the same-initialization exact-recovery
  difference is $+0.67$ targets ($[-0.67,\,2.33]$), and the
  solve-AUC difference is $+0.0029$ ($[-0.0013,\,0.0081]$).
  These intervals include zero, so the small observed differences
  are not statistically significant. Against RNSS-adapted DesiRNA
  with native initialization, KnitNet achieves a significant
  solve-AUC improvement of $0.0077$ ($[0.0013,\,0.0159]$),
  while the exact-recovery difference remains unresolved.


  \par
  \endgroup

  \begin{table}[htbp]
      \centering
      \scriptsize
      \setlength{\tabcolsep}{4pt}
      \renewcommand{\arraystretch}{1.08}

      \caption{%
      \textbf{Reference outcomes for the paired RNSS-adapted comparison.}
      Exact counts use the full benchmark denominators.
      TTS ratio is the comparator's capped mean TTS divided by
      KnitNet's capped mean TTS within the same panel.
      These values are the reference outcomes for
      Table~\ref{tab:rnss-paired-differences}.}
      \label{tab:rnss-paired-reference}

      \begin{adjustbox}{max width=\textwidth}
      \begin{tabular}{@{}lcccc@{}}
          \toprule
          Method
          & Exact $\uparrow$
          & Solve AUC $\uparrow$
          & \shortstack{Capped TTS\\(min) $\downarrow$}
          & TTS ratio \\
          \midrule

          \multicolumn{5}{l}{\textit{OpenKnot57}} \\
          KnitNet
          & $56.67/57$ & $0.97424$ & $0.773$ & $1.00\times$ \\
          DesiRNA--RNSS, KnitNet init.
          & $55.00/57$ & $0.93060$ & $2.082$ & $2.69\times$ \\
          DesiRNA--RNSS, native init.
          & $54.33/57$ & $0.91921$ & $2.424$ & $3.14\times$ \\

          \midrule
          \multicolumn{5}{l}{\textit{PseudoBase++-339}} \\
          KnitNet
          & $335.67/339$ & $0.98821$ & $0.354$ & $1.00\times$ \\
          DesiRNA--RNSS, KnitNet init.
          & $335.00/339$ & $0.98534$ & $0.440$ & $1.24\times$ \\
          DesiRNA--RNSS, native init.
          & $334.67/339$ & $0.98046$ & $0.586$ & $1.66\times$ \\

          \midrule
          \multicolumn{5}{l}{\textit{500-target test set}} \\
          KnitNet
          & $275/500$ & $0.52572$ & $14.228$ & $1.00\times$ \\
          DesiRNA--RNSS, KnitNet init.
          & $260/500$ & $0.49316$ & $15.205$ & $1.07\times$ \\
          DesiRNA--RNSS, native init.
          & $255/500$ & $0.48308$ & $15.507$ & $1.09\times$ \\
          Off-the-shelf DesiRNA
          & $189/500$ & $0.36302$ & $19.109$ & $1.34\times$ \\

          \bottomrule
      \end{tabular}
      \end{adjustbox}
  \end{table}

  \begin{table}[htbp]
      \centering
      \scriptsize
      \setlength{\tabcolsep}{3pt}
      \renewcommand{\arraystretch}{1.15}

      \caption{%
      \textbf{Paired differences between KnitNet and DesiRNA.}
      Differences are KnitNet minus the comparator; brackets give
      $95\%$ confidence intervals. Positive exact-count and solve-AUC
      differences favor KnitNet, as do negative capped-TTS differences.
      Reference outcomes appear in
      Table~\ref{tab:rnss-paired-reference}.}
      \label{tab:rnss-paired-differences}

      \begin{adjustbox}{max width=\textwidth}
      \begin{tabular}{@{}lccc@{}}
          \toprule
          Comparator
          & $\Delta$ exact (95\% CI)
          & $\Delta$ AUC (95\% CI)
          & $\Delta$ TTS, min (95\% CI) \\
          \midrule

          \multicolumn{4}{l}{\textit{OpenKnot57}} \\
          DesiRNA--RNSS, KnitNet init.
          & $+1.67\ [0.33,\,3.67]$
          & $+0.0436\ [0.0072,\,0.0914]$
          & $-1.309\ [-2.742,\,-0.216]$ \\
          DesiRNA--RNSS, native init.
          & $+2.33\ [0.33,\,5.00]$
          & $+0.0550\ [0.0151,\,0.1055]$
          & $-1.651\ [-3.164,\,-0.453]$ \\

          \midrule
          \multicolumn{4}{l}{\textit{PseudoBase++-339}} \\
          DesiRNA--RNSS, KnitNet init.
          & $+0.67\ [-0.67,\,2.33]$
          & $+0.0029\ [-0.0013,\,0.0081]$
          & $-0.086\ [-0.242,\,0.038]$ \\
          DesiRNA--RNSS, native init.
          & $+1.00\ [-0.67,\,3.67]$
          & $+0.0077\ [0.0013,\,0.0159]$
          & $-0.232\ [-0.476,\,-0.039]$ \\

          \midrule
          \multicolumn{4}{l}{\textit{500-target test set}} \\
          DesiRNA--RNSS, KnitNet init.
          & \shortstack{$+15\ [4,\,27]$\\$p=0.0167$}
          & $+0.03256\ [0.01347,\,0.05225]$
          & $-0.977\ [-1.567,\,-0.404]$ \\
          DesiRNA--RNSS, native init.
          & \shortstack{$+20\ [7,\,34]$\\$p=0.0055$}
          & $+0.04264\ [0.02057,\,0.06469]$
          & $-1.279\ [-1.941,\,-0.617]$ \\
          Off-the-shelf DesiRNA
          & \shortstack{$+86\ [70,\,103]$\\$p=5.75\times10^{-25}$}
          & $+0.16270\ [0.13330,\,0.19358]$
          & $-4.881\ [-5.807,\,-3.999]$ \\

          \bottomrule
      \end{tabular}
      \end{adjustbox}
  \end{table}

%% file: tables/appendix/planar_benchmark.tex
\begingroup
      \scriptsize
      \setlength{\tabcolsep}{1.4pt}
      \renewcommand{\arraystretch}{0.97}
      \setlength{\LTpre}{0.4em}
      \setlength{\LTpost}{0.4em}
      \setlength{\LTcapwidth}{\textwidth}

      \begin{longtable}{@{}llccccc@{}}
        \caption{\textbf{Complete ViennaRNA benchmark results and matched
        native-generation controls.}
        Entries are mean $\pm$ sample SD over seeds 42, 731, and 2718.
        Exact is the endpoint number of ViennaRNA-2.6.4 MFE-exact targets.
        Solve AUC is the normalized continuous first-exact solve curve over the
        60-minute post-initialization horizon; equivalently, it is
        $1-\operatorname{mean}[
        \min(T_{\mathrm{first\ exact}},60\,\mathrm{min})]/60\,\mathrm{min}$.
        Jaccard and MCC are endpoint structural diagnostics where terminal
        sequences are available; dashes indicate unavailable measurements.
        Capped TTS is the corresponding full-panel mean time to first exact, in
        minutes. Native B96 is an untimed initialization result.
        The native-generation rows continue the identical frozen B96 pool using the
        source generator; this matched control was registered for Struct2SeQ,
        Struct2SeQ OA, RIFT-VAE, and KnitNet.}
        \label{tab:h131-full-metrics}\\

        \toprule
        Method & Variant &
        Exact $\uparrow$ &
        Solve AUC $\uparrow$ &
        Jaccard $\uparrow$ &
        MCC $\uparrow$ &
        \shortstack{Capped TTS\\(min) $\downarrow$} \\
        \midrule
        \endfirsthead

        \multicolumn{7}{c}{%
          \tablename\ \thetable\ continued from previous page
        } \\
        \addlinespace[2pt]
        \toprule
        Method & Variant &
        Exact $\uparrow$ &
        Solve AUC $\uparrow$ &
        Jaccard $\uparrow$ &
        MCC $\uparrow$ &
        \shortstack{Capped TTS\\(min) $\downarrow$} \\
        \midrule
        \endhead

        \midrule
        \multicolumn{7}{r}{Continued on next page} \\
        \endfoot

        \bottomrule
        \endlastfoot

        \multicolumn{7}{l}{\textit{A. Eterna100 (100 targets)}} \\*
        \addlinespace[1pt]
        \multicolumn{7}{l}{\textit{A.1 Generator baselines}} \\*

        \multirow{3}{*}{Struct2SeQ}
          & Generation only
          & $33.0 \pm 1.00$
          & --
          & $0.8093 \pm 0.0058$
          & $0.8619 \pm 0.0059$
          & -- \\*
          & + Continued sampling
          & $49.7 \pm 0.58$
          & $0.4718 \pm 0.0043$
          & $0.8846 \pm 0.0030$
          & $0.9123 \pm 0.0042$
          & $31.69 \pm 0.26$ \\*
          & + KnitNet search
          & $81.0 \pm 0.00$
          & $0.7620 \pm 0.0047$
          & $0.9723 \pm 0.0018$
          & $0.9780 \pm 0.0020$
          & $14.28 \pm 0.28$ \\
        \methodrule

        \multirow{3}{*}{Struct2SeQ OA}
          & Generation only
          & $40.0 \pm 1.73$
          & --
          & $0.8136 \pm 0.0041$
          & $0.8535 \pm 0.0044$
          & -- \\*
          & + Continued sampling
          & $56.7 \pm 0.58$
          & $0.5421 \pm 0.0095$
          & $0.9000 \pm 0.0032$
          & $0.9185 \pm 0.0035$
          & $27.47 \pm 0.57$ \\*
          & + KnitNet search
          & $82.0 \pm 1.00$
          & $0.7781 \pm 0.0042$
          & $0.9670 \pm 0.0022$
          & $0.9727 \pm 0.0015$
          & $13.32 \pm 0.25$ \\
        \methodrule

        \multirow{3}{*}{RIFT-VAE}
          & Generation only
          & $38.0 \pm 0.00$
          & --
          & $0.7521 \pm 0.0000$
          & $0.8452 \pm 0.0000$
          & -- \\*
          & + Continued sampling
          & $59.7 \pm 0.58$
          & $0.5343 \pm 0.0008$
          & $0.8816 \pm 0.0006$
          & $0.9305 \pm 0.0005$
          & $27.94 \pm 0.05$ \\*
          & + KnitNet search
          & $85.7 \pm 1.53$
          & $0.8061 \pm 0.0076$
          & $0.9805 \pm 0.0012$
          & $0.9811 \pm 0.0008$
          & $11.63 \pm 0.46$ \\
        \methodrule
        \addlinespace[2pt]

        \multicolumn{7}{l}{\textit{A.2 Iterative solvers}} \\*

        \multirow{2}{*}{DesiRNA}
          & Original init.
          & $87.7 \pm 2.52$
          & $0.8461 \pm 0.0184$
          & $0.9936 \pm 0.0008$
          & $0.9944 \pm 0.0011$
          & $9.23 \pm 1.11$ \\*
          & KnitNet init.
          & $80.7 \pm 2.08$
          & $0.7773 \pm 0.0160$
          & $0.9627 \pm 0.0018$
          & $0.9720 \pm 0.0021$
          & $13.36 \pm 0.96$ \\
        \methodrule

        DRAG$^\dagger$
          & Original init.
          & $62.0 \pm 1.00$
          & $0.6015 \pm 0.0119$
          & $0.9327 \pm 0.0015$
          & $0.9421 \pm 0.0022$
          & $23.91 \pm 0.71$ \\
        \methodrule

        \multirow{2}{*}{FastDesign}
          & Original init.
          & $80.3 \pm 1.53$
          & $0.7793 \pm 0.0147$
          & $0.9816 \pm 0.0009$
          & $0.9856 \pm 0.0006$
          & $13.24 \pm 0.88$ \\*
          & KnitNet init.
          & $75.3 \pm 3.21$
          & $0.7373 \pm 0.0333$
          & $0.9744 \pm 0.0004$
          & $0.9803 \pm 0.0005$
          & $15.76 \pm 2.00$ \\
        \methodrule

        LEARNA$^\dagger$
          & Original init.
          & $56.0 \pm 1.73$
          & $0.5039 \pm 0.0117$
          & $0.8581 \pm 0.0064$
          & $0.8786 \pm 0.0077$
          & $29.77 \pm 0.70$ \\
        \methodrule

        Meta-LEARNA$^\dagger$
          & Original init.
          & $62.3 \pm 1.15$
          & $0.5980 \pm 0.0029$
          & $0.9268 \pm 0.0018$
          & $0.9549 \pm 0.0015$
          & $24.12 \pm 0.17$ \\
        \methodrule

        Meta-LEARNA-Adapt$^\dagger$
          & Original init.
          & $63.3 \pm 1.53$
          & $0.5962 \pm 0.0051$
          & $0.9234 \pm 0.0034$
          & $0.9544 \pm 0.0025$
          & $24.23 \pm 0.31$ \\
        \methodrule

        SamplingDesign$^\dagger$
          & Original init.
          & $67.3 \pm 1.15$
          & $0.6215 \pm 0.0126$
          & $0.9034 \pm 0.0031$
          & $0.9461 \pm 0.0021$
          & $22.71 \pm 0.76$ \\
        \methodrule

        GREED-RNA$^\dagger$
          & Original init.
          & $80.7 \pm 1.15$
          & $0.7821 \pm 0.0096$
          & $0.9846 \pm 0.0016$
          & $0.9853 \pm 0.0017$
          & $13.07 \pm 0.57$ \\
        \methodrule

        SAMFEO$^\dagger$
          & Original init.
          & $80.3 \pm 2.08$
          & $0.7678 \pm 0.0102$
          & $0.9862 \pm 0.0031$
          & $0.9875 \pm 0.0017$
          & $13.93 \pm 0.61$ \\
        \methodrule

        \multirow{2}{*}{aRNAque}
          & Original init.
          & $72.0 \pm 2.00$
          & $0.6970 \pm 0.0120$
          & $0.9582 \pm 0.0045$
          & $0.9721 \pm 0.0024$
          & $18.18 \pm 0.72$ \\*
          & KnitNet init.
          & $74.0 \pm 0.00$
          & $0.7184 \pm 0.0038$
          & $0.9696 \pm 0.0017$
          & $0.9777 \pm 0.0029$
          & $16.90 \pm 0.23$ \\
        \methodrule

        \multirow{2}{*}{NEMO}
          & Original init.
          & $89.0 \pm 1.00$
          & $0.8579 \pm 0.0060$
          & $0.9893 \pm 0.0010$
          & $0.9915 \pm 0.0007$
          & $8.53 \pm 0.36$ \\*
          & KnitNet init.
          & $89.7 \pm 0.58$
          & $0.8620 \pm 0.0055$
          & $0.9897 \pm 0.0011$
          & $0.9917 \pm 0.0002$
          & $8.28 \pm 0.33$ \\
        \methodrule

        SentRNA$^\dagger$
          & Original init.
          & $51.7 \pm 1.15$
          & $0.4548 \pm 0.0122$
          & $0.8715 \pm 0.0034$
          & $0.9167 \pm 0.0057$
          & $32.71 \pm 0.73$ \\
        \methodrule

        RNAiFold-LNS$^\dagger$
          & Original init.
          & $66.7 \pm 1.15$
          & $0.6378 \pm 0.0072$
          & --
          & --
          & $21.73 \pm 0.43$ \\
        \methodrule
        \addlinespace[2pt]

        \multicolumn{7}{l}{\textit{A.3 Our method}} \\*

        \multirow{3}{*}{\textbf{KnitNet (ours)}}
          & Generation only
          & $36.0 \pm 1.73$
          & --
          & $0.8292 \pm 0.0064$
          & $0.8810 \pm 0.0093$
          & -- \\*
          & + Continued sampling
          & $59.7 \pm 2.08$
          & $0.5684 \pm 0.0074$
          & $0.9176 \pm 0.0031$
          & $0.9426 \pm 0.0004$
          & $25.90 \pm 0.44$ \\*
          & + KnitNet search
          & $82.0 \pm 1.00$
          & $0.7826 \pm 0.0024$
          & $0.9789 \pm 0.0025$
          & $0.9819 \pm 0.0022$
          & $13.05 \pm 0.14$ \\

        \midrule
        \multicolumn{7}{l}{\textit{B. Rfam-Learn-Test (100 targets)}} \\*
        \addlinespace[1pt]
        \multicolumn{7}{l}{\textit{B.1 Generator baselines}} \\*

        \multirow{3}{*}{Struct2SeQ}
          & Generation only
          & $18.0 \pm 3.61$
          & --
          & $0.9739 \pm 0.0005$
          & $0.9668 \pm 0.0003$
          & -- \\*
          & + Continued sampling
          & $62.7 \pm 0.58$
          & $0.5154 \pm 0.0248$
          & $0.9942 \pm 0.0006$
          & $0.9925 \pm 0.0008$
          & $29.07 \pm 1.49$ \\*
          & + KnitNet search
          & $100.0 \pm 0.00$
          & $0.9803 \pm 0.0008$
          & $1.0000 \pm 0.0000$
          & $1.0000 \pm 0.0000$
          & $1.18 \pm 0.05$ \\
        \methodrule

        \multirow{3}{*}{Struct2SeQ OA}
          & Generation only
          & $65.0 \pm 2.65$
          & --
          & $0.9926 \pm 0.0003$
          & $0.9900 \pm 0.0002$
          & -- \\*
          & + Continued sampling
          & $96.0 \pm 0.00$
          & $0.9404 \pm 0.0061$
          & $0.9996 \pm 0.0000$
          & $0.9995 \pm 0.0000$
          & $3.57 \pm 0.37$ \\*
          & + KnitNet search
          & $100.0 \pm 0.00$
          & $0.9966 \pm 0.0006$
          & $1.0000 \pm 0.0000$
          & $1.0000 \pm 0.0000$
          & $0.21 \pm 0.04$ \\
        \methodrule

        \multirow{3}{*}{RIFT-VAE}
          & Generation only
          & $94.0 \pm 0.00$
          & --
          & $0.9983 \pm 0.0000$
          & $0.9981 \pm 0.0000$
          & -- \\*
          & + Continued sampling
          & $99.0 \pm 0.00$
          & $0.9865 \pm 0.0001$
          & $0.9999 \pm 0.0000$
          & $0.9999 \pm 0.0000$
          & $0.81 \pm 0.01$ \\*
          & + KnitNet search
          & $100.0 \pm 0.00$
          & $0.9997 \pm 0.0000$
          & $1.0000 \pm 0.0000$
          & $1.0000 \pm 0.0000$
          & $0.02 \pm 0.00$ \\
        \methodrule
        \addlinespace[2pt]

        \multicolumn{7}{l}{\textit{B.2 Iterative solvers}} \\*

        \multirow{2}{*}{DesiRNA}
          & Original init.
          & $100.0 \pm 0.00$
          & $0.9827 \pm 0.0015$
          & $1.0000 \pm 0.0000$
          & $1.0000 \pm 0.0000$
          & $1.04 \pm 0.09$ \\*
          & KnitNet init.
          & $100.0 \pm 0.00$
          & $0.9707 \pm 0.0011$
          & $1.0000 \pm 0.0000$
          & $1.0000 \pm 0.0000$
          & $1.76 \pm 0.07$ \\
        \methodrule

        DRAG$^\dagger$
          & Original init.
          & $100.0 \pm 0.00$
          & $0.9956 \pm 0.0004$
          & $1.0000 \pm 0.0000$
          & $1.0000 \pm 0.0000$
          & $0.26 \pm 0.02$ \\
        \methodrule

        \multirow{2}{*}{FastDesign}
          & Original init.
          & $100.0 \pm 0.00$
          & $0.9654 \pm 0.0010$
          & $1.0000 \pm 0.0000$
          & $1.0000 \pm 0.0000$
          & $2.07 \pm 0.06$ \\*
          & KnitNet init.
          & $100.0 \pm 0.00$
          & $0.9722 \pm 0.0003$
          & $1.0000 \pm 0.0000$
          & $1.0000 \pm 0.0000$
          & $1.67 \pm 0.02$ \\
        \methodrule

        LEARNA$^\dagger$
          & Original init.
          & $76.0 \pm 2.00$
          & $0.6130 \pm 0.0326$
          & $0.9959 \pm 0.0006$
          & $0.9946 \pm 0.0007$
          & $23.22 \pm 1.96$ \\
        \methodrule

        Meta-LEARNA$^\dagger$
          & Original init.
          & $88.3 \pm 1.15$
          & $0.8075 \pm 0.0124$
          & $0.9985 \pm 0.0001$
          & $0.9980 \pm 0.0001$
          & $11.55 \pm 0.74$ \\
        \methodrule

        Meta-LEARNA-Adapt$^\dagger$
          & Original init.
          & $89.0 \pm 1.00$
          & $0.8097 \pm 0.0149$
          & $0.9986 \pm 0.0001$
          & $0.9982 \pm 0.0001$
          & $11.42 \pm 0.90$ \\
        \methodrule

        SamplingDesign$^\dagger$
          & Original init.
          & $67.0 \pm 1.73$
          & $0.5401 \pm 0.0113$
          & $0.9928 \pm 0.0006$
          & $0.9910 \pm 0.0011$
          & $27.59 \pm 0.68$ \\
        \methodrule

        GREED-RNA$^\dagger$
          & Original init.
          & $99.7 \pm 0.58$
          & $0.9787 \pm 0.0043$
          & $1.0000 \pm 0.0000$
          & $1.0000 \pm 0.0001$
          & $1.28 \pm 0.26$ \\
        \methodrule

        SAMFEO$^\dagger$
          & Original init.
          & $99.7 \pm 0.58$
          & $0.9837 \pm 0.0056$
          & $0.9999 \pm 0.0002$
          & $0.9999 \pm 0.0002$
          & $0.98 \pm 0.34$ \\
        \methodrule

        \multirow{2}{*}{aRNAque}
          & Original init.
          & $99.3 \pm 1.15$
          & $0.8937 \pm 0.0078$
          & $0.9999 \pm 0.0001$
          & $0.9999 \pm 0.0002$
          & $6.38 \pm 0.47$ \\*
          & KnitNet init.
          & $100.0 \pm 0.00$
          & $0.9700 \pm 0.0008$
          & $1.0000 \pm 0.0000$
          & $1.0000 \pm 0.0000$
          & $1.80 \pm 0.05$ \\
        \methodrule

        \multirow{2}{*}{NEMO}
          & Original init.
          & $100.0 \pm 0.00$
          & $0.9908 \pm 0.0004$
          & $1.0000 \pm 0.0000$
          & $1.0000 \pm 0.0000$
          & $0.55 \pm 0.03$ \\*
          & KnitNet init.
          & $100.0 \pm 0.00$
          & $0.9989 \pm 0.0006$
          & $1.0000 \pm 0.0000$
          & $1.0000 \pm 0.0000$
          & $0.07 \pm 0.04$ \\
        \methodrule

        SentRNA$^\dagger$
          & Original init.
          & $91.7 \pm 1.53$
          & $0.8151 \pm 0.0249$
          & $0.9984 \pm 0.0005$
          & $0.9978 \pm 0.0006$
          & $11.10 \pm 1.49$ \\
        \methodrule

        RNAiFold-LNS$^\dagger$
          & Original init.
          & $97.3 \pm 0.58$
          & $0.9584 \pm 0.0096$
          & --
          & --
          & $2.49 \pm 0.58$ \\
        \methodrule
        \addlinespace[2pt]

        \multicolumn{7}{l}{\textit{B.3 Our method}} \\*

        \multirow{3}{*}{\textbf{KnitNet (ours)}}
          & Generation only
          & $29.7 \pm 1.53$
          & --
          & $0.9689 \pm 0.0013$
          & $0.9628 \pm 0.0014$
          & -- \\*
          & + Continued sampling
          & $74.7 \pm 1.53$
          & $0.6764 \pm 0.0070$
          & $0.9942 \pm 0.0005$
          & $0.9928 \pm 0.0006$
          & $19.42 \pm 0.42$ \\*
          & + KnitNet search
          & $100.0 \pm 0.00$
          & $0.9759 \pm 0.0032$
          & $1.0000 \pm 0.0000$
          & $1.0000 \pm 0.0000$
          & $1.45 \pm 0.19$ \\

        \midrule
        \multicolumn{7}{l}{\textit{C. Rfam-Taneda (29 targets)}} \\*
        \addlinespace[1pt]
        \multicolumn{7}{l}{\textit{C.1 Generator baselines}} \\*

        \multirow{3}{*}{Struct2SeQ}
          & Generation only
          & $8.7 \pm 1.15$
          & --
          & $0.9052 \pm 0.0031$
          & $0.9214 \pm 0.0024$
          & -- \\*
          & + Continued sampling
          & $16.0 \pm 0.00$
          & $0.5127 \pm 0.0157$
          & $0.9599 \pm 0.0038$
          & $0.9682 \pm 0.0025$
          & $29.24 \pm 0.94$ \\*
          & + KnitNet search
          & $23.0 \pm 0.00$
          & $0.7827 \pm 0.0011$
          & $0.9946 \pm 0.0002$
          & $0.9954 \pm 0.0002$
          & $13.04 \pm 0.06$ \\
        \methodrule

        \multirow{3}{*}{Struct2SeQ OA}
          & Generation only
          & $17.0 \pm 1.00$
          & --
          & $0.9434 \pm 0.0041$
          & $0.9563 \pm 0.0048$
          & -- \\*
          & + Continued sampling
          & $20.0 \pm 0.00$
          & $0.6872 \pm 0.0010$
          & $0.9881 \pm 0.0012$
          & $0.9903 \pm 0.0011$
          & $18.77 \pm 0.06$ \\*
          & + KnitNet search
          & $23.7 \pm 0.58$
          & $0.7982 \pm 0.0176$
          & $0.9950 \pm 0.0007$
          & $0.9957 \pm 0.0007$
          & $12.11 \pm 1.06$ \\
        \methodrule

        \multirow{3}{*}{RIFT-VAE}
          & Generation only
          & $17.0 \pm 0.00$
          & --
          & $0.9245 \pm 0.0000$
          & $0.9459 \pm 0.0000$
          & -- \\*
          & + Continued sampling
          & $21.0 \pm 0.00$
          & $0.7091 \pm 0.0007$
          & $0.9798 \pm 0.0013$
          & $0.9836 \pm 0.0011$
          & $17.45 \pm 0.04$ \\*
          & + KnitNet search
          & $24.3 \pm 0.58$
          & $0.7982 \pm 0.0091$
          & $0.9953 \pm 0.0002$
          & $0.9959 \pm 0.0002$
          & $12.11 \pm 0.55$ \\
        \methodrule
        \addlinespace[2pt]

        \multicolumn{7}{l}{\textit{C.2 Iterative solvers}} \\*

        \multirow{2}{*}{DesiRNA}
          & Original init.
          & $24.0 \pm 0.00$
          & $0.8243 \pm 0.0007$
          & $0.9945 \pm 0.0009$
          & $0.9954 \pm 0.0007$
          & $10.54 \pm 0.04$ \\*
          & KnitNet init.
          & $23.3 \pm 1.15$
          & $0.7768 \pm 0.0265$
          & $0.9946 \pm 0.0011$
          & $0.9952 \pm 0.0010$
          & $13.39 \pm 1.59$ \\
        \methodrule

        DRAG$^\dagger$
          & Original init.
          & $23.7 \pm 0.58$
          & $0.8125 \pm 0.0238$
          & $0.9938 \pm 0.0003$
          & $0.9944 \pm 0.0002$
          & $11.25 \pm 1.43$ \\
        \methodrule

        \multirow{2}{*}{FastDesign}
          & Original init.
          & $24.0 \pm 0.00$
          & $0.8101 \pm 0.0016$
          & $0.9953 \pm 0.0004$
          & $0.9959 \pm 0.0004$
          & $11.39 \pm 0.10$ \\*
          & KnitNet init.
          & $24.0 \pm 0.00$
          & $0.8132 \pm 0.0006$
          & $0.9947 \pm 0.0002$
          & $0.9955 \pm 0.0005$
          & $11.21 \pm 0.03$ \\
        \methodrule

        LEARNA$^\dagger$
          & Original init.
          & $21.3 \pm 0.58$
          & $0.6877 \pm 0.0087$
          & $0.9883 \pm 0.0020$
          & $0.9920 \pm 0.0023$
          & $18.74 \pm 0.52$ \\
        \methodrule

        Meta-LEARNA$^\dagger$
          & Original init.
          & $23.7 \pm 0.58$
          & $0.7927 \pm 0.0194$
          & $0.9934 \pm 0.0005$
          & $0.9945 \pm 0.0008$
          & $12.44 \pm 1.17$ \\
        \methodrule

        Meta-LEARNA-Adapt$^\dagger$
          & Original init.
          & $23.0 \pm 0.00$
          & $0.7754 \pm 0.0100$
          & $0.9930 \pm 0.0002$
          & $0.9941 \pm 0.0006$
          & $13.48 \pm 0.60$ \\
        \methodrule

        SamplingDesign$^\dagger$
          & Original init.
          & $22.3 \pm 0.58$
          & $0.6904 \pm 0.0086$
          & $0.9861 \pm 0.0019$
          & $0.9895 \pm 0.0010$
          & $18.58 \pm 0.52$ \\
        \methodrule

        GREED-RNA$^\dagger$
          & Original init.
          & $24.0 \pm 0.00$
          & $0.8246 \pm 0.0011$
          & $0.9951 \pm 0.0005$
          & $0.9956 \pm 0.0006$
          & $10.52 \pm 0.07$ \\
        \methodrule

        SAMFEO$^\dagger$
          & Original init.
          & $24.0 \pm 0.00$
          & $0.8268 \pm 0.0003$
          & $0.9906 \pm 0.0015$
          & $0.9947 \pm 0.0003$
          & $10.39 \pm 0.02$ \\
        \methodrule

        \multirow{2}{*}{aRNAque}
          & Original init.
          & $23.3 \pm 0.58$
          & $0.7847 \pm 0.0034$
          & $0.9937 \pm 0.0017$
          & $0.9952 \pm 0.0005$
          & $12.92 \pm 0.20$ \\*
          & KnitNet init.
          & $23.3 \pm 0.58$
          & $0.7705 \pm 0.0188$
          & $0.9945 \pm 0.0004$
          & $0.9951 \pm 0.0003$
          & $13.77 \pm 1.13$ \\
        \methodrule

        \multirow{2}{*}{NEMO}
          & Original init.
          & $24.7 \pm 0.58$
          & $0.8399 \pm 0.0130$
          & $0.9957 \pm 0.0004$
          & $0.9963 \pm 0.0004$
          & $9.61 \pm 0.78$ \\*
          & KnitNet init.
          & $24.0 \pm 0.00$
          & $0.8264 \pm 0.0006$
          & $0.9954 \pm 0.0002$
          & $0.9960 \pm 0.0002$
          & $10.42 \pm 0.03$ \\
        \methodrule

        SentRNA$^\dagger$
          & Original init.
          & $21.7 \pm 1.53$
          & $0.7083 \pm 0.0445$
          & $0.9878 \pm 0.0013$
          & $0.9899 \pm 0.0009$
          & $17.50 \pm 2.67$ \\
        \methodrule

        RNAiFold-LNS$^\dagger$
          & Original init.
          & $24.0 \pm 0.00$
          & $0.8083 \pm 0.0152$
          & --
          & --
          & $11.50 \pm 0.91$ \\
        \methodrule
        \addlinespace[2pt]

        \multicolumn{7}{l}{\textit{C.3 Our method}} \\*

        \multirow{3}{*}{\textbf{KnitNet (ours)}}
          & Generation only
          & $11.0 \pm 1.00$
          & --
          & $0.8804 \pm 0.0120$
          & $0.8992 \pm 0.0146$
          & -- \\*
          & + Continued sampling
          & $14.3 \pm 1.53$
          & $0.4685 \pm 0.0343$
          & $0.9537 \pm 0.0031$
          & $0.9592 \pm 0.0014$
          & $31.89 \pm 2.06$ \\*
          & + KnitNet search
          & $23.7 \pm 0.58$
          & $0.7780 \pm 0.0228$
          & $0.9949 \pm 0.0006$
          & $0.9955 \pm 0.0006$
          & $13.32 \pm 1.37$ \\

      \end{longtable}

      \noindent
      \begin{minipage}{\textwidth}
        \scriptsize
        $^\dagger$ The released DRAG, LEARNA, Meta-LEARNA,
        Meta-LEARNA-Adapt, SamplingDesign, GREED-RNA, SAMFEO, SentRNA, and
        RNAiFold-LNS implementations expose no admitted external-sequence
        initialization interface. Their native results are retained as
        source-faithful baselines. RNAiFold-LNS emits a terminal sequence only
        upon finding an exact solution, so Jaccard and MCC are unavailable for
        unsolved runs and are shown as dashes. DRAG-EVO is unavailable from the
        official source, and RNA-FMQA was not evaluated because the official
        Amplify credential was unavailable. All TTS values measure
        post-initialization search rather than strict process-launch-to-answer wall
        time.
      \end{minipage}

      \par
  \endgroup

  \clearpage
  \subsection{Extended Eterna100-v2 campaign}
  \label{app:eterna_extended}

  \begingroup
  \normalsize

  To assess achievable coverage beyond the matched 60-minute evaluation, we continued only the seed-2718 KnitNet state, which had solved 35 targets at initialization and 47 more during the original search hour. The extended campaign used deterministic, RNG-divergent continuations and restarts from this frozen state, stopping each target at its first independently verified ViennaRNA-exact solution. It added 14 solutions through extended search and one through a target-specific redesign of KnitNet's own near-misses, yielding $97/100$ cumulative single-seed coverage. The campaign used up to 576 one-thread CPU workers and reached the 97th solution after approximately 149 active allocation hours, including the source hour. All 97 sequences were frozen before independent-predictor evaluation; the three unresolved targets were excluded rather than replaced by external solutions.

  \begin{figure}[htbp]
      \centering
      \color{violet}
      \captionsetup{
          font=footnotesize,
          justification=justified,
          singlelinecheck=false,
          skip=6pt
      }
      \includegraphics[
          width=1\linewidth,
          height=0.5\textheight,
          keepaspectratio
      ]{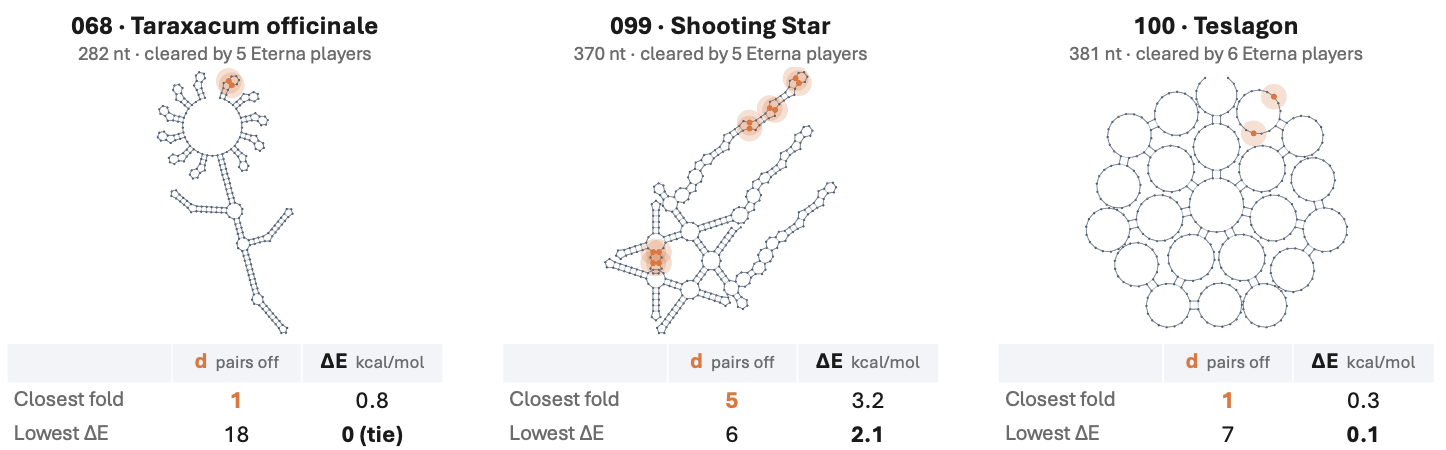}
      \captionsetup{}
        \caption{%
            \textbf{Residual errors for the three Eterna100-v2 targets unresolved by the extended KnitNet campaign.} Orange marks pairing discrepancies in the closest design under the pinned ViennaRNA 2.6.4
            setting. Here $d=|T\triangle F(x)|$ is the number of missing or unwanted pairs, and $\Delta E=G(x,T)-G(x,F(x))$ is the target's energy above the returned MFE structure. For each target, the first row
            shows the closest verified fold and the second shows the design with the smallest observed energy gap. Target 068 reaches an MFE tie, but ViennaRNA's deterministic traceback returns another co-optimal
            structure, so it does not satisfy the strict exact-match criterion.}
        \label{fig:eterna_remaining_targets}
    \end{figure}

  The remaining targets are close in pair space but retain different
  thermodynamic obstacles under the pinned oracle
  (Figure~\ref{fig:eterna_remaining_targets}). Target 068 is one missing pair
  from the target in its closest fold; a separate design reaches
  $\Delta E=0$, but the target is one of six co-optimal structures and is not
  the structure returned by ViennaRNA. Target 099 remains five missing pairs
  away, while target 100 differs by one unwanted pair and lies only
  $0.3\,\mathrm{kcal\,mol^{-1}}$ above its competing MFE fold. Diagnostic
  analyses associate the residual errors in 068 and 099 with isolated target
  pairs adjacent to one-nucleotide bulges, and identify a competing hairpin
  near Teslagon's terminal interior-loop module. These observations localize
  the remaining discrepancies but do not establish a unique causal mechanism
  or prove that the targets are intrinsically unsolvable.

  Future work will prospectively examine how alternative thermodynamic settings and co-optimal-structure policies affect these residual cases. Such evaluations will be reported separately from the benchmark result, which remains defined by the frozen ViennaRNA 2.6.4, Turner-2004, $37\,^{\circ}\mathrm{C}$, dangles-2, first-traceback criterion.

  \par
  \endgroup
  \FloatBarrier